\documentclass[10pt]{article} % For LaTeX2e
\usepackage[accepted]{tmlr}
\usepackage[utf8]{inputenc}
\usepackage{amsmath,amsfonts}
\usepackage{amsthm}
\usepackage{latexsym}
\usepackage{bm}
\usepackage{graphicx}
\usepackage{booktabs}
\usepackage{multirow}
\usepackage{tabularx}
\usepackage{xltabular}
\usepackage{enumitem}
\usepackage{wasysym}
\usepackage{microtype}
\usepackage{forest}
\useforestlibrary{edges}
\usepackage[most]{tcolorbox}
\usepackage{hyperref}
\usepackage{url}

\usepackage{makecell}

\def\openreview{\url{https://openreview.net/forum?id=GRqtQL3pnd}}
\def\month{11}
\def\year{2026}

\definecolor{claimMath}{RGB}{0,90,180}
\definecolor{claimEmp}{RGB}{0,128,96}
\definecolor{claimInterp}{RGB}{180,90,0}
\DeclareRobustCommand{\Mmark}[1]{{\color{claimMath}#1}\textsuperscript{M}}
\DeclareRobustCommand{\Emark}[1]{{\color{claimEmp}#1}\textsuperscript{E}}
\DeclareRobustCommand{\Amark}[1]{{\color{claimInterp}#1}\textsuperscript{A}}
\DeclareRobustCommand{\Msym}{\Mmark{\CIRCLE}}
\DeclareRobustCommand{\Esym}{\Emark{\CIRCLE}}
\DeclareRobustCommand{\Asym}{\Amark{\CIRCLE}}
\DeclareRobustCommand{\claimlegend}{Colored circle marks distinguish claim type:
  \Msym\ = mathematically derived;
  \Esym\ = empirically supported by the surveyed literature;
  \Asym\ = authors' analytical interpretation.}

\let\cite\citep

\definecolor{defaccent}{RGB}{0,90,180}
\definecolor{deftint}{RGB}{247,249,252}
 
\makeatletter
\newtheoremstyle{defstyle}{6pt}{6pt}{\normalfont}{0pt}
  {\bfseries\color{defaccent}}{.}{ }
  {\thmname{#1}\thmnumber{ #2}\thmnote{ \normalfont\bfseries\color{defaccent}(#3)}}
\makeatother
 
\newcommand{\highlight}[1]{\textbf{\textit{#1}}}
\newcommand{\provenance}[1]{%
  \vspace{-0.4em}%
  {\small\noindent\textbf{\textit{Provenance.}}~#1\par}%
  \vspace{0.4em}%
}

\title{Toward Fair Speech Technologies: A Comprehensive Survey of Bias and Fairness in Speech AI}

\author{%
  \name Yi-Cheng~Lin \email f12942075@ntu.edu.tw \\
  \name Yun-Shao~Tsai$^\dagger$ \email r14942093@ntu.edu.tw \\
  \name Kuan-Yu~Chen$^\dagger$ \email f13942135@ntu.edu.tw \\
  \addr Graduate Institute of Communication Engineering, National Taiwan University
  \AND
  \name Hsiao-Ying~Huang$^\dagger$ \email monicahuangml1@gmail.com \\
  \addr Graduate Institute of Electrical Engineering, National Taiwan University
  \AND
  \name Huang-Cheng~Chou$^\dagger$ \email huangchengchou@gmail.com \\
  \name Shrikanth Narayanan \email shri@usc.edu \\
  \addr SAIL, Ming Hsieh Department of Electrical and Computer Engineering, University of Southern California
  \AND
  \name Yu~Tsao \email yu.tsao@citi.sinica.edu.tw \\
  \addr Research Center for Information Technology Innovation, Academia Sinica
  \AND
  \name Jian-Jiun~Ding \email jjding@ntu.edu.tw \\
  \addr Graduate Institute of Communication Engineering, National Taiwan University
  \AND
  \name Hung-yi~Lee \email hungyilee@ntu.edu.tw \\
  \addr AI Center of Research Excellence, National Taiwan University}

\begin{document}

\maketitle
{\renewcommand{\thefootnote}{\fnsymbol{footnote}}\footnotetext[2]{Equal second contribution.}}
{\renewcommand{\thefootnote}{}\footnotetext{The authors used large language models (Claude, Gemini) to assist with language editing.}}

\begin{abstract}
Speech technologies are deployed in high-stakes settings, yet fairness concerns remain fragmented across tasks and disciplines.
Existing surveys either adopt a general machine-learning perspective that overlooks speech-specific properties or focus on a single task, missing failure patterns shared across the speech domain. 
Synthesizing over 400 studies spanning generation and perception tasks and emerging speech large language models (SpeechLLMs), this survey presents a unified framework that links formal fairness definitions to evaluation, diagnosis, and mitigation. 
We formalize seven fairness definitions adapted to the speech modality and organize the field's conceptual expansion through three paradigms: Robustness, Representation, and Governance. 
We then ground evaluation metrics in the mathematical cores of these definitions, organizing them into six families and mapping each family back to the definitions it operationalizes.
We diagnose bias sources along the speech processing pipeline, surfacing speech-specific mechanisms such as channel bias as a demographic proxy and annotation subjectivity in emotion labels. 
We systematize mitigation strategies across four intervention stages, mapping each to the diagnosed sources. Finally, we identify open challenges and propose directions for future research.
\end{abstract}

\section{Introduction}
\label{sec:intro}
Over the past few decades, speech processing technologies, encompassing Automatic Speech Recognition (ASR) \cite{yu2016automatic, li2022recent, su-etal-2024-task, lin2025pseudo2real}, Text-to-Speech (TTS)\cite{lin2025you, huang2022meta, kaur2023conventional}, Speech Emotion Recognition (SER) \cite{madanian2023speech, wu2024emo}, Speaker Verification (SV) \cite{1454425, zhang22h_interspeech}, speech large language models (SpeechLLMs) \cite{cui-etal-2025-recent, arora2025on, huang2025dynamicsuperb, yang-etal-2025-towards-holistic}, and so on, have transitioned from controlled laboratory environments to pervasive deployment in real-world settings. 
Traditionally, progress in this field has been driven by engineering-centric metrics, such as minimizing Word Error Rate (WER) \cite{chen2025cantoasr}, optimizing Real-Time Factor (RTF), or enhancing Mean Opinion Score (MOS) \cite{huang2025advancing, lin2025mmmos, ren2025highratemos, ritter2025astar}. 
However, as these technologies increasingly permeate high-stakes domains such as healthcare \cite{larasati2025inclusivity}, criminal justice \cite{loakes2022does}, recruitment \cite{hickman2024automated}, and personal assistance \cite{10.1145/3637337}, purely technical performance indicators are no longer sufficient to capture the complex impacts these systems have on human society \cite{10.1145/3287560.3287598}.

This inadequacy stems from a fundamental shift in the nature of deployment: speech technologies have evolved from supportive tools into institutional gatekeepers. 
When speech systems are deployed in high-stakes environments, they do not merely process data; they distribute resources and recognition \cite{7163127, cheng2011automatic, doi:10.1177/20539517241297889}.
A failure in these contexts is not random noise but often a reproduction of existing societal inequities, privileging dominant acoustic profiles while systematically erasing or pathologizing marginalized voices.
Beyond mere failure, models that enforce rigid, standard-language norms inadvertently demand acoustic assimilation, treating diverse cultural identities as deviations to be corrected \cite{choi2025fairness}.
Therefore, fairness cannot be treated as a post-hoc optimization step but must be understood as a core reliability requirement \cite{Cunningham_Adjagbodjou_Basoah_Jawara_Kadoma_Lewis_2025, zolnoori2024decoding}. 
Crucially, achieving this in the speech domain requires navigating physical constraints absent in other modalities; it demands addressing the unique properties of the speech signal that make fairness uniquely difficult to enforce \cite{BENZEGHIBA2007763}.

\subsection{Uniqueness of speech modality in fairness}
Speech presents distinct challenges for fairness because sensitive attributes are temporally and acoustically entangled with the linguistic content \cite{10603395, 9003979}. 
In text, linguistic units are discrete and editable, allowing sensitive terms to be removed at the token level \cite{he-etal-2021-detect-perturb, 10.1145/3287560.3287572, sun-etal-2019-mitigating}. 
In images, sensitive cues such as faces or symbols may be partially localized, enabling region-level masking without necessarily destroying the entire visual context \cite{meister2023gender}. 
Speech, by contrast, encodes lexical content, speaker identity, accent, affect, and other attributes jointly across time and frequency.  
Because these features are fused in the continuous signal, naive attempts to remove sensitive variables often degrade downstream utility or eliminate essential pragmatic cues, leaving no direct analogue to strategies such as removing a gender token or cropping a face.

% Speech also uniquely introduces interactional fairness challenges that are largely absent in text and image modalities. 
% Spoken interaction requires models to infer turn-taking, handle overlapping speech, and perform speaker diarization to determine who spoke and when \cite{castillo-lopez-etal-2025-survey}. 
% In text, authorship and turn boundaries are typically explicit, while images are static and do not require temporal attribution of speakers \cite{li-etal-2017-dailydialog}. 
% In contrast, errors in speech diarization can systematically misattribute, merge, or omit speakers \cite{ryant19_interspeech}. 
% As a result, fairness failures in speech systems can occur prior to content recognition, marginalizing speakers before their speech is even processed, evaluated, or acted upon.

Speech is also fundamentally temporal.
Spoken interaction unfolds over time, so a system must decide who is speaking and when before it can interpret what was said \cite{castillo-lopez-etal-2025-survey}.
Diarization and endpointing rely on assumptions about turn structure and pause length that hold for typical adult conversation but not for everyone \cite{ryant19_interspeech}: a person who pauses mid-utterance for cognitive reasons is cut off as if finished, and children or overlapping talkers are merged or dropped.
Because these assumptions track particular populations, the resulting errors are not random but concentrate on the same groups.
These failures concern who is heard at all, not just how accurately their words are transcribed, so a speaker can be disadvantaged even when the system would have recognized their speech correctly.

\subsection{Difference from Prior Surveys}
Existing surveys on fairness and bias fall into three broad categories: \textit{general}, \textit{Natural Language Processing (NLP)-focused}, and \textit{speech-focused}. General fairness surveys \cite{mehrabi2021survey, 10.1145/3494672, Oneto2020, 10.1145/3597199} provide foundational definitions and mitigation taxonomies but do not account for speech-specific properties such as acoustic entanglement or cascading errors across speech pipelines. NLP-focused fairness surveys \cite{blodgett-etal-2020-language, gallegos2024bias, bansal2022survey, sun-etal-2019-mitigating} address language-related bias yet treat text as the primary modality, overlooking the additional challenges introduced by continuous acoustic signals. Speech-focused reviews, in turn, often concentrate on a single task (e.g., ASR \cite{hinsvark2021accented, 10.1007/978-3-031-21707-4_30, Cunningham_Adjagbodjou_Basoah_Jawara_Kadoma_Lewis_2025, 10.1145/3769089, martin2022}, health care~\cite{larasati2025inclusivity}) or a narrow subset of risk dimensions (e.g., the privacy-fairness interplay \cite{leschanowsky2024examining}). 

Our survey differs from these in two respects. In \textit{scope}, we jointly address speech production and perception tasks, as well as emerging SpeechLLMs, enabling a cross-task perspective that reveals shared failure patterns invisible to single-task reviews. In \textit{depth}, rather than descriptively cataloging known biases, we formalize seven definitions of fairness for the speech modality and build each subsequent layer of analysis on the preceding one: evaluation metrics are derived primarily from the mathematical cores of these definitions, bias sources are diagnosed along the speech processing pipeline, and mitigation strategies are mapped back to the identified sources.

\begin{figure}[tbp]
    \centering
    \includegraphics[width=1.00\linewidth]{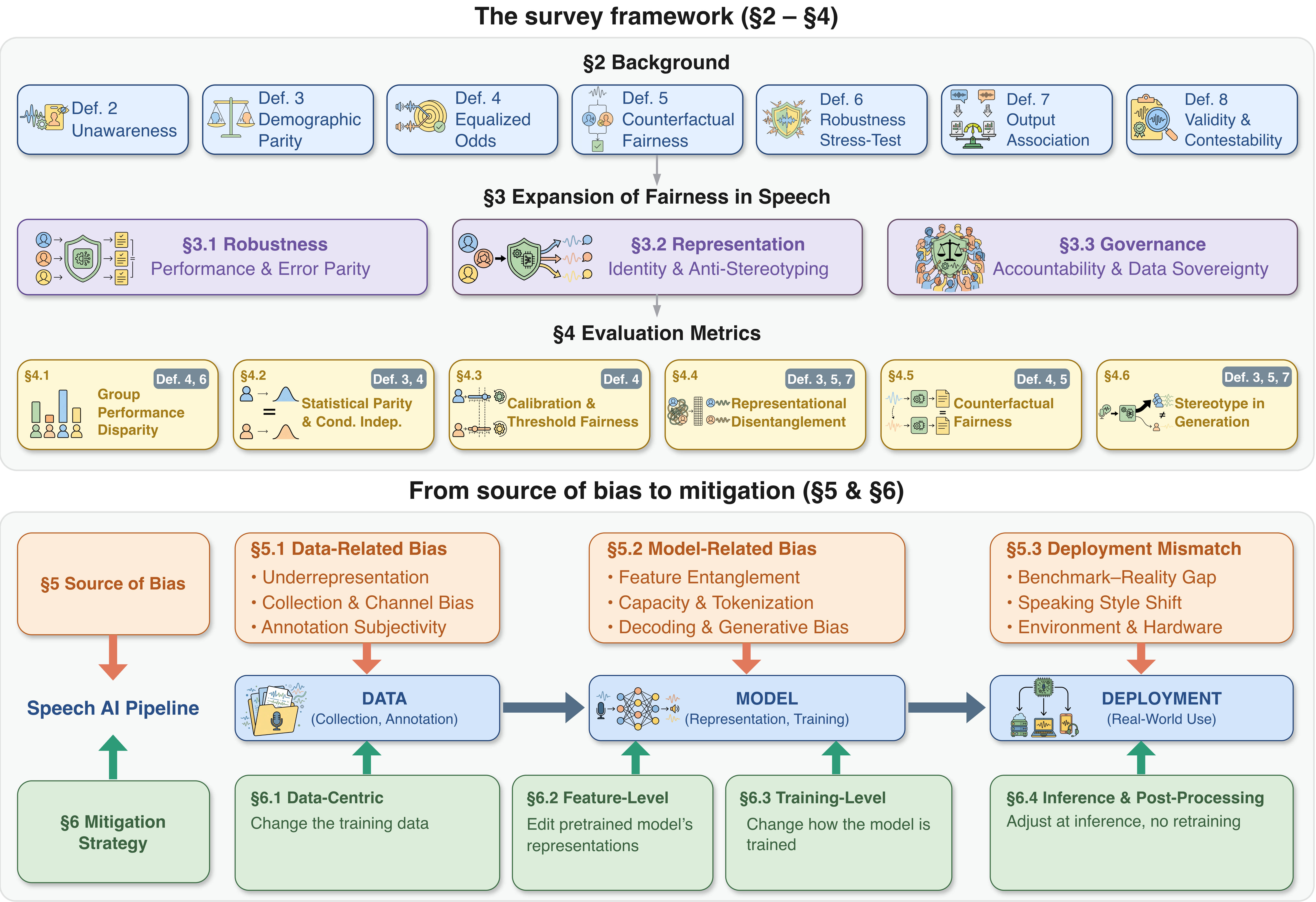}
    \caption{Overview of the survey, read top to bottom. The \textbf{upper half} is the framework: seven fairness definitions (\S\ref{sec:background}), three paradigms (\S\ref{sec:expansion}), and six metric families (\S\ref{sec:evaluation}). The \textbf{lower half} follows the speech processing pipeline, where each bias source (\S\ref{sec:source}) sits directly above the mitigation that addresses it (\S\ref{sec:mitigation}); \S\ref{sec:challenge} lists the open problems.}
    \label{fig:overview}
\end{figure}

\subsection{Main Contributions}
To address the fragmentation in existing literature, this survey constructs a unified framework that connects formal definitions of fairness to computable metrics, bias diagnosis, and mitigation strategies across speech technologies. The four contributions below are organized as a coherent analytical pipeline: we first formalize \textit{what fairness means} in the speech context, then specify \textit{how to measure} it, \textit{where bias originates}, and \textit{how to mitigate} it, as summarized in Figure~\ref{fig:overview}.
\begin{itemize}
    \item \textbf{Formal Definitional Framework} (Sections~\ref{sec:background} and~\ref{sec:expansion}): We formalize seven definitions of fairness adapted to the speech modality, spanning classical statistical notions and speech-specific desiderata. We further organize the conceptual expansion of speech fairness through three paradigms: Robustness, Representation, and Governance.

    % \item \textbf{Grounded Evaluation Metrics} (Section~\ref{sec:evaluation}): We organize evaluation metrics into six families defined by their shared mathematical objective, and map each family back to the fairness definitions it operationalizes, making explicit which metric answers which fairness question.
    \item \textbf{Grounded Evaluation Metrics} (Section~\ref{sec:evaluation}): We organize evaluation metrics into six families, and map each family back to the fairness definitions it operationalizes, making explicit which metric answers which fairness question.

    \item \textbf{Speech-Specific Bias Diagnosis} (Section~\ref{sec:source}): We instantiate a pipeline-aware bias taxonomy, covering data, model, and deployment stages, with speech-specific mechanisms such as channel bias as a demographic proxy, annotation subjectivity in emotion labels, and decoding assumptions that penalize atypical speech.

    \item \textbf{Mitigation Strategies} (Section~\ref{sec:mitigation}): We systematize intervention techniques across four stages (data-centric, feature-level, training-level, and inference-stage), mapping each to the bias sources diagnosed above.
\end{itemize}

% §1.4 Survey Scope and Methodology
% 新增文字以藍色標示，方便共同作者辨識
% 需要在 preamble 加入：\usepackage{xcolor}

\subsection{Survey Scope and Methodology}

This survey follows an iterative, framework-driven search strategy.
We describe the process below to support transparency and reproducibility of scope.

\textbf{Search procedure.}
We queried Google Scholar, the ACL Anthology, IEEE Xplore, the ISCA Archive, and arXiv using keyword combinations drawn from three axes:
\emph{(i)}~task terms (\emph{speech recognition, ASR, text-to-speech, TTS, speech synthesis, speaker verification, speaker recognition, speaker diarization, speech emotion recognition, voice conversion, speech translation, spoken dialogue, speech enhancement, pronunciation assessment, pathological speech, dysarthric speech, voice disorder detection, speech language model});
\emph{(ii)}~fairness terms (\emph{bias, fairness, equity, disparity, discrimination, inclusive, underrepresented});
and \emph{(iii)}~demographic terms (\emph{accent, gender, age, race, dialect, disability, dysarthria, child speech, non-native}).
The search covered publications available before June 2026.
From the initial results we performed forward and backward citation tracking (snowballing) to capture works not surfaced by keyword search alone.

\textbf{Inclusion and exclusion criteria.}
A study was included if it satisfies both of the following conditions:
\begin{itemize}
  \item \textbf{Speech signal involvement.}
  The study must address a task in which a speech signal is either the primary input or the primary output.
  We exclude text-only NLP studies even when they address language-related fairness.
  \item \textbf{Fairness relevance.}
  The study must address fairness, bias, or equity in speech technologies, whether through analyzing performance disparities across speaker populations, examining how speech systems encode or perpetuate social stereotypes, or investigating the ethical and governance dimensions of speech data collection and use.
  Work that benchmarks a speech system without examining any of these concerns falls outside our scope.
\end{itemize}

% The rest of this survey is structured to guide the reader from theoretical foundations to practical interventions. Section~\ref{sec:background} establishes the theoretical foundations by defining bias and fairness, while Section~\ref{sec:evolution} traces the historical evolution of these concepts in speech processing. Moving to practice, Section~\ref{sec:evaluation} surveys metrics for quantifying disparities, followed by Section~\ref{sec:source}, which analyzes the root causes of bias across the pipeline. Section~\ref{sec:mitigation} systematizes mitigation strategies into data, model, and inference-level approaches. Finally, Section~\ref{sec:challenge} identifies open challenges and directions for future research.

\section{Background}
\label{sec:background}
Current academic and industrial trends indicate a \highlight{sociotechnical turn} in speech technology research \cite{10.1145/3287560.3287598}. 
Scholars and practitioners are increasingly recognizing that technical systems do not operate in a vacuum \cite{blodgett-etal-2020-language}; rather, they are deeply embedded within social structures \cite{10.1145/3442188.3445896, BIRHANE2021100205}.
Consequently, these systems often exhibit significant performance disparities when interacting with diverse groups defined by race, gender, accent, or disability \cite{Koenecke_2020, Markl_2022}. 
To address these challenges, the academic community has begun to introduce and rigorously redefine two core concepts: Bias and Fairness.

\subsection{Social Bias}
\label{ssec:social_bias}
While \textit{bias} is a ubiquitous term in machine learning research, its definition remains frequently ambiguous and under-theorized. 
In traditional engineering contexts, the term \textit{bias} typically refers to the error caused by simplifying a complex real-world problem into a model \cite{Brynjarsdóttir_2014, 6797087, hastie2009elements}. 
However, this technical definition is insufficient for capturing the ethical stakes of speech technology. 
Technical literature often fails to articulate exactly the harmful system behaviors, the populations they impact, and the normative ethical frameworks that justify these concerns \cite{blodgett-etal-2020-language}. 
Without this normative grounding, \textit{bias} risks being treated merely as a dataset imbalance to be corrected rather than a mechanism of social power that requires structural intervention. 
Therefore, this paper adopts the specific term \highlight{Social Bias} to describe systematic and unfair discrimination against individuals or groups based on socially salient characteristics such as race, gender, accent, or ability. 
By focusing on social bias, we move beyond technical errors to examine how speech technologies reflect and reinforce existing societal hierarchies.

\subsubsection{The Mechanisms of Social Bias}
% Social bias in speech systems is not a singular artifact, but rather the result of complex mechanisms operating in tandem \cite{lopez2021bias, dobbe2018broader}. 
% We utilize the sociotechnical framework established by Friedman and Nissenbaum, categorizing social biases into the interaction between three mechanisms \cite{10.1145/230538.230561}. 

% \textbf{Pre-existing Bias.} This refers to social biases that exist in the world independently of the system but are absorbed into it. 
% In speech processing, this is most visible in the \textit{Standard Language Ideology}, the societal norm that elevates certain dialects (e.g., General American English) as correct while stigmatizing others (e.g., African American English (AAE) or regional accents \cite{wolfram2015american}) as improper \cite{lippi2012english}. 
% When training data reflects these societal prejudices, the resulting models inherit the belief that non-standard speech is merely noise rather than a valid form of communication \cite{tatman2017gender}.
Following the sociotechnical framework of Friedman and Nissenbaum \cite{10.1145/230538.230561}, we distinguish three mechanisms by which social bias enters a speech system \cite{lopez2021bias, dobbe2018broader}, ordered by when they act in the system's life cycle.
\textbf{Pre-existing Bias} is absorbed from society before any engineering begins.
Its clearest instance in speech is the \textit{Standard Language Ideology}, the societal norm that treats dominant varieties such as General American English as correct while stigmatizing others, such as African American English (AAE), as improper \cite{lippi2012english, wolfram2015american}. 
Speech technology inherits this norm through training corpora that favor prestigious varieties~\cite{markl-mcnulty-2022-language}, and ASR systems often make more errors for speakers of stigmatised varieties \cite{sanabria2023edinburgh, scott-etal-2026-sociophonetic} such as AAE~\cite{chang24d_interspeech, Markl_2022}.

\textbf{Technical Bias} arises during design, when seemingly neutral engineering choices carry discriminatory effects. One example is telephony codecs, which degrade the quality of female speech more than male speech~\cite{Altwlkany_2025}.
\textbf{Emergent Bias} appears after deployment, when the usage context diverges from what designers assumed, as with a system trained on read speech encountering spontaneous or dysarthric speech \cite{doi:10.1044/2024_JSLHR-24-00045}.
Section~\ref{sec:source} traces how these mechanisms materialize along the speech pipeline, from data collection through model design to deployment.

% \textbf{Technical Bias.} This arises from technical constraints or engineering decisions that may seem neutral but have discriminatory effects. For example, standard audio compression algorithms (codecs) used in telephony and VoIP have been found to systematically degrade the signal quality of female voices more than male voices \cite{Altwlkany_2025}. Because these codecs are often optimized for the lower fundamental frequencies typical of adult males, they struggle to encode the higher pitch and harmonic density of female speech efficiently \cite{ma2010electroglottographic, fitch1990consistency}. This technical constraint creates a performance ceiling that no amount of downstream model training can fully correct, as the raw input signal itself is compromised.

% \textbf{Emergent Bias.} This occurs when a system is deployed in a context significantly different from its design environment. 
% A speech recognition system trained on read speech may exhibit emergent bias when used by populations who rely on spontaneous, conversational speech patterns, or when used by individuals with speech impairments (e.g., dysarthria) whose acoustic characteristics were never anticipated during the system's training phase \cite{doi:10.1044/2024_JSLHR-24-00045}.

\subsubsection{The Nature of Harm}
To understand the impact of these mechanisms, it is necessary to distinguish between the types of damage they cause. Following ~\citet{barocas2017problem}, we distinguish two types of harm.

{\small
\renewcommand{\arraystretch}{1.5}
\begin{xltabular}{\textwidth}{@{}l p{0.22\textwidth} X@{}}
\caption{Taxonomy of Allocative Harms in Speech Processing}
\label{tab:speech_allocative_harms}\label{tab:taxonomy_allocative_harms}\\
\toprule
\textbf{Harm Type} & \textbf{Definition} & \textbf{System Manifestations in Speech Technology} \\
\midrule
\endfirsthead
\toprule
\textbf{Harm Type} & \textbf{Definition} & \textbf{System Manifestations in Speech Technology} \\
\midrule
\endhead
\bottomrule
\endfoot
        \textbf{Service Disparity} &
        The system functions with significantly lower accuracy, reliability, or utility for specific social groups, imposing a functional tax or effective denial of service. &
        \begin{itemize}[leftmargin=0pt, nosep, before=\vspace{-\baselineskip}]
            \item \textbf{High Error Rates:} ASR systems exhibit significantly higher WERs for racial minorities (e.g., AAE speakers) and regional dialects compared to ``Standard'' speakers, rendering the technology unusable \cite{Koenecke_2020, tatman2017gender}.
            \item \textbf{Disability Exclusion:} Voice interfaces such as voice assistants often fail catastrophically for speakers who stutter, which effectively locks people with speech disabilities out of the ``voice-first'' digital ecosystem \cite{sridhar2025jjj, 10.1145/3544548.3581224, mitra21_interspeech}.
        \end{itemize} \\
        \midrule
        \textbf{Opportunity Loss} &
        The system serves as a barrier to formal opportunities, such as employment, housing, or financial services, by unfairly filtering candidates based on vocal characteristics. &
        \begin{itemize}[leftmargin=0pt, nosep, before=\vspace{-\baselineskip}]
            \item \textbf{Hiring Discrimination:} In automatically scored job interviews, commercial ASR transcribes second-language speakers less accurately than first-language speakers~\cite{hickman2024automated}.
            \item \textbf{Educational Gatekeeping:} English proficiency tests have incorporated automated speaking assessments. Test takers’ first-language backgrounds affect automated scores \cite{loukina-etal-2019-many, kwako-etal-2023-bert}; such biases can make it more difficult for certain groups to meet thresholds required for admission or scholarships.
        \end{itemize} \\
        \midrule
        \textbf{Economic Loss} &
        The system's failure or bias results in the deprivation of existing assets, direct financial costs, or reduced productivity (efficiency tax) for specific groups. &
        \begin{itemize}[leftmargin=0pt, nosep, before=\vspace{-\baselineskip}]
            \item \textbf{Financial Lockout:} Voice authentication systems assume vocal stability over time. As voices change with age or health, verification accuracy degrades over the interval between enrollment and testing~\cite{kelly14_interspeech, 10.1007/978-3-642-19530-3_11, ai2025voxaging}, triggering false rejections that lock users out of essential financial accounts.
            \item \textbf{Labor Displacement:} Generative voice technologies (TTS/VC) decouple vocal identity from the physical speaker.
            The unauthorized extraction of professional voice actors' data to train high-fidelity clones leads to direct labor substitution, as market-ready synthetic voices replace human labor in media production, devaluing the original speakers' biological assets and rights to monetization \cite{sharma2025prac3, hutiri2024notmyvoice, berkowitz2025look}.
        \end{itemize} \\
        \midrule
        \textbf{Physical Risk} &
        The system's failure to function in critical scenarios can result in physical harm, compromised health, or a delayed emergency response. &
        \begin{itemize}[leftmargin=0pt, nosep, before=\vspace{-\baselineskip}]
            \item \textbf{Clinical Misdocumentation:} ASR used for medical dictation exhibits higher error rates for accented speech~\cite{10.1162/tacl_a_00627, Fatapour2025.08.29.25333548}, and such errors include wrong drug names and dosages that could lead to patient harm~\cite{hodgson2016risks}.
        \end{itemize} \\
\end{xltabular}
}

\textbf{Allocative Harms.} Allocative harms occur when a system unfairly distributes or withholds resources, opportunities, or information among different social groups. We categorize these harms based on the specific nature of the deprivation. A detailed taxonomy of these allocative failures specifically within the speech domain is summarized in Table~\ref{tab:speech_allocative_harms}.

{\small
\renewcommand{\arraystretch}{1.5}
\begin{xltabular}{\textwidth}{@{}l p{0.22\textwidth} X@{}}
\caption{Taxonomy of Representational Harms in Speech Processing. This framework encompasses emerging harms associated with generative voice and recognition systems.}
\label{tab:speech_representational_harms}\\
\toprule
\textbf{Harm Type} & \textbf{Definition} & \textbf{System Manifestations in Speech Technology} \\
\midrule
\endfirsthead
\toprule
\textbf{Harm Type} & \textbf{Definition} & \textbf{System Manifestations in Speech Technology} \\
\midrule
\endhead
\bottomrule
\endfoot
        \textbf{Stereotyping} &
        The system reinforces essentialist, negative, or limiting generalizations about specific social groups through voice design or association. &
        \begin{itemize}[leftmargin=*, nosep, before=\vspace{-\baselineskip}, ]
            \item \textbf{Digital Authority Bias:} Voice assistants were shipped female by default and scripted as helpful, while authority is read as male~\cite{west2019d}. \citet{ross2026sexualised} report the same split in commercial voice catalogues, with male personas read as dominant and female ones as submissive.
            \item \textbf{Speaker Assignment Bias:} synthesis models implicitly infer speaker gender from text prompts~\cite{Puhach_2025}. When no speaker is specified, these systems tend to assign voices that align with occupational gender stereotypes (e.g., defaulting to female voices for "Nurse" prompts)\cite{kuan-lee-2025-gender}.
        \end{itemize} \\
        \midrule
        \textbf{Demeaning} &
        The system generates content that mocks, degrades, or diminishes the dignity of specific groups (Denigration). &
        \begin{itemize}[leftmargin=*, nosep, before=\vspace{-\baselineskip}, ]
            \item \textbf{Accent Caricature:} TTS systems generating "accented" speech that relies on offensive stereotypes or "mimicry" rather than authentic linguistic patterns, creating a representation that users feel mocks their identity \cite{Michel_2025}.
            \item \textbf{Algorithmic Stigmatization:} ASR systems misrecognize African American English grammatical features such as habitual ``be''~\cite{martin20_interspeech}, and \citet{martin2022} argue that automation bias compounds this, since those acting on biased output are more likely to blame the speaker than the system that produced the harm.
        \end{itemize} \\
        \midrule
        \textbf{Erasure} &
        The system systematically fails to recognize or represent the existence of specific groups, rendering them invisible. &
        \begin{itemize}[leftmargin=*, nosep, before=\vspace{-\baselineskip}, ]
            \item \textbf{Misgendering:} Automatic Gender Recognition (AGR) systems that enforce a rigid male/female binary, actively invalidating non-binary and transgender identities \cite{sanchez2024beyond}.
            \item \textbf{Linguistic Erasure:} ASR failures force users to alter their speech to obtain a usable response, and users read that need as a sign that the technology was not built for them~\cite{mengesha2021culturally}.
        \end{itemize} \\
        \midrule
        \textbf{Homogenization} &
        The system collapses diverse sub-groups into a single, uniform representation, erasing intra-group cultural nuances. &
        \begin{itemize}[leftmargin=*, nosep, before=\vspace{-\baselineskip}, ]
            \item \textbf{Accent Flattening:} Data curation practices that label diverse regional varieties simply as "Non-Native," leading to models that treat these distinct acoustic cultures as a monolith \cite{10.1145/3630106.3658969}.
            \item \textbf{Global English Convergence:} Multilingual and global ASR systems disproportionately optimize for a narrow set of dominant English varieties, treating linguistic diversity across World Englishes as noise rather than as structurally meaningful variation \cite{del2023accents}.
        \end{itemize} \\
        % \midrule
        % \textbf{Cultural Appropriation} &
        % The unauthorized extraction and commodification of a marginalized group's vocal identity for the benefit or entertainment of dominant groups without consent, often stripping the voice of its cultural context. &
        % \begin{itemize}[leftmargin=*, nosep, before=\vspace{-\baselineskip}, ]
        %      \item \textbf{Unauthorized Model Training:} The extraction of audio data from professional voice actors, musicians, or social media users to train TTS/VC models without consent or compensation, effectively commodifying their vocal identity \cite{hutiri2024notmyvoice}.
        %      \item \textbf{Deepfake Impersonation:} The use of voice cloning to facilitate consumer fraud (e.g., "vishing" scams) or synthesize non-consensual political speech, exploiting the trust associated with a specific person's voice to spread disinformation \cite{Mirsky_2021}.
        % \end{itemize} \\
        \midrule
        \textbf{Normative Bias} &
The system treats certain voices, speech styles, or interaction patterns as the default or normative standard, implicitly framing deviations as marked, deficient, or less appropriate. &
\begin{itemize}[leftmargin=*, nosep, before=\vspace{-\baselineskip}, ]
    \item \textbf{Unmarked Voice Standard:} ASR systems implicitly center \textit{neutral} or \textit{standard} accents (often aligned with dominant social groups) as unaccented defaults, while labeling other speech varieties as \textit{accented,} positioning them as deviations from an assumed norm \cite{10.1145/3630106.3658969}.
    \item \textbf{Interactional Norm Bias.} Endpointing models optimized for low latency on typical speech cut off individuals with dysfluency far more frequently~\cite{10.1145/3544548.3581224}, while a single timing budget cannot accommodate speakers who need more time to formulate an utterance~\cite{wu2020seewhat}.
\end{itemize} \\
\midrule
\textbf{Toxicity} &
The system generates content that is abusive, threatening, or hateful, or conversely, it systematically misidentifies neutral speech as having these characteristics due to bias against a specific identity.  &
\begin{itemize}[leftmargin=*, nosep, before=\vspace{-\baselineskip}]
    \item \textbf{Identity-Based False Positives.} Speech toxicity classifiers flag utterances containing demographic terms at higher rates than neutral speech~\cite{bell-etal-2025-role}, creating a higher risk that benign references to marginalized groups are penalized.

\end{itemize} \\
\midrule
\textbf{\makecell[l]{Cultural\\ Appropriation}} &
The unauthorized extraction and commodification of a marginalized group's vocal identity for the benefit or entertainment of dominant groups, stripping the voice of its original cultural context. &
\begin{itemize}[leftmargin=*, nosep, before=\vspace{-\baselineskip}]
    \item \textbf{Racialized Voice Personas.} Commercial voice products have been packaged using minstrel imagery, and listeners hear race in synthetic speech in the same way they do in human voices~\cite{eidsheim2019race}, allowing vendors to sell a racial identity without including members of that group.
    \item \textbf{Extractive Voice Data Use.} Speech generation models have been trained and released on Indigenous language recordings without consulting the communities or speakers involved~\cite{pine2025speech}, treating those languages as data ready for the taking~\cite{bird2020decolonising}.
\end{itemize} \\
\end{xltabular}
}

\textbf{Representational Harms.} Representational harms arise when technical systems reinforce the subordination of specific social groups along the lines of identity. Unlike allocative harms, which concern the distribution of tangible resources or opportunities, representational harms affect the social standing of a group. These harms manifest when systems stereotype, demean, or erase a community, thereby influencing how that group is perceived, understood, and valued by society at large. Table~\ref{tab:speech_representational_harms} instantiates this taxonomy for speech, integrating harm categories from~\cite{blodgett-etal-2020-language, shelby2023sociotechnical, weidinger2021ethical,  gallegos2024bias}.

\newcommand{\E}{\mathbb{E}}
\newcommand{\cX}{\mathcal{X}}
\newcommand{\cO}{\mathcal{O}}
\newcommand{\cG}{\mathcal{G}}
\newcommand{\cD}{\mathcal{D}}
\newcommand{\DeltaDist}{\Delta} % distribution space
\newcommand{\eps}{\varepsilon}

% ---- Definition environments ----
% `defstyle' (defined in tmlr.tex) sets an upright body with a bold, coloured
% run-in heading. \tcolorboxenvironment then wraps the environment in a tinted
% panel with a left accent rule, without touching any \begin{definition} body
% or any \ref{def:...} cross-reference.
\theoremstyle{defstyle}
\newtheorem{definition}{Definition}
\theoremstyle{definition}
\newtheorem{constraint}{Constraint}
\newtheorem{desideratum}{Desideratum}
 
\tcolorboxenvironment{definition}{
  enhanced, breakable, sharp corners, boxrule=0pt,
  colback=deftint, colframe=deftint,
  borderline west={2.5pt}{0pt}{defaccent},
  left=9pt, right=9pt, top=7pt, bottom=7pt,
  before skip=11pt, after skip=11pt,
}

\subsection{Fairness}
\label{ssec:fairness}
If \textit{social bias} constitutes the diagnosis of a system’s pathology, \highlight{Fairness} represents the normative goal for its cure. 
In the machine learning literature, fairness is often reduced to a mathematical constraint (such as “equal performance across demographics”). 
Yet sociotechnical scholarship cautions that treating fairness purely as a technical specification can lead to abstraction traps, where designers narrow the problem in ways that erase the social context that gives the technology its meaning \cite{10.1145/3287560.3287598}.
Related work on language technologies further emphasizes that fairness claims must be explicit about \highlight{what harm is being prevented, who is harmed, and which normative values justify the constraint}, because “fairness” is inherently value- and context-dependent rather than a single universal property \cite{gallegos2024bias}. 

Accordingly, this survey treats fairness for speech systems not as the mere absence of statistical bias but as a \highlight{sociotechnical practice of justice} \cite{10.1145/3287560.3287598}.
Because such a normative goal cannot be tested directly, it must be translated into a set of \highlight{operational constraints} (desiderata) stating which statistics, associations, or invariances a system should satisfy in a given application, and the selection of these constraints is itself a value-laden choice \cite{mehrabi2021survey}.

% Below, we formalize seven such constraints (Definitions~\ref{def:unawareness}--\ref{def:validity}), drawing on \cite{mehrabi2021survey, 10.1145/3287560.3287598, dwork2012fairness, verma2018fairness} and ordered from weak to strong; Def.~\ref{def:sys} fixes the shared notation.
% The weakest, Unawareness (Def.~\ref{def:unawareness}), requires only that protected attributes not appear as explicit inputs, but in speech, acoustic features serve as strong proxies for demographics, so this alone is insufficient.
% Def.\ref{def:dp}--\ref{def:robustness} therefore progressively strengthen the constraints from output-level statistical tests to causal and environmental invariance.
% Def.~\ref{def:output} shift from system performance to the content of generated outputs, and Def.~\ref{def:validity} asks whether the task formulation itself is valid and ethically defensible.
Definition~\ref{def:sys} first fixes the notation, and Definitions~\ref{def:unawareness}--\ref{def:validity} then formalize seven such constraints.
We begin with classical criteria from the fairness literature (Definitions~\ref{def:unawareness}--\ref{def:counterfactual}), extend them to settings specific to speech (Definitions~\ref{def:robustness} and~\ref{def:output}), and end with a constraint on how the task itself is formulated (Definition~\ref{def:validity}).
Each of these constraints is followed by a \textit{provenance} note marking it as \highlight{established} (imported from previous fairness literature), \highlight{adapted} (an established notion reformulated for speech), or \highlight{proposed} (introduced here), and naming the prior work it builds on.

%: speech technologies should not deepen existing social hierarchies and should instead reduce barriers that prevent marginalized communities from accessing reliable and respectful AI \cite{10.1145/3287560.3287598}. Fairness

\begin{comment}
\begin{definition}[Speech System and Outcomes]
\label{def:sys}
Let a speech system be a model $M_\theta$ that maps an input $X \in \cX$ to an outcome
$\hat{Y} = M_\theta(X) \in \cO$ (such as transcript, score, class label). More generally, let $\mathcal{L}(\cdot)$ denote the law (distribution) of a random variable. When $M_\theta$ is stochastic, it induces an outcome distribution $\mathcal{L}(\hat{Y})\in \DeltaDist(\cO).$ 
%, or more generally to an
% \textit{outcome distribution} $\mathcal{L}:=\DeltaDist(\cO)=M_\theta(\cX)$. % \to \DeltaDist(\cO)$.
Let $G \in \cG$ denote a \highlight{social group} variable induced by one or more protected attributes.
Let $Y \in \cO$ denote the ground-truth outcome (reference transcript, true label, or target score).
% (contextual, often socially constructed) Fairness is specified by \highlight{operational constraints} (desiderata) that state what statistics,
% associations, or invariances the system should satisfy in a given application \cite{10.1145/3287560.3287598, mehrabi2021survey}.
% (See also: LLM fairness desiderata formalization in \cite{gallegos2024biasfairnessllm}.)
\end{definition}    
\end{comment}

\begin{definition}[Speech System and Outcomes]
\label{def:sys}
Let a speech system be a model $M_\theta$ that maps an input $X \in \mathcal{X}$ to an outcome $\hat{Y} = M_\theta(X) \in \mathcal{O}$. 
The outcome space $\mathcal{O}$ covers both discriminative outputs, which are discrete or low-dimensional (a transcript, score, or class label), and generative outputs, which are variable-length speech signals produced by TTS, voice conversion, speech-to-speech translation, and speech-output speech-language models.
More generally, let $\mathcal{L}(\cdot)$ denote the law (distribution) of a random variable. 
When $M_\theta$ is stochastic, it induces an outcome distribution $\mathcal{L}(\hat{Y}) \in \Delta(\mathcal{O})$. 
Let $G \in \mathcal{G}$ denote a social group variable induced by one or more protected attributes. 
Let $Y \in \mathcal{O}$ denote the ground-truth outcome (reference transcript, true label, target score, or reference signal).
\end{definition}

% Below we provide some of the most widely used definitions, along with their explanations inspired from  \cite{mehrabi2021survey, 10.1145/3287560.3287598, dwork2012fairness, verma2018fairness} and adopt the following desiderata as speech fairness constraints/definitions: %(a system may satisfy some but not others, depending on the application and stakeholder priorities):
% A: protected attribute
% -----------------------------
% Definition 2: Fairness through Unawareness (Speech)
% -----------------------------
\par
\begin{definition}[Fairness Through Unawareness]
\label{def:unawareness}
A speech system satisfies fairness through unawareness if protected attributes are not explicitly used in the decision process.
Operationally, if $X=(X_{\text{audio}}, G)$ includes audio and explicit protected-attribute metadata $G$, unawareness requires that the system's output does not change when $G$ is removed:
\begin{equation}
\label{eq:unaware}
M_\theta(X_{\text{audio}}, G) = M_\theta(X_{\text{audio}}).
\end{equation}
% This is a weak constraint when protected attributes can be inferred from acoustic proxies (such as accent or prosody), motivating stronger invariance/robustness constraints.
In speech, withholding $G$ as an explicit input fails to prevent the model from relying on it, because accent, prosody, and voice quality often make $G$ recoverable from the audio itself \cite{singh2019profiling}; the decision can still depend on $G$ through these acoustic proxies, motivating the stronger constraints below.
\end{definition}
\provenance{Established ~\cite{dwork2012fairness, verma2018fairness,mehrabi2021survey} and adapted: we write it as invariance to removing explicit metadata.}
% (Fairness through unawareness in classic ML surveys; LLM analogue FTU: \cite{Mehrabi2021,Gallegos2024})

% -----------------------------
% Definition 3: Demographic Parity / Statistical Parity (Score/Decision Tasks)
% -----------------------------
\begin{definition}[Demographic Parity] %/ Statistical Parity
\label{def:dp}
Demographic parity requires that a system's decisions are not affected by protected attributes.
Let $\tau$ be a decision threshold chosen by the system operator, and define the binary decision $\hat{Z} = \mathbb{I}[\hat{Y} > \tau]\in\{0,1\}$ (such as accept/reject in SV or a class decision in SER). Demographic parity requires:
\begin{equation}
P(\hat{Z}=1 \mid G=g) = P(\hat{Z}=1 \mid G=g') \quad \forall g,g'\in\mathcal{G}.
\end{equation}
More generally, parity can also be defined by distributional equality: %for a score $S$ (verification score)
\begin{equation}
\label{eq:dp-dist}
\mathcal{L}(\hat{Y}\mid G=g)=\mathcal{L}(\hat{Y}\mid G=g') \quad \forall g,g'\in\mathcal{G},
\end{equation}
or by bounded discrepancy under a distance $D$ between distributions.
\end{definition}
\provenance{Established~\cite{calders2010three, dwork2012fairness} and adapted: the thresholded decision and the distributional form (Eq.~\ref{eq:dp-dist}) accommodate speech scores.}
% Classic definition: \cite{Mehrabi2021}

% -----------------------------
% Definition 4: Equalized Odds / Equal Opportunity (Score/Decision Tasks)
% -----------------------------
\begin{definition}[Equalized Odds and Equal Opportunity]
\label{def:eo}
% Let $Z =  \mathbb{I}[Y > \tau] \in \{0,1\}$ be the ground-truth binary label (derived from the same threshold $\tau$ as the predicted decision $\hat{Z}$ in \textit{Def.\ref{def:dp}}). %$Z\in\{0,1\}$ be the ground-truth label and $\hat{Z}\in\{0,1\}$ the predicted decision.
Let $Z \in \{0,1\}$ be the ground-truth binary label, and let $\hat{Z} = \mathbb{I}[\hat{Y} > \tau]$ be the predicted decision as in Definition~\ref{def:dp}.
Equalized odds requires conditional independence of $\hat{Z}$ and $G$ given $Z$:
% \begin{align*}
% P(\hat{Z}=1 \mid Z=z, G=g) = P(\hat{Z}=1 \mid Z=z, G=g') \quad 
% \forall z\in\{0,1\},\ \forall g,g'.
% \end{align*}
\[
P(\hat{Z}=1 \mid Z=z, G=g) = P(\hat{Z}=1 \mid Z=z, G=g') \quad
\]
\begin{equation}
\forall z\in\{0,1\},\ \forall g,g'.
\end{equation}
Equal opportunity is the special case for the positive class:
\[
P(\hat{Z}=1 \mid Z=1, G=g) = P(\hat{Z}=1 \mid Z=1, G=g') \quad
\]
\begin{equation}
\forall g,g'.
\end{equation}
These constraints are particularly relevant for thresholded biometric verification systems
that trade off false accepts and false rejects.
\end{definition}
\textit{Definitions~\ref{def:dp}--\ref{def:eo}} apply directly to binary or categorical classification tasks.
For sequence-to-sequence tasks such as ASR, they require adaptation, for instance by defining the output as a thresholded WER indicator or applying the constraints at the token level.
\S~\ref{sec:evaluation_metrics} provides the concrete metrics that operationalize these adapted constraints.

\provenance{Established~\cite{10.5555/3157382.3157469}; we add the remark on adapting it to sequence-to-sequence tasks.}
% Classic definition: \cite{Mehrabi2021}; SV fairness frameworks: \cite{Toussaint2021SVEvaFair}; pitfalls: \cite{Hutiri2024Pitfalls}

% -----------------------------
% Definition 5: Counterfactual / Invariance Fairness for Speech (Audio-level)
% -----------------------------
\begin{definition}[Counterfactual Fairness and its Transformation-Based Test]
\label{def:counterfactual}
% Counterfactual fairness requires that a speaker's outcome would be unchanged had that speaker's protected attribute been $g'$ rather than $g$, with all attribute-independent factors held fixed. So stated, the criterion presupposes a structural causal model linking $G$ to the generative factors of the acoustic signal \citep{kusner2017counterfactual}, and is not identifiable without one. No such model is available for speech, so the literature tests a transformation-based surrogate instead.

Counterfactual fairness requires that a model's prediction remain unchanged if a speaker's protected attribute were hypothetically shifted from $g$ to $g'$, holding all attribute-independent factors constant. 
This criterion is not directly testable on speech, so we state an operational surrogate alongside it. 
Let $T_{g\to g'}(X)$ be a transformation that replaces the $G$-linked characteristics of $X$ with those of group $g'$ while preserving task-relevant content,
and let $M_\theta : \mathcal{X} \to \mathcal{O}$ be a model mapping acoustic inputs to an outcome space, and $\psi:\mathcal{O}\times\mathcal{O}\to[0,\infty)$ be a dissimilarity function satisfying $\psi(a,b)=0$ iff $a=b$. 
The model $M_\theta$ satisfies transformation invariance up to $\varepsilon$ if:
\begin{equation}\label{eq:cf}
\psi\!\left(M_\theta(X),\, M_\theta(T_{g\to g'}(X))\right) \le \varepsilon
\qquad \forall g,g'\in\mathcal{G}.
\end{equation}
When $M_\theta$ outputs probability distributions, $\psi$ can be replaced by a distributional distance $D$:
\begin{equation}\label{eq:cf-dist}
D\!\left(M_\theta(\cdot\mid X),\, M_\theta(\cdot\mid T_{g\to g'}(X))\right)
\le \varepsilon'.
\end{equation}
\end{definition}

Counterfactual fairness presupposes a structural causal model linking $G$ to the acoustic properties of the voice~\citep{kusner2017counterfactual}.
However, no such causal model exists for speech, since the demographic traits are physically entangled with speaker identity and vocal tract acoustics, so no current method can manipulate $G$ in isolation.
In practice, $T_{g\to g'}$ is therefore instantiated by voice conversion or signal-level manipulation.
These transformations alter more than $G$ and leave detectable artifacts in the waveform.
Consequently, a violation of Eq.~\ref{eq:cf} may reflect the model's sensitivity to these conversion artifacts rather than a genuine causal dependence on $G$.

\provenance{The causal criterion is established~\cite{kusner2017counterfactual, pearl2009causal}. The Transformation-based test was defined for text~\cite{garg2019counterfactual}, and \citet{Sarı_2021} applied one to ASR as a training objective; we state it as a general criterion for any speech system and present it separately from counterfactual fairness, as a test that can be run when counterfactual fairness itself cannot be checked.}

Definitions~\ref{def:dp}--\ref{def:counterfactual} assess fairness under the recording conditions of a given dataset.
However, the same background noise or channel distortion can degrade performance more for some groups than others~\cite{Rajan_2023, shah2025srb}, so a system that is fair under one set of conditions may be unfair under another.
Definition~\ref{def:robustness} checks whether fairness still holds when the same perturbations are applied to every group's input.
% Formalized for ASR as counterfactual fairness: \cite{Sari2021CounterfactuallyFairASR}

% -----------------------------
% Definition 6: Robustness-as-Fairness (Stress-test parity)
% -----------------------------
\begin{definition}[Robustness-as-Fairness Stress-Test Constraint]
\label{def:robustness}
Let $\phi_\delta$ denote a realistic perturbation (task-irrelevant environmental/channel variations, noise) indexed by
$\delta\sim \Pi$, where $\Pi$ is a distribution over perturbation parameters. For an error/performance measure $m$, define the perturbed conditional risk:
\begin{equation}
R_m^{\text{pert}}(g)=\mathbb{E}_{\delta\sim \Pi}\ \mathbb{E}\!\left[m(M_\theta(\phi_\delta(X)),Y)\mid G=g\right].
\end{equation}
Robustness-as-fairness requires bounded group disparities under perturbations:
\begin{equation}
\left|R_m^{\text{pert}}(g)-R_m^{\text{pert}}(g')\right|\le \varepsilon \quad \forall g,g'\in\mathcal{G}, 
\end{equation}
or, as a complementary criterion, bounded disparity in degradation: 
given $R_m(g)= \mathbb{E}\!\left[m(M_\theta(X),Y)\mid G=g\right]$ and $\Delta_m(g)=R_m^{\text{pert}}(g)-R_m(g)$,
\begin{equation}
\label{eq:degrad}
\left|\Delta_m(g)-\Delta_m(g')\right|\le \varepsilon' \quad \forall g,g'\in\mathcal{G}.
\end{equation}
\end{definition}
\provenance{Adapted from the robustness disparity of \citet{nanda2021fairness}, a fairness notion that compares how robust a model is for each demographic group under worst-case adversarial perturbations; we instead consider realistic acoustic and channel perturbations, as earlier speech studies do~\cite{Rajan_2023, shah2025srb}.}
% Operationalized via automated fairness testing for ASR: \cite{Rajan2021AequeVox}

% -----------------------------
% Definition 7: Transcript/Output Association Equality (Speech-to-Text pipelines)
% -----------------------------
\begin{definition}[Output Association Equality]
\label{def:output} % for Speech-to-Text Pipelines
%When a speech system outputs text or feeds downstream language modules, 
Let $s(\hat{Y})$ be a scalar association function measured on the output, such as toxicity score, sentiment, synthesis quality, or perceived speaker characteristics. Association equality requires: % let $T\sim M_\theta(\cdot\mid X)$ denote the transcript random variable.
\begin{equation}
\left|\mathbb{E}[s(\hat{Y})\mid G=g]-\mathbb{E}[s(\hat{Y})\mid G=g']\right| \le \varepsilon \quad \forall g,g'\in\mathcal{G},
\end{equation}
or the distance $D$ between the groups' distributions of the read-out,
\begin{equation}
D(\mathcal{L}(s(\hat{Y})\mid G=g),\mathcal{L}(s(\hat{Y})\mid G=g'))\le \varepsilon \quad \forall g,g'\in\mathcal{G}.
\end{equation}
\end{definition}
While \textit{Definition~\ref{def:dp}} concerns the primary output itself, this definition targets secondary attributes derived from $\hat{Y}$ via an association function $s(\cdot)$.

\provenance{Established in NLP for text outputs~\cite{gallegos2024bias}; we extend it to speech outputs, where $s(\cdot)$ can also measure properties of the generated audio such as synthesis quality or perceived speaker characteristics.}
\begin{definition}[Validity and Contestability Constraint]
\label{def:validity}
Let $C$ denote the intended construct, and let $L$ be the operational label/target.
A minimal validity desideratum is measurement invariance across groups:
\begin{equation}
P(L \mid C, G=g)=P(L \mid C, G=g') \quad \forall g,g'\in\mathcal{G}.
\end{equation}
More broadly, the task specification should be contestable: stakeholders can interrogate whether $L$
is an ethically and scientifically defensible operationalization of $C$ even under high accuracy.
\end{definition}
\provenance{Components established~\cite{jacobs2021measurement, alfrink2023contestable}; we join them into a single desideratum over the task formulation.}

Because several of these constraints can conflict, no single metric captures ``fairness''. 
The conflict between Definitions~\ref{def:dp} and~\ref{def:eo} is the canonical case, and it holds under two non-degeneracy conditions: the base rates differ across groups ($Z \not\perp G$) and the predictor is informative about the label ($\hat{Z} \not\perp Z$)~\cite{ barocas2023fairness}. 
Both conditions are necessary: a degenerate predictor that ignores its input, e.g. an SV system that accepts every trial, or an SER classifier that emits a constant label, satisfies both criteria trivially.
When base rates are equal, any predictor that satisfies equalized odds also satisfies demographic parity, although the converse does not hold.
In practice, a useful predictor is always informative, so whenever base rates differ across groups, the two criteria cannot be satisfied at the same time.
Practitioners must therefore decide which constraints to prioritize and justify that choice based on the application context and stakeholder priorities.
% Both conditions are necessary. A degenerate predictor that ignores its input, e.g. an SV system that accepts every trial, or an SER classifier that emits a constant label, satisfies demographic parity and equalized odds simultaneously and trivially, and the two criteria also coincide when base rates are equal across groups. 
% The incompatibility is therefore a statement about \textit{useful} predictors rather than a logical contradiction; it is precisely in the regime practitioners operate in that a choice between the criteria must be made and justified. 
% Practitioners must select and justify a subset appropriate to their application context and stakeholder priorities.

\paragraph{Generative speech systems.}
Evaluating Definitions~\ref{def:unawareness} through~\ref{def:robustness} requires a way to compare outcomes.
Transcripts and labels compare directly, whereas waveforms require a summary metric, since two generations of the same sentence differ as signals while carrying identical content.
The read-out $s(\hat{Y})$ of Definition~\ref{def:output} provides this by mapping each output to a scalar, and stochastic generation makes the comparison one between distributions of $s(\hat{Y})$.
Restating a constraint on $s(\hat{Y})$ yields its generative form.
For example, Definition~\ref{def:counterfactual} (Eq.~\ref{eq:cf-dist}) requires that repeated generations from $X$ and $T_{g\to g'}(X)$ yield nearly identical distributions of $s(\hat{Y})$.
However, not every definition can be restated this way.
Definitions~\ref{def:dp} and~\ref{def:counterfactual} require only the read-out, whereas Definition~\ref{def:eo} and~\ref{def:robustness} conditions on a ground-truth label that free-form generation lacks.
The focus therefore shifts from the form of the constraint to the choice of $s$.

The choice of read-out determines which harm is tested.
When $s$ is a quality metric such as MOS, speaker similarity, or intelligibility, the constraints bound how synthesis fidelity varies across groups, so that a cloning system maintains its similarity on under-represented accents.
When $s$ reflects a perceived speaker characteristic, such as gender estimated by listeners or a classifier, the interpretation of Definition~\ref{def:output} turns on how $G$ is specified.
In voice conversion and cloning, $G$ is the target speaker's group.
Text prompt-driven generation has no target speaker, so $G$ indexes the demographic association of the prompt, requiring that the assigned voice be independent of cues such as gender-typed job terms.

Definitions~\ref{def:dp}--\ref{def:output} each require some quantity to be equal across speakers, and each leaves three choices to whoever applies them: which quantity is compared, how that comparison is structured, and whether the task definition is itself open to question. The first choice forms the core of \S\ref{sec:evaluation}, while the remaining two are discussed in the next subsection.

% If \textit{social bias} constitutes the diagnosis of a system's pathology, \highlight{Fairness} represents the normative goal for its cure. In the machine learning literature, fairness is frequently operationalized as \highlight{statistical parity}, a mathematical state where a system yields equivalent performance metrics (such as WER or Equal Error Rate (EER)) across distinct demographic groups. While parity is a necessary baseline, recent scholarship in sociotechnical systems argues that treating fairness purely as a mathematical constraint often leads to abstraction traps—failures that occur when engineers define a problem too narrowly, stripping away the social context that gives the technology its meaning \cite{10.1145/3359221}.

% Therefore, in this paper, we define fairness not merely as the absence of statistical bias, but as a sociotechnical practice of justice. This practice requires aligning the system's design with the goal of anti-subordination, ensuring that speech technologies do not deepen existing social hierarchies but actively work to dismantle the barriers that prevent marginalized communities from accessing the benefits of AI.

% \begin{figure*}
%     \centering
%     \includegraphics[width=1.0\linewidth]{figures/dimensions/temp1.png}
%     \caption{Relations between Fairness Dimensions}
%     \label{fig:fairness_dimensions}
% \end{figure*}

\subsection{Fairness Dimension}
\label{ssec:fairness_dimension}
% \label{sec:dimention}
% Speech systems pose fairness risks at multiple layers of the pipeline, so no single metric suffices. At the system-performance layer, disparities appear as \highlight{Group} gaps across demographics, \highlight{Individual} violations where similar speakers/utterances are treated differently, and \highlight{Intersectional} harms that emerge at the interaction of identities and conditions.
% Beyond outcomes, fairness can also fail upstream in how we define and justify what a speech system should predict: \highlight{Problem Formulation and Validity} asks whether the chosen target is a valid operationalization of the intended construct, or whether the task itself is ethically contested even under perfect accuracy.

% Accordingly, we organize this section into \highlight{System Performance}, and \highlight{Problem Formulation and Validity}, with system performance further structured by \highlight{Group}, \highlight{Individual}, and \highlight{Intersectional} fairness.

We organize fairness concerns along two axes.
% \highlight{System Performance} asks whether outcome quality is uneven across speakers, covering \highlight{Group} gaps across subpopulations or conditions, \highlight{Individual} violations where similar inputs receive different treatment, and \highlight{Intersectional} harms at the combination of multiple attributes.
% \highlight{Problem Formulation and Validity} asks a prior question: whether the prediction target is a valid operationalization of the intended construct, or whether the task is ethically contested even under perfect accuracy.
\highlight{System Performance} asks whether outcome quality is uneven across speakers; \highlight{Problem Formulation and Validity} asks whether the task itself should be optimized at all.

\subsubsection{System Performance Fairness}
Here we characterize how disparities are compared across speakers.
The evaluated quantity can be a conventional performance metric such as error rate, or a property measured on the output itself, such as toxicity or perceived speaker traits.
In either case, the comparison structures described below apply in the same way, differing primarily in how they partition or relate speaker populations. 
The literature examines \highlight{Group} and \highlight{Intersectional} fairness, which differ in how finely the grouping variable $G$ partitions the speakers, and \highlight{Individual} fairness, which does not group speakers at all but requires that similar speakers or utterances receive similar outcomes.

\textbf{Group Fairness:} Group fairness, the most widely studied granularity in the speech literature~\cite{mehrabi2021survey, Chen_2025, Chien_2024}, instantiates $G$ coarsely, partitioning speakers by demographic attributes (gender, age, race), linguistic attributes (accent, dialect, language). 
Because a system can serve some of these groups much worse than others, aggregate metrics such as overall WER can hide large gaps between them. 
Prior ASR evaluations show that break results down by speaker group reveals disparity patterns invisible in corpus-level summaries~\cite{feng2021quantifying, 10.1016/j.csl.2023.101567}. 
This motivates evaluation practices that report per-group performance and dispersion summaries, so that the structure of disparity is visible before any mitigation is attempted~\cite{patel2025evaluate, liu2022towards, tevissen2023towards, lin2024social}.

\textbf{Intersectional Fairness:}
% \label{ssec:intersectional_fairness}
Intersectional fairness examines combinations of attributes, recognizing that harms may concentrate in subgroups that are small, underrepresented, or acoustically distinct~\cite{gohar2023survey, zee2024group, Chen_2025}.
Intersectional gaps often arise when \highlight{multiple distribution shifts compound}, placing a subgroup at the margins of several groups that are underrepresented in or mismatched with the training distribution, where generalization is weakest. 
The resulting small sample sizes then make these concentrated errors harder to estimate and thus easier to overlook~\cite{gohar2023survey}.

Recent multilingual ASR studies highlight this compounding: disparities can persist or even widen in worst-case slices despite improving average WER, and model scaling does not guarantee better fairness when the hardest intersectional groups remain under-covered or systematically mismatched~\cite{zee2024group}.
Similarly, cross-language audits show that the same demographic factor can manifest differently across languages and architectures; phoneme-level analyses link intersectional error patterns to articulation and phone confusions that are not captured by coarse group averages~\cite{10.1016/j.csl.2023.101567}.

%     Individual fairness requires that \textit{similar} speakers or utterances receive similar outcomes\cite{Chen_2025, mehrabi2021survey, Chien_2024, dwork2012fairness, feng2023review}. In speech, violations often arise because models latch onto nuisance \highlight{variability} (such as speaking style, channel conditions, or speaker-specific expression), which is not intended to influence the decision, creating instability even within the same speaker or for the same linguistic content\cite{dwork2012fairness, verma2018fairness}. 

% This bias is evident across tasks: in SV, decisions can be unintentionally influenced by phonetic content, so two utterances from the same speaker may differ in score primarily because their phoneme inventories differ\cite{baali2024pdaf}. % (e.g., frequent/long phones)
% In SER, emotional expression is typically individual and context-dependent; population-level models can therefore work well on average while systematically failing on particular speakers whose expression patterns differ from the training norm\cite{triantafyllopoulos2024enrolment, chou2024inter}.
% In ASR, when protected attributes correlate with acoustic realizations in the data, changing those attributes (while holding linguistic content fixed) can still shift the transcription distribution, revealing that the system has entangled content with demographic-correlated cues\cite{Sarı_2021}. 
\textbf{Individual Fairness:}
Individual fairness in its general form requires that \textit{similar} speakers or utterances receive similar outcomes, replacing the partition with a similarity measure \cite{dwork2012fairness, Chen_2025, mehrabi2021survey, feng2023review}.
Speech studies operationalize this either through similarity in a learned embedding space~\cite{Chien_2024} or, more coarsely, by comparing outcomes across individual speakers~\cite{ triantafyllopoulos2024enrolment, chou2024inter}.
In SV, performance varies widely across individual speakers~\cite{hughes24_odyssey}. In SER, a model's overall accuracy can conceal very low accuracy for some individual speakers~\cite{triantafyllopoulos2024enrolment}. 
A model that is fair across groups is also not necessarily fair across individual speakers~\cite{chou2024inter}.

\subsubsection{Problem Formulation and Validity Fairness}
\label{sssec:validity}
% System performance fairness treats the task definition as fixed and asks whether outcome quality is uneven across speakers or conditions. 
% In contrast, problem formulation and validity fairness interrogate whether the target itself is a valid and ethically defensible construct to predict from speech, and whether labels/metrics faithfully operationalize that construct. 
% Many fairness harms arise when an unobservable construct (hirability, communication skill, professionalism) is inferred from convenient speech proxies, creating a mismatch between what is meant and what is measured. 
% \cite{jacobs2021measurement} formalizes this concern as \highlight{construct validity}: harms often emerge because the operationalization of a construct departs from its theoretical meaning, so “improving accuracy” can simply entrench an invalid measurement. 

% The preceding subsection treats the task definition as fixed and asks whether outcome quality is uneven across groups.
% This subsection asks a prior question: Is the task itself a valid and ethically defensible thing to optimize?
% Two failure modes arise: the task may pursue a legitimate goal through an invalid proxy (\highlight{construct validity}), or the task may target a construct that is inherently contested (\highlight{intrinsically unfair tasks}).
Definition~\ref{def:validity} asks whether the task itself is a valid and ethically defensible thing to optimize, and its failure takes two forms in speech.
A task may pursue a legitimate goal through an invalid proxy (\highlight{construct validity}), or it may target a construct that is contested in itself (\highlight{intrinsically unfair tasks}).

\textbf{Construct Validity:}
\citet{jacobs2021measurement} describe construct validity as the extent to which an operationalization yields measurements that are meaningful and useful for the construct it is meant to capture, and they argue that many fairness-related harms stem from mismatches between the two.
When such a mismatch exists, making a model more accurate only reproduces the flawed operationalization more closely.
A concrete example in speech is automated interview scoring, where ``hirability'' prediction models use prosodic cues such as pauses and intonation, sometimes explicitly recommending that candidates ``speak more fluently''~\cite{rasipuram2016automatic, naim2016automated}.
Yet fluency is not a neutral skill: employment contexts already penalize disfluent speakers, so optimizing for fluency risks laundering an existing discriminatory norm into an algorithmic target rather than measuring genuine job competence~\cite{klein2004impact}.
A similar issue exists in ASR, where evaluating systems against transcripts stripped of disfluencies \citep[e.g.,][]{mujtaba-etal-2024-lost, 10.1145/3772318.3790405} can fail to capture how people who stutter want their speech to be represented.

\textbf{Intrinsically Unfair Tasks:}
Some speech tasks are contested not because the proxy is poor but because the construct itself is ethically problematic.
\textit{Automatic Gender Recognition} (AGR) from voice reduces gender to a binary physiological inference, which systematically conflicts with transgender and non-binary identities and can impose representational harm through misgendering, even when a system reports strong aggregate accuracy~\cite{sanchez2024beyond}.
Here, improving performance does not resolve the fairness issue: the harm stems from inferring and enforcing an identity label that many people do not consent to, a problem rooted in the task definition rather than in model quality.

\medskip
The concepts introduced in this section provide the formal vocabulary for the remainder of this survey.
However, the speech fairness literature did not adopt these concepts simultaneously; different research communities have foregrounded different diagnoses of what makes a system unfair.
The next section traces three such paradigms and how they coexist in current research.

% \subsection{Inclusion}
% While fairness provides a normative target for model performance by demanding that systems function equitably across groups, \highlight{Inclusion} addresses the broader sociotechnical context of how those systems are built, whose knowledge they value, and who holds power over their deployment \cite{10.1145/3287560.3287598, 10.7551/mitpress/12255.001.0001}. We therefore define \highlight{Inclusion} as a procedural \cite{10.1145/3359284, decker2025procedural} and epistemic \cite{10.1093/acprof:oso/9780198237907.003.0002, Kay_Kasirzadeh_Mohamed_2024} commitment. It distinguishes sharply between the \textit{outcome} of the system and the \textit{process} of its creation. While fairness asks whether the system works equally well for everyone, inclusion poses a more fundamental set of questions: \textit{"Who defined how the system works? Whose values does it encode? And does it empower or extract from the communities it serves?"} \cite{liao2019enabling, 10.1145/3613904.3642810} 

% For example, an ASR system might achieve statistical parity for a marginalized dialect by scraping data without consent \cite{longpre2025bridging}, yet it remains exclusionary because it extracts value without empowering the community \cite{sharma2025prac3, doi:10.1177/10506519251348462}. In contrast, projects like Te Hiku Media \cite{Spano_Zhang_2025} demonstrate inclusion by ensuring Indigenous data sovereignty, where the community retains governance over the technology's development and deployment.

\section{Expansion of Fairness in Speech}
\label{sec:expansion}
To synthesize the surveyed literature, we propose a conceptual framework structured around three coexisting, complementary paradigms, distinguished by what each treats as the primary source of unfairness.
\highlight{Robustness} attributes unfairness to distributional mismatch, seeking to close error gaps across speaker demographics, recording conditions, and deployment environments through broader data coverage and model adaptation.
\highlight{Representation} attributes unfairness to the encoding of social meaning, asking whether learned representations and model outputs perpetuate stereotypes or erase marginalized identities.
\highlight{Governance} attributes unfairness to power asymmetries in the sociotechnical context, addressing who collects the data, who defines the task, and who is accountable for downstream harms.

\subsection{Fairness as Robustness: Bridging Performance Gaps in Recognition}
\label{ssec:robustness}

In the initial paradigm identified through our analysis (emerging prominently from the mid-2010s, and still widely practiced today), fairness was most commonly operationalized as \highlight{performance parity}~\cite{tatman2017gender}. Research focused on auditing discriminative tasks such as ASR and SV, quantifying fairness through disparities in standard error metrics, including WER and EER~\cite{Shen_2022}. Seminal work highlighted performance gaps for AAE speakers~\cite{tatman17_interspeech, Koenecke_2020}, and similar challenges were documented across other languages and dialects~\cite{yang2022open}. These disparities were largely framed as \highlight{robustness failures}: demographic differences were treated as distributional shifts that impaired generalization~\cite{hinsvark2021accented}, and the primary remedy was data diversification to narrow inter-group error gaps\footnote{We use ``robustness'' throughout this section as an overarching conceptual paradigm referring to model invariance and performance parity across demographic shifts, distinct from its narrow operationalization as a perturbation stress-test constraint in Definition~6.}.

At a deeper conceptual level, however, this robustness-oriented framing carried a strong implicit assumption: inter-group variation was treated as noise to be reduced rather than as a meaningful expression of speaker identity.
While this body of work was instrumental in establishing baseline benchmarks and making algorithmic bias empirically visible, it operationalized fairness primarily as accuracy-based parity. This framing centers on measurable error gaps rather than the social meaning or structural conditions of unequal outcomes.
As a result, broader questions concerning deployment contexts, social meaning, and representational harm remained largely outside the scope of analysis, even in cases where aggregate error rates appeared to be equalized.
The paradigms that follow take up these broader questions, widening what fairness is taken to include rather than correcting robustness.

\subsection{Fairness as Representation: Mitigating Essentialism in Profiling and Generation}
\label{ssec:representation}

As speech technologies expanded beyond transcription into tasks that infer or generate speaker characteristics, a different category of fairness risk emerged: rather than merely failing to serve certain speakers, systems may actively mischaracterize or impose speaker identities.
Accordingly, the focus of fairness research expanded beyond accuracy-centric evaluation to address essentialism, profiling, and representational harm.

\textbf{Profiling and anti-essentialism:}
The rise of paralinguistic tasks, including SER and speech-based clinical assessment, introduces new risks associated with \highlight{profiling}: systems infer latent personal attributes (such as emotional state, health conditions, or employability) from observable vocal characteristics~\cite{singh2019profiling}.
These inferences rest on an \highlight{essentialist} assumption, namely that the mapping from acoustic features to personal attributes is fixed and universal across speakers~\cite{hanna2020towards}.
In practice, however, vocal expression is culturally situated and individually variable~\cite{van2023modelling}, so a mapping learned from one population does not generalize as a neutral ground truth.
Evidence of systemic bias spans modalities: automated affective analysis has been found to misinterpret Black individuals as angrier~\cite{rhue2018racial}, while ASR performance degrades because training corpora underrepresent distinctive features of African American Language, such as habitual `be'~\cite{martin2021spoken}.
The problem is not merely lower accuracy for AAE speakers, but that the model encodes a culturally specific norm as if it were an objective link between voice and emotion.
In high-stakes contexts such as automated hiring and medical decision support, such essentialist mappings can lead to allocative harms through the systematic misinterpretation of socially meaningful vocal variation~\cite{srivastava2026cost, wu2025evaluating, kim2025domain}.

\textbf{Representation and stereotyping:}
Concurrently, the increasing ubiquity of TTS and VC technologies prompted the speech research community to initiate a critical discourse regarding representation in generated speech~\cite{hutiri2024notmyvoice}.
Unlike discriminative models, generative models actively shape the voices and interaction styles presented to users, raising concerns about representational harms.
Fairness considerations in this context, therefore, extend to whether synthesized voices reproduce or amplify existing social hierarchies and stereotypes.
For example, early deployments of voice assistants were frequently~\cite{strengers2021smart,phan2019amazon} designed with compliant female-coded vocal personas, reinforcing gendered assumptions about servitude and care work~\cite{west2019d}.
Similarly, VC and accent-altering systems have raised concerns about identity erasure, as the vocal distinctiveness of minority speakers can be normalized to standardized target voices, resulting in suppression of socially and culturally significant variation~\cite{Michel_2025, payne2024beyond}.

Profiling and generative stereotyping are two directions of the same representational failure. In both, the system treats the link between vocal signal and social identity as a fixed natural fact rather than something socially constructed. Profiling does this on the way in, reading identity from a voice, while generation does it on the way out, writing identity into a voice.

\subsection{Fairness as Governance: Navigating Systemic Risks in Sociotechnical Pipelines}
\label{ssec:governance}

Current research on fairness in speech technologies has shifted from focusing on isolated model components to examining the \highlight{sociotechnical ecosystems} in which these technologies are deployed.
Because technical architectures and design choices also shape who holds agency over a system's behavior~\cite{lessig2000code, winner1980artifacts}, we treat \textit{governance} broadly, beyond institutional regulation alone.
We organize the resulting fairness risks across three interrelated scales: the pipeline scale of accountability, the interaction scale of human--AI power dynamics, and the institutional scale of data governance.

\textbf{Pipeline scale: Composite systems and the accountability gap.}
In real-world deployments, speech technologies rarely operate as standalone modules. They are assembled into multi-stage pipelines that link recognition, language processing, and synthesis, and these components are frequently built and owned by different parties. This assembly, not any single model, is where a distinct fairness concern arises. When a harm emerges from the interaction of many components, responsibility for it is spread across \highlight{many hands}~\cite{nissenbaum1996accountability}, so no single party is answerable and the affected speaker has no clear place to seek a correction or hold anyone responsible~\cite{cooper2022accountability}. This is a part of fairness that no amount of technical accuracy can resolve~\cite{10.1145/3287560.3287598}. Cascading errors make the harm concrete. A high word error rate in an upstream transcription component can constrain a downstream language model's interpretation of spoken intent~\cite{ezema2025feels}, leaving the speaker unable to make themselves understood, yet the resulting failure belongs to no component in isolation. Improving the accuracy of any one stage does not settle who is accountable when the assembled system fails a speaker.

\textbf{Interaction scale: Digital authority and automation bias.}
Beyond system architecture, recent work also highlights the \highlight{power dynamics} embedded in human--AI interaction.
Whereas previous research emphasized stereotyping in model outputs, current analyses focus on the \highlight{digital authority} of speech-based systems, understood as users’ tendency to defer to automated voice interfaces due to their perceived neutrality or expertise.
Scholars argue that interaction design choices, including the use of standardized accents, authoritative speaking styles, or gendered personas, can function as sociotechnical control mechanisms that discourage questioning or critical engagement~\cite{10.1145/3711039}, thereby eroding user agency.
In high-stakes settings, such dynamics contribute to automation bias~\cite{cummings2017automation}, whereby system outputs are treated as directives rather than fallible recommendations, reinforcing existing social hierarchies.

\textbf{Institutional scale: From extraction to sovereignty.}
At the institutional level, early efforts to improve coverage focused on collecting speech from underrepresented accents and speaker groups.
Such collection, however, was sometimes carried out without community consent or reciprocal benefit, turning data diversification into an \highlight{extractive practice}~\cite{bird2020decolonising}.
More recent scholarship responds by emphasizing \highlight{data sovereignty}: marginalized and Indigenous communities have a right to determine how their linguistic data are collected, represented, and used~\cite{leoni2024solving, rajab2025esethu}.
Fairness is thus reframed not as a problem of maximizing data diversity alone, but as a question of reclaiming community agency through institutional legitimacy, consent, and participation.

This institutional tension is one instance of a broader pattern. The three paradigms are not mutually exclusive, and a single practice can sit differently under each. Data diversification is the clearest example. What the robustness paradigm (\S\ref{ssec:robustness}) treats as the remedy for demographic error gaps, the governance paradigm sees as an extractive practice when it is done without community consent. A system is therefore best examined through all three lenses at once, since an intervention that answers one paradigm can raise a problem under another.

% \textcolor{blue}{\textbf{Scope demarcation with Responsible AI.}
% Finally, we demarcate this framework relative to the broader Responsible-AI landscape. While Responsible AI serves as an overarching umbrella encompassing safety, privacy, and accountability, our survey centers squarely on fairness. Rather than imposing an artificial hierarchy, we distinguish between \textit{fairness-specific} concerns (e.g., error parity, group performance dispersion, and representational stereotyping) and \textit{fairness-relevant} concerns (e.g., data governance, biometric privacy, and labor displacement) that intersect with adjacent Responsible-AI pillars. We incorporate the latter because technical fairness in speech cannot be disentangled from the institutional conditions of data stewardship and deployment power dynamics.}
\subsection{Boundary with Responsible AI}
Responsible AI encompasses safety, privacy, and accountability alongside fairness, and here we set out what this survey covers.
A concern is \textit{fairness-specific} if it is about group disparities in how a system treats or portrays speakers.
Table~\ref{tab:speech_allocative_harms} and Table~\ref{tab:speech_representational_harms} collect these as allocative and representational harms.
A concern is \textit{fairness-relevant} when it influences these disparities without constituting a group difference itself, such as data governance, consent, and power asymmetries in \S\ref{ssec:governance}.
These two categories serve distinct roles in this survey.
Sections~\ref{sec:evaluation} through~\ref{sec:mitigation} focus on the metrics, bias sources, and mitigation methods for fairness-specific concerns.
Fairness-relevant concerns appear where they explain why disparities arise or constrain how they may be addressed, and \S\ref{sec:challenge} revisits them as open problems.
Broader Responsible AI topics such as adversarial robustness and model security lie outside our scope.

\section{Evaluation Metrics}
\label{sec:evaluation}
\label{sec:evaluation_metrics}

% To operationalize the fairness paradigms of \S~\ref{sec:expansion}, concrete evaluation metrics are needed.
% Because metrics that appear unrelated often share the same mathematical structure, while similar formulas can serve different normative goals, we organize metrics by their \textit{mathematical objective} rather than by paradigm or task, yielding six \textit{mathematical families} (Table~\ref{tab:fairness_metrics_master}).
% Five families group metrics around a shared formal quantity; the sixth (\S~\ref{ssec:generative_bias}) collects metrics whose forms overlap with earlier families but whose evaluation target, the content of generative output, warrants separate treatment.
% Each family is annotated with its primary paradigm affiliation (Robustness, Representation, or Governance), making explicit where the same formula serves different normative purposes.
To operationalize the fairness paradigms of \S~\ref{sec:expansion}, concrete evaluation metrics are needed.
Because metrics that appear unrelated often share the same mathematical structure, while similar formulas can serve different normative goals, we organize metrics primarily by their \textit{mathematical objective}, yielding six families (Table~\ref{tab:fairness_metrics_master}).
% Five families group metrics around a shared formal quantity; the sixth (\S\ref{ssec:generative_bias}) collects metrics whose forms overlap with earlier families but whose evaluation target, the content of generative output, warrants separate treatment.
\S\ref{ssec:group_disparity}-\ref{ssec:counterfactual} are each organized around a single target quantity, such as per-group error rates or demographic information in a representation, with metrics differing primarily in how they aggregate or condition that measure.
\S\ref{ssec:generative_bias} is unified instead by its evaluation goal, examining what a system's outputs convey about different demographic groups.
Classifying them by formula would scatter them across \S\ref{ssec:group_disparity} and \S\ref{ssec:statistical_parity} and obscure the goal they share.
Each family is annotated with its primary paradigm affiliation (Robustness, Representation, or Governance), making explicit where the same formula serves different normative purposes.

Table~\ref{tab:def_family_mapping} maps these six families back to the fairness desiderata introduced in \S\ref{sec:background}, showing which definition each family directly operationalizes (\CIRCLE) or partially reflects (\LEFTcircle).
Definitions~\ref{def:unawareness} and~\ref{def:validity} are omitted because they operate at a meta-level.
Definition~\ref{def:unawareness} (Unawareness) requires only that $G$ not be used explicitly, but in speech, acoustic features are strong proxies for demographics, so every family in this section implicitly goes beyond that baseline.
Definition~\ref{def:validity} (Validity) questions whether the task formulation and label taxonomy are themselves defensible, a concern that no single metric can operationalize but that every family should be evaluated against.
Each of the six family subsections closes with a paragraph explaining its specific connections to the desiderata.
These mappings are marked by claim type, separating what follows from the definitions, what the surveyed literature attests, and what is our own reading.

\begin{table}[t]
\centering
\small
\caption{Mapping between the fairness desiderata of \S\ref{sec:background} (Definitions~\ref{def:dp}--\ref{def:output}) and the mathematical families of \S\ref{sec:evaluation_metrics}.
\CIRCLE\ = the family directly operationalizes the definition;
\LEFTcircle\ = partial or indirect connection.
Definitions~\ref{def:unawareness} and \ref{def:validity} are meta-level constraints discussed in the text.
\claimlegend}
\vspace{8pt}
\label{tab:def_family_mapping}
\setlength{\tabcolsep}{7pt}
\renewcommand{\arraystretch}{1.2}
\begin{tabular}{@{} l c c c c c @{}}
\toprule
 & \textbf{\shortstack{Def\,\ref{def:dp}\\(DP)}}
 & \textbf{\shortstack{Def\,\ref{def:eo}\\(EO)}}
 & \textbf{\shortstack{Def\,\ref{def:counterfactual}\\(CF)}}
 & \textbf{\shortstack{Def\,\ref{def:robustness}\\(Rob.)}}
 & \textbf{\shortstack{Def\,\ref{def:output}\\(Assoc.)}} \\
\midrule
\ref{ssec:group_disparity}: Group Perf.\ Disparity   &  & \Mmark{\CIRCLE} &             & \Mmark{\LEFTcircle} &             \\
\ref{ssec:statistical_parity}: Statistical Parity       & \Mmark{\CIRCLE}     & \Mmark{\CIRCLE}     &             &             &             \\
\ref{ssec:calibration}: Calibration \& Threshold &  & \Mmark{\CIRCLE}     &             &             &             \\
\ref{ssec:disentanglement}: Representational Disent. & \Mmark{\LEFTcircle} &             & \Amark{\LEFTcircle} &             & \Amark{\LEFTcircle} \\
\ref{ssec:counterfactual}: Counterfactual Fairness  &             & \Emark{\LEFTcircle} & \Mmark{\CIRCLE}     &             &             \\
\ref{ssec:generative_bias}: Stereotyping in Generation     & \Mmark{\LEFTcircle} &  & \Mmark{\LEFTcircle} &             & \Mmark{\CIRCLE}     \\
\bottomrule
\end{tabular}
\end{table}

\begin{table*}[t]
\centering
\small
\caption{Taxonomy of fairness evaluation metrics in speech, organized by mathematical family.
Each row groups metrics sharing the same formal objective; the paradigm column indicates the normative affiliation(s) from \S\ref{sec:expansion}: Robustness (Rob.), Representation (Rep.), or Governance (Gov.).
Claim type is marked by the colored circle under each header, using the same symbols as Table~\ref{tab:def_family_mapping}.
}
\vspace{8pt}
\label{tab:fairness_metrics_master}
\setlength{\tabcolsep}{5pt}
\renewcommand{\arraystretch}{1.25}
\begin{tabularx}{\textwidth}{@{}
  >{\raggedright\arraybackslash}p{2.7cm}
  >{\raggedright\arraybackslash}p{2.9cm}
  >{\raggedright\arraybackslash}X
  >{\raggedright\arraybackslash}p{3.1cm}
  >{\raggedright\arraybackslash}p{1.35cm} @{}}
\toprule
\makecell[tl]{\textbf{Math Family}}
  & \makecell[tl]{\textbf{Unified}\\\textbf{Objective}\\[2pt]\Msym}
  & \makecell[tl]{\textbf{Representative}\\\textbf{Metrics}\\[2pt]\Esym}
  & \makecell[tl]{\textbf{Target}\\\textbf{Tasks}\\[2pt]\Esym}
  & \makecell[tl]{\textbf{Paradigm}\\[2pt]\Asym} \\
\midrule
Group Performance Disparity (\S\ref{ssec:group_disparity})
  & Quantify dispersion of $\{\mathcal{E}(g)\}_{g \in \mathcal{G}}$
  & $\Delta_{max}$,\; SNSV,\; Fairness ($F$),\; Gini,\; GARBE,\; FAAS,\; $\Delta_{\text{MOS}}$;\; \textit{indiv.:}\; Consistency,\; $L_R$
  & ASR, SV, SER, Diarization, Deepfake, Clinical, TTS\(\dagger\)
  & Rob. \\
\midrule
Statistical Parity \& Cond.\ Independence (\S\ref{ssec:statistical_parity})
  & $\hat{Y} \!\perp\! G$ (DP) or $\hat{Y} \!\perp\! G \mid Y$ (EO)
  & $\Delta_{DP}$,\; $\Delta_{EO}$,\; Equal Opportunity,\; PPR
  & SER, Clinical/Health
  & Rep. \\
\midrule
Calibration \& Threshold Fairness (\S\ref{ssec:calibration})
  & Fairness invariance across threshold~$\tau$
  & $C_{cal}$,\; ECE,\; NXE,\; FaDR,\; auFaDR-FAR,\; IR,\; GARBE
  & SV, Voice Disorder, Depression/Anxiety
  & Rob. \\
\midrule
Representational Disentanglement (\S\ref{ssec:disentanglement})
  & Quantify demographic information in learned representations
  & MI (DV bound),\; SpEAT(Cohen's\,$d$),\; Silhouette/DB/CH indices
  & SSL Embeddings, SER, SV, Pathological Speech
  & Rep. \\
\midrule
Counterfactual Fairness (\S\ref{ssec:counterfactual})
  & $\hat{Y}$ invariance under $\mathrm{do}(G)$
  & C3T fairness
  & ASR, SpeechLLMs\(\dagger\)
  & Rep. \\
\midrule
Stereotyping in Generation(\S\ref{ssec:generative_bias})
  & Content neutrality w.r.t.\ demographics
  & slbs,\; VoiceBBQ,\; $s_\mathrm{DIS}$/$s_\mathrm{AMB}$,\; $\Delta G$/$\Delta S$,\; SAS/BAAS,\; $J$/$\Delta$,\; GUS/SNSR
  & SpeechLLMs, Spoken Dialogue, ST, TTS, Text-to-Audio
  & Rep. \\
\bottomrule
\end{tabularx}
\end{table*}
%% ====================================================================
%%  IV-A  Group Performance Disparity
%% ====================================================================
\subsection{Group Performance Disparity}
\label{ssec:group_disparity}
 
This family measures how a scalar performance indicator varies across demographic groups.
Let $\mathcal{E}(g)$ denote a task-specific performance measure (e.g., WER, EER, or MOS) for group $g \in \mathcal{G}$.
The unified mathematical objective is to quantify the spread of $\{\mathcal{E}(g)\}_{g \in \mathcal{G}}$; different aggregation norms yield different metrics within the same family.
 
\textbf{Per-Group Performance Reporting.}
The most common evaluation practice in the current literature is to report $\mathcal{E}(g)$ for each demographic group without applying an explicit aggregation function.
In ASR, per-group WER, Match Error Rate (MER) \cite{Morris_2004}, Character Error Rate (CER), and Phoneme Error Rate (PER) are compared across speaking rates \cite{feng2021quantifying, Mirghafori_1996, 9747897}, non-native accents \cite{10096836, feng2021quantifying, Markl_2022, Zhang_2023, Yixuan_2024, maison23_interspeech}, regional dialects \cite{tatman2017gender, 10096836, zhang22n_interspeech, Serditova_2026}, speech impairments \cite{10584335, Wang_2023}, diverse languages  \cite{emezue2025naijavoices,Chauhan_2026}, age \cite{1255448, feng2021quantifying, zhang24d_interspeech, zhao23c_interspeech, 10.1007/978-3-642-35292-8_15, Wang_2023, S_2024, maison23_interspeech}, gender \cite{tatman2017gender, 1255448, Sarı_2021, feng2021quantifying, Zhang_2023, 10096836, 10.1145/3347449.3357480, S_2024, maison23_interspeech}, cultural backgrounds \cite{feng2021quantifying, Zhang_2023}, race \cite{Koenecke_2020, 10096836, ezema2025feels, Si_2021, lai23_interspeech}, disfluency rate \cite{Li_2024}, and socioeconomic status \cite{riviere2021asr4real}.

Per-group comparisons extend well beyond ASR.
Among speaker-centric tasks, SV reports per-group EER, false match rate (FMR), and false non-match rate (FNMR) \cite{Jin_2022, fenu2022demographic}; speaker diarization has been evaluated using group-specific Diarization Fairness Rates (DFR), measuring the proportion of single-speaker recordings correctly identified as containing one speaker \cite{tevissen2023towards}; and deepfake detection reports EER and false positive rate (FPR) differentials across gender, age, accent, and language \cite{yadav2024fairssd, stanvek2025scdf, katamneni2022demographic}.
For atypical and clinical speech, per-group reporting covers CER, deletion rate, and diagnostic accuracy gaps \cite{Li_2024, gulzar2026bias, pahar2026trust}.
The practice even extends to speech synthesis: in dysarthric speech cloning, Parity Difference and Disparate Impact quantify per-severity-level deviations from healthy-speaker baselines~\cite{m25_interspeech}.
Beyond system-side metrics, per-group comparisons extend to the \textit{user-perception} side, uncovering accent-dependent gaps in perceived success and error-recovery cost \cite{ezema2025feels, Michel_2025}.
Several of the studies above further enhance per-group reporting with significance tests and effect sizes; these statistical validation tools are discussed in \S~\ref{ssec:statistical_validation}.
While per-group reporting is informative, it leaves the aggregation step implicit; the explicit disparity metrics defined below ($\Delta_{max}$, $\Delta_{L^2}$, $\Delta_{L^1}$) improve cross-study comparability.
 
\textbf{Worst-Case Disparity ($L^\infty$).}
The Maximal Performance Gap ($\Delta_{max}$) formalizes the worst-case pairwise comparison:
\begin{equation}
\label{eq:delta_max}
    \Delta_{max} = \max_{g, g^\prime \in \mathcal{G}} \big| \mathcal{E}(g) - \mathcal{E}(g^\prime) \big| .
\end{equation}
A large $\Delta_{max}$ indicates that at least one pair of groups exhibits a substantial performance gap; the metric therefore prioritizes worst-case equity over average performance.
$\Delta_{max}$ is widely adopted across ASR fairness evaluations, either in absolute form \cite{zee2024group, swain2025towards, Swain_2024} or normalized by a reference WER to yield a scale-invariant relative gap suitable for cross-lingual comparison \cite{attanasio2024twists}.
WER Parity further binarizes this ratio against a tolerance threshold $\varepsilon$ for pass/fail auditing \cite{Raes_2024}.
Alternatively, a \textit{population-referenced} variant measures each subgroup's deviation from the overall mean, identifying which groups fall above or below the population baseline \cite{Koudounas_2025}.
All metrics above index disparity by the speaker's demographic group; however, the same $L^\infty$ structure can also index the evaluator's identity.
The annotator-gender MOS Gap ($\Delta_{\text{MOS}}$), the difference in mean MOS between male and female raters, captures systematic rating bias that propagates into automated MOS prediction models trained on human annotations~\cite{ren2026mos}.
 
\textbf{Variance-Based Disparity ($L^2$).}
Variance-based metrics penalize the dispersion of performance across all groups: % [EDIT] was misattributed as "Group Unfairness Score (GUS)"; GUS in wu2025evaluating is an L1 pairwise metric
\begin{equation}
\label{eq:delta_l2}
    \Delta_{L^2} = \sqrt{\frac{1}{|\mathcal{G}|} \sum_{g \in \mathcal{G}} \big(\mathcal{E}(g) - \mu_{\mathcal{E}}\big)^2},
\end{equation}
where $\mu_{\mathcal{E}}$ is the global mean across all groups.
Unlike $\Delta_{max}$, which captures only the worst-case pair, $\Delta_{L^2}$ reflects the performance dispersion across the full set of groups. 
The choice of $\mathcal{E}$ is task-dependent: the Sensitive-to-Sensitive Similarity Variance (SNSV) of Wu et al. \cite{wu2025evaluating} sets $\mathcal{E} = \text{PRAG}^*@K$ (top-$K$ ranking agreement) to audit spoken-dialogue SpeechLLMs, while EMO-Debias \cite{Lin_2025_2} instantiates as the true positive rate (TPR), FPR, or F1 for SER.
 
\textbf{Mean-Deviation Disparity ($L^1$).} 
The mean-deviation disparity measures the mean absolute deviation (MAD) of group performance:
\begin{equation}
\label{eq:delta_l1}
    \Delta_{L^1} = \frac{1}{|\mathcal{G}|} \sum_{g \in \mathcal{G}} \big|\mathcal{E}(g) - \mu_{\mathcal{E}}\big|.
\end{equation}
Several domain-specific metrics instantiate $\Delta_{L^1}$ with different $\mathcal{E}$:
the Fairness score $F$ averages $\Delta_{L^1}$ over multiple error rates for biometric verification \cite{Fenu_2020};
FairLENS sets $\mathcal{E} = \text{WER}$ for law-enforcement ASR \cite{Wang_2024};
Overall Bias uses the mean signed WER deviation from a reference group \cite{patel2025evaluate};
and Emo-bias applies $\Delta_{L^1}$ to SER accuracy across gender and racial subgroups \cite{Lin_2024_2}.
The Sum of Group Error Differences ($\mathrm{SED}_G$) takes a pairwise view, summing $|\mathcal{E}(g) - \mathcal{E}(g')|$ over all group pairs rather than deviations from the mean \cite{Elobaid_2024}.
 
\textbf{Composite and Welfare-Theoretic Metrics.}
The $\ell_p$ metrics above are purely \textit{descriptive}: they quantify how large the disparity is, but not how it should be traded off against overall utility.
The metrics below go further by either embedding normative preferences into the aggregation or combining utility and fairness into a single score.
On the normative side, the Gini Coefficient \cite{gini1912variabilita} measures inequality via the mean pairwise $L^1$ distance between group performances, normalized by the overall mean:
\begin{equation}
\label{eq:gini}
    G = \frac{\sum_{g \in \mathcal{G}} \sum_{g' \in \mathcal{G}} \big|\mathcal{E}(g) - \mathcal{E}(g')\big|}{2\,|\mathcal{G}|^2\,\mu_{\mathcal{E}}} .
\end{equation}
While $\Delta_{L^1}$ measures deviations from the mean, $G$ captures the full pairwise structure and is scale-invariant due to the $\mu_{\mathcal{E}}$ denominator.
The Isoelastic Social Welfare Function further introduces an inequality-aversion parameter $\rho$ that interpolates between a utilitarian objective and a Rawlsian maximin objective \cite{triantafyllopoulos2024enrolment}.
On the composite side, the Gini Aggregation Rate for Biometric Equitability (GARBE), originally developed for face recognition \cite{howard2022evaluating}, aggregates Gini-based inequality across operating points and has since been adapted to SV \cite{Oubaïda_2024}.
For ASR, the Fairness Adjusted ASR Score (FAAS) folds a regression-based fairness penalty into WER to rank systems on leaderboards \cite{Rai_2025}. % [EDIT] dropped Veliche_2024 (Fair-Speech dataset) here; consider citing it in Sec. 5 data resources instead
Because these frameworks embed explicit value judgments, researchers should make their parameter choices transparent when reporting results.

\textbf{Individual-Level Analogues.}
All preceding metrics index disparity by demographic group $g \in \mathcal{G}$; individual fairness instead requires that \textit{similar samples receive similar predictions}, regardless of group membership~\cite{dwork2012fairness}.
The same $\ell_p$ norm structure extends naturally to this setting.
The Consistency Score~\cite{zemel2013learning} measures the average $L^1$ deviation between each sample's prediction and the mean prediction of its $k$-nearest neighbors in embedding space, mirroring $\Delta_{L^1}$ at the sample level:
\begin{equation}
\label{eq:consistency}
    \mathrm{Consistency} = 1 - \frac{1}{n}\sum_{i=1}^{n} \Big|\hat{y}_i - \frac{1}{k}\sum_{j \in \mathrm{kNN}(i)} \hat{y}_j\Big| ,
\end{equation}
where the ideal value is 1; this metric is used in SER~\cite{chien2023achieving, Chien_2024}.
The Lipschitz constant $L_R = \sup_{x \neq x'} |M_\theta(x) - M_\theta(x')|\,/\,d(x,x')$ captures worst-case prediction sensitivity, analogous to $\Delta_{max}$ but normalized by input distance; a lower $L_R$ indicates smoother decisions between acoustically similar utterances~\cite{chien2024subsets}.
Empirical evidence in SER shows a trade-off: optimizing group-level parity can degrade individual-level consistency, particularly when attribute distributions between groups are distant~\cite{Chien_2024, chien2024subsets}.

% \textbf{Connection to Fairness Desiderata.}
% This family quantifies the \textit{spread} of $\mathcal{E}(g)$ across groups rather than testing a distributional independence condition, so the connections to Def.~\ref{def:dp} and~\ref{def:eo} are both partial.
% Def.~\ref{def:eo} (EO, \LEFTcircle): because $\mathcal{E}(g)$ is usually computed against the ground truth (e.g., WER, EER), a nonzero gap often signals an EO violation when $P(Y \mid G)$ is similar across groups, but a gap can also arise from differing base rates under perfect EO.
% Def.~\ref{def:dp} (DP, \LEFTcircle): when $\mathcal{E}$ is an unconditional statistic such as the overall acceptance rate the test is closer to a DP check, though most instantiations in the literature use ground-truth-referenced error metrics, limiting this reading to a subset of use cases.
% Def.~\ref{def:robustness} (Robustness, \LEFTcircle): the metrics of this family capture performance gaps under naturalistic variability, while Def.~\ref{def:robustness}'s stress-test criterion is operationalized by robustness benchmarks that compare error-rate degradation across demographic groups under injected noise, channel, and codec distortions \cite{Rajan_2023, shah2025srb, baali2025sveritas}.
\textbf{Connection to Fairness Desiderata.}
Because this family measures the spread of a group-level quantity $\mathcal{E}(g)$, its connection to each fairness definition depends on the choice of $\mathcal{E}$.
\begin{itemize}
    \item Def.~\ref{def:eo} (EO): when $\mathcal{E}$ is an error rate conditioned on the true label, zero spread corresponds directly to EO.
    Examples include the true- and false-positive rates in EMO-Debias~\cite{Lin_2025_2} as well as FMR and FNMR in GARBE.
    By contrast, overall performance metrics such as WER or accuracy reflect EO only partially.
    In binary classification, a group's overall error rate is a base-rate-weighted combination of its false-positive and false-negative rates.
    When base rates are equal across groups, EO implies equal overall error rates, so a gap in overall error then shows that EO fails.
    When base rates differ, however, a gap in overall error can arise even under perfect EO.
    \item Def.~\ref{def:robustness} (Robustness): the two criteria of this definition bound the inter-group spread of perturbed risk and of performance degradation (Eq.~\ref{eq:degrad}), that is, this family's disparity measure computed under perturbation.
    Robustness benchmarks measure such degradation gaps across demographic groups under acoustic noise, channel effects, and codec compression~\cite{Rajan_2023, shah2025srb, baali2025sveritas}.
\end{itemize}
%metrics such as the population-referenced variant and $\Delta_{\text{MOS}}$ measure performance gaps under naturalistic recording variability, echoing the stress-test disparity in Def.~\ref{def:robustness}, but this family does not apply controlled perturbations.

\textbf{Limitations.}
All metrics in this family treat demographic identity as an extraneous source of variance to minimize, without questioning whether the performance measure $\mathcal{E}$ itself is culturally appropriate or whether the group taxonomy $\mathcal{G}$ captures the relevant social categories.
A narrow inter-group gap does not imply fairness if absolute performance remains poor for all groups, or if the evaluation metric fails to reflect the construct it purports to measure (see Def.~\ref{def:validity}).

%% ====================================================================
%%  IV-B  Statistical Parity & Conditional Independence
%% ====================================================================
\subsection{Statistical Parity and Conditional Independence}
\label{ssec:statistical_parity}
 
Moving beyond the scalar performance comparisons of \S~\ref{ssec:group_disparity}, this family tests whether the model's \textit{prediction distribution} $P(\hat{Y} \mid G)$ is independent of the sensitive attribute $G$.
The gap metrics there reduce each group to one aggregate score, fusing per-class behavior with the group's base rates $P(Y \mid G)$.
DP instead requires $\hat{Y} \perp G$ while ignoring $Y$; EO requires the same within each true class.
Aggregate gaps and these independence tests therefore need not agree. Per-class disparities may cancel out in the aggregate, and even under perfect EO, differing base rates alone can produce a nonzero gap.
% The gap metrics of \S~\ref{ssec:group_disparity} reduce each group to one aggregate score, fusing per-class behavior with the group's base rates $P(Y \mid G)$.
% This family instead tests independence at the distribution level: DP requires $\hat{Y} \perp G$ while ignoring $Y$, and EO requires it within each true class.
% Aggregate gaps and these independence tests therefore need not agree. Per-class disparities may cancel out in the aggregate, and even under perfect EO, differing base rates alone can produce a nonzero gap.
 
\textbf{Core Metrics: $\Delta_{DP}$ and $\Delta_{EO}$.}
For a classification task with predicted output $\hat{Y} \in \mathcal{Y}$ and sensitive attribute $G$, the per-class gap between two groups is:
\begin{equation}
    d_{DP}(y,g,g') = \big| P(\hat{Y}{=}y \mid G{=}g) - P(\hat{Y}{=}y \mid G{=}g') \big|.
\end{equation}
Demographic Parity (DP) requires this gap to vanish for all classes and group pairs.
When base rates naturally vary across groups (e.g., in SER, where emotional expressiveness differs by culture), enforcing DP can penalize legitimate variance \cite{10.5555/3157382.3157469, dwork2012fairness}.
Equalized Odds (EO) addresses this by conditioning on the ground-truth label $Y$ to ensure full conditional independence $\hat{Y} \perp G \mid Y$ across all predicted classes $\hat{y}$ and true classes $y$:
\begin{equation}
\begin{split}
    d_{EO}(\hat{y}, y, g, g') &= \big| P(\hat{Y}{=}\hat{y} \mid G{=}g,\, Y{=}y) \\
    &\quad - P(\hat{Y}{=}\hat{y} \mid G{=}g',\, Y{=}y) \big| ,
\end{split}
\end{equation}
requiring equal outcome probabilities across groups for both correct predictions ($\hat{y}=y$) and misclassification errors ($\hat{y} \neq y$), thereby preserving natural variance while prohibiting demographic shortcuts \cite{10.5555/3157382.3157469, lin25c_interspeech}.
To obtain a scalar summary $\Delta_{DP}$ (or $\Delta_{EO}$), these per-class gaps must be aggregated over the label space and $\mathcal{G}$.
Common choices include worst-case ($\max_y \max_{g,g'}$ for DP; $\max_{\hat{y}, y} \max_{g,g'}$ for EO), macro-averaging over classes. Restricting the EO comparison to $\hat{y}=y$ leaves only the probability that a class is predicted correctly, so EO reduces to a comparison of per-class true-positive rates across groups.
This is the multiclass counterpart of Equal Opportunity, which Definition~\ref{def:eo} states for the binary positive class.
The choice of aggregation affects sensitivity: worst-case highlights the single most unfair (class, group) pair, while macro-averaging reflects disparity across the full label space.

\textbf{Domain Applications.}
In SER, macro-averaged $\Delta_{EO}$ and $\Delta_{DP}$ jointly quantify the trade-off between debiasing and accuracy \cite{Lin_2025_2};
cross-corpus evaluations use $\Delta_{DP}$ and $\Delta_{EO}$ to test whether gender fairness transfers across datasets \cite{Upadhyay_2025}.
In clinical settings, the predictive positive rate (PPR) and Statistical Parity deconstruct biases across intersecting demographic factors in health predictions \cite{Yang_2024}, while Equal Opportunity and Equalized Odds are applied to multi-class cognitive impairment detection across gender, age, education, and language subgroups \cite{Haghbin_2026}.

% \textbf{Connection to Fairness Desiderata.}
% This family directly operationalizes both Def.~\ref{def:dp} (Demographic Parity, \CIRCLE) and Def.~\ref{def:eo} (Equalized Odds, \CIRCLE).
% $\Delta_{DP}$ tests $\hat{Y} \perp G$, which is exactly the constraint stated in Def.~\ref{def:dp}; $\Delta_{EO}$ tests $\hat{Y} \perp G \mid Y$, matching Def.~\ref{def:eo}.
% The two metrics correspond one-to-one with the mathematical statements in Def.~\ref{def:dp} and~\ref{def:eo}, making this family the most direct bridge between the formal desiderata and computable evaluation.
\textbf{Connection to Fairness Desiderata.}
This family directly operationalizes both Def.~\ref{def:dp} (Demographic Parity) and Def.~\ref{def:eo} (Equalized Odds).
$\Delta_{DP}$ tests $\hat{Y} \perp G$, which is exactly the constraint stated in Def.~\ref{def:dp}; $\Delta_{EO}$ tests $\hat{Y} \perp G \mid Y$, matching Def.~\ref{def:eo}.
The two metrics correspond one-to-one with the mathematical statements in Def.~\ref{def:dp} and~\ref{def:eo}, making this family the most direct bridge between the formal desiderata and computable evaluation.

\textbf{Limitations.}
Both $\Delta_{DP}$ and $\Delta_{EO}$ are defined over discrete predicted classes; extending them to continuous outputs such as MOS or confidence scores requires discretization, and empirical evidence shows that fairness conclusions can reverse when the operating point shifts~\cite{Hutiri_2024}.
Moreover, $\Delta_{EO}$ conditions on the ground-truth label $Y$. Any bias present in the annotations propagates directly into the fairness assessment. In SER, where labels reflect annotator subjectivity and speaker-rater demographic interactions~\cite{10447167}, apparent compliance with EO may mask rather than resolve unfairness (see \S~\ref{sec:annotation}).
 
%% ====================================================================
%%  IV-C  Calibration & Threshold Fairness
%% ====================================================================
\subsection{Calibration and Threshold Fairness}
\label{ssec:calibration}
 
\S~\ref{ssec:group_disparity} and \S~\ref{ssec:statistical_parity} evaluate disparity at a fixed operating point.
This family treats the decision threshold~$\tau$ as a \textit{free variable}, asking whether a fairness conclusion reached at one operating point still holds at another.
Two complementary perspectives arise: \textit{score calibration}, which tests whether output scores carry the same probabilistic meaning across groups regardless of~$\tau$, and \textit{threshold sweep}, which measures fairness as a function of~$\tau$ across the full operating range.
 
\textbf{Score Calibration.}
The cost of log-likelihood ratios ($C_{llr}$) decomposes into a discrimination term and a calibration term~$C_{cal}$, where $C_{cal}$ isolates how well output scores are calibrated independently of the system's ability to separate target from non-target trials \cite{BRUMMER2006230}.
In SV, $C_{cal}$ degrades dramatically for underrepresented accents even when discrimination performance remains stable~\cite{Estevez_2023}.
In voice disorder detection, Normalized Cross-Entropy (NXE) reveals similar calibration gaps across gender and age subgroups, which group-dependent Platt scaling substantially reduces~\cite{Estevez_2025}.
In speech-based depression and anxiety prediction, Expected Calibration Error (ECE) likewise varies across demographic groups~\cite{norbury2026multimodal}.
Across all three tasks, the pattern is consistent: discrimination and calibration are separable, and a system can rank speakers correctly yet produce scores whose confidence is systematically misleading for certain populations.
 
\textbf{Threshold Sweep.}
The Fairness Discrepancy Rate $\text{FaDR}(\tau) = 1 - [\alpha \cdot \Delta_{\text{FMR}}(\tau) + (1-\alpha) \cdot \Delta_{\text{FNMR}}(\tau)]$~\cite{pereira2022FaDR} combines the largest inter-group FMR gap ($\Delta_{\text{FMR}}$) and FNMR gap ($\Delta_{\text{FNMR}}$) at each threshold, with $\text{FaDR} = 1$ indicating perfect fairness.
Integrating FaDR over the FMR axis yields a threshold-free summary, auFaDR-FAR~\cite{peri2022train}.
FaDR, IR \cite{grother2022face}, and GARBE have been compared head-to-head for SV~\cite{Oubaïda_2024}, using the functional fairness measure criteria of Howard et al.~\cite{howard2022evaluating} as the yardstick.
These criteria ask whether the contributions of the two error rates remain intuitive, whether the score stays within defined bounds, and whether it remains computable when a subgroup achieves zero errors.
Only GARBE satisfies all three.

\textbf{Connection to Fairness Desiderata.}
This family relates only to Def.~\ref{def:eo} (EO), but its threshold-sweep and calibration metrics do so in different ways.
The threshold-sweep metrics FaDR, auFaDR-FAR, and GARBE evaluate group-level disparities in FMR and FNMR across thresholds~$\tau$, directly testing the EO condition $P(\hat{Z}{=}1 \mid Z{=}z,\, G{=}g) = P(\hat{Z}{=}1 \mid Z{=}z,\, G{=}g')$.
For calibration metrics, the connection depends on whether scores represent likelihood ratios or posterior probabilities.
Because $C_{cal}$ evaluates log-likelihood ratios and weights target and non-target trials equally, within each group it depends solely on the score distributions over target and non-target trials, not on the proportion of each trial type.
If EO holds at every threshold, these two distributions are the same in every group, and so is $C_{cal}$.
Groups with markedly worse $C_{cal}$, such as the underrepresented accents in \citet{Estevez_2023}, therefore show that EO fails at some threshold.
Equal $C_{cal}$, however, does not establish EO.
ECE and NXE instead treat the score as the probability that the condition is present.
Under this reading, calibration and EO conflict whenever base rates differ across groups.
In such cases, the impossibility result of \citet{kleinberg2016inherent} shows that no score can be calibrated within every group while keeping error rates equal across groups at every threshold, unless it classifies every case correctly.

%% ====================================================================
%%  IV-D  Representational Disentanglement
%% ====================================================================
\subsection{Representational Disentanglement}
\label{ssec:disentanglement}
 
The preceding families evaluate bias at the output level.
This family moves upstream, quantifying how much demographic information leaks into learned speech representations before any decision is made.

The central metric is Mutual Information ($MI(H;G)$) between the embedding $H$ and the sensitive attribute $G$: a high value indicates that the representation encodes demographic identity, enabling downstream classifiers to exploit group membership even when the primary task is unrelated to demographics. 
Because exact MI is intractable for high-dimensional speech vectors, practical estimation relies on variational lower bounds such as the Donsker--Varadhan (DV) bound, which trains a neural discriminator to tighten a tractable approximation. 
In the speech domain, WavShape applies this strategy to self-supervised learning (SSL) embeddings~\cite{baser25_interspeech}.

Beyond MI, several complementary evaluation strategies have emerged.
The Speech Embedding Association Test (SpEAT) adapts the Implicit Association Test from psychology to speech embeddings~\cite{slaughter-etal-2023-pre}.
It quantifies, via Cohen's $d$, the differential association between two groups of speaker stimuli (e.g., male vs.\ female voices) and two sets of evaluative attributes (e.g., pleasant vs.\ unpleasant); applied to major self-supervised architectures, SpEAT reveals that these models amplify social biases across gender, age, and nationality relative to classic acoustic features~\cite{lin2024social}.
From a geometric perspective, clustering quality indices such as the Silhouette score \cite{Rousseeuw_1987}, Davies--Bouldin index \cite{Davies_1979}, and Calinski--Harabasz index \cite{Calinski_1974} can quantify how tightly embeddings group by demographic category rather than by task-relevant class.
{%
In pathological speech detection, \citet[][preprint]{kashyap2026geometric} report that these indices track how strongly accent variation confounds clinical markers across multilingual populations.
}
% On the statistical side, mixed-effects regression on $\ell_2$-normalized embedding distances separates the variance attributable to group membership from that attributable to channel or phonetic content in speaker verification~\cite{Jahan_2023}.

% \textbf{Connection to Fairness Desiderata.}
% Definition~\ref{def:dp} (DP, \LEFTcircle): if $MI(Z;G) = 0$, then $Z \perp G$, and any downstream prediction $\hat{Y} = h(Z)$ must also satisfy $\hat{Y} \perp G$ by the data-processing inequality.
% Therefore, disentanglement provides an upstream sufficient condition for DP, though the connection is approximate in practice because the DV bound yields only a lower bound on MI.
% Def.~\ref{def:counterfactual} (Counterfactual Fairness, \LEFTcircle): SpEAT asks whether group membership produces asymmetric associations in embedding space, which relates to Def.~\ref{def:counterfactual}'s output invariance requirement, but the connection is indirect because SpEAT quantifies association asymmetry rather than invariance under an explicit transformation $T_{g \to g'}$.
% Def.~\ref{def:output} (Output Association Equality, \LEFTcircle): when applied to generative speech models, SpEAT's association measurement overlaps with Def.~\ref{def:output}'s $s(\hat{Y})$, but operates in embedding space rather than on the final output.
\textbf{Connection to Fairness Desiderata.}
\begin{itemize}
\item Def.~\ref{def:dp} (DP): if $MI(H;G) = 0$, then $H \perp G$, and any downstream prediction $\hat{Y} = h(H)$ must also satisfy $\hat{Y} \perp G$ by the data-processing inequality.
Therefore, disentanglement provides an upstream sufficient condition for DP, though the connection is approximate in practice because the DV bound yields only a lower bound on MI.
\item Def.~\ref{def:counterfactual} (Counterfactual Fairness): most of SpEAT's original tests use recordings of the same read passage by speakers of different groups~\cite{slaughter-etal-2023-pre}, so content is held fixed while the group varies, as in Eq.~\ref{eq:cf}.
\item Def.~\ref{def:output} (Output Association Equality): SpEAT's association measurement overlaps with Def.~\ref{def:output}'s $s(\hat{Y})$, but operates in embedding space rather than on the final output.
\end{itemize}
\textbf{Limitation.}
These metrics share two important limitations.
First, the DV bound provides only a lower bound on MI, so a value near zero does not guarantee that the representation is free of demographic information; the true leakage may be higher.
Second, low demographic information in the representation does not ensure downstream fairness: a large-scale evaluation of SER shows that upstream representational bias and downstream performance disparity can diverge~\cite{Lin_2024_2}.

%% ====================================================================
%%  IV-F  Counterfactual & Individual Fairness
%% ====================================================================
\subsection{Counterfactual Fairness}
\label{ssec:counterfactual}
 
The preceding families evaluate fairness at the group level; this family adopts a causal perspective: changing a speaker's demographic attribute while holding linguistic content constant should not alter the model's output~\cite{kusner2017counterfactual}.
In speech, operationalizing this principle requires a perturbation mechanism that modifies perceived demographics without changing lexical content.
 
For ASR, \citet{Sarı_2021} formulate counterfactual fairness in three forms.
Each compares an utterance with a version regenerated for a different gender, age, or education group while holding lexical content fixed.
The first form requires the probability of producing the correct transcript to remain invariant across the pair.
The second requires invariance of the output distribution over characters at each frame.
The third requires invariance of the probability of aligning each frame to each position in the correct transcript.
For SpeechLLMs, the Cross-modal Capabilities Conservation Test (C3T) benchmark defines a counterfactual fairness metric as $\mathrm{fairness} = \#\mathrm{fair\_tasks}\,/\,\#\mathrm{tasks}$, where a task is fair if it yields the same answer for every speaker regardless of demographics; conditional variants further decompose this score by specific attributes such as age, gender, and dialect~\cite{kubis2025preservation}.

% \textbf{Connection to Fairness Desiderata.}
% This family directly operationalizes Definition~\ref{def:counterfactual} (Counterfactual / Invariance Fairness, \CIRCLE).
% Both ICEO and C3T's fairness metric formalize the requirement that $M_\theta(X)$ remains invariant when the speaker's demographic attributes are counterfactually altered via $T_{g \to g'}$, which is the defining equation of Def~\ref{def:counterfactual}.
% Definition~\ref{def:eo} (Equalized Odds, \LEFTcircle) is partially reflected.
% ICEO is explicitly named ``Individualized Counterfactual \textit{Equal Odds},'' embedding the EO constraint within a causal framework: it requires not only that error rates be equal across groups, but that this equality hold at the individual level under counterfactual demographic change.
\textbf{Connection to Fairness Desiderata.}
This family directly operationalizes the transformation-based test of Definition~\ref{def:counterfactual} (Counterfactual).
C3T's fairness metric formalize the requirement that $M_\theta(X)$ remains invariant when the speaker's demographic attributes are counterfactually altered via $T_{g \to g'}$, as Def.~\ref{def:counterfactual} states.
Definition~\ref{def:eo} (Equalized Odds) has an utterance-level analogue in the first form of \citet{Sarı_2021}.
While EO requires different demographic groups to have equal accuracy on inputs sharing the same ground truth, this form requires an utterance and its regenerated counterpart to be equally likely to produce their shared correct transcript.
Because this constraint enforces invariance for each individual pair rather than across whole populations, it does not by itself guarantee equal error rates across groups.

\textbf{Limitation.}
Counterfactual fairness evaluation in speech remains an open direction: the perturbation mechanisms available today, including adversarial auto-encoders and voice conversion, inevitably alter acoustic properties beyond the target demographic attribute, making it difficult to isolate causal effects cleanly.

%% ====================================================================
%%  IV-E  Bias in Generative Speech Systems
%% ====================================================================
% \subsection{Bias in Generative Speech Systems}
% \subsection{{Stereotyping and Demographic Preference}}
\subsection{Stereotyping and Demographic Preference in Generation}
\label{ssec:generative_bias}

{
Whereas \S\ref{ssec:group_disparity}--\ref{ssec:statistical_parity} compare system \textit{performance} across demographic groups of input speakers, the metrics here evaluate whether what a system produces encodes social stereotypes or demographic preferences.
This covers systems that generate speech as well as those that take speech and return text.
Mathematically, most metrics in this section build on the parity and proportion comparisons introduced in \S\ref{ssec:statistical_parity}, though several adopt alternative formulations such as set overlap, log-odds interaction, or composite scoring.
The composite scores come largely from broad trustworthiness benchmarks such as AudioTrust~\cite{li2025audiotrust} and AHELM~\cite{lee2025ahelm}, where fairness sits alongside safety and robustness as one dimension among several.
}

% Recent holistic benchmarks such as AudioTrust~\cite{li2025audiotrust} and AHELM~\cite{lee2025ahelm} treat fairness as one of several trustworthiness dimensions for audio language models, each introducing its own composite scoring. In parallel, a structural taxonomy of audio LLMs highlights the need for architecture-aware evaluation across cascaded, multimodal, and audio-native designs~\cite{aloufi2026evaluation}.
 
\textbf{SpeechLLMs and Spoken Dialogue.}
At the language-modeling level, the Speech Language Bias Score (slbs) measures the percentage of instances in which a model prefers a stereotypical over an anti-stereotypical continuation~\cite{lin2024spokenstereoset}.
In question answering, a key design axis is whether the context uniquely determines the correct answer (disambiguated) or leaves it open (ambiguous).
VoiceBBQ reports accuracy on disambiguated items and a bias score on ambiguous items, additionally separating content bias from acoustic bias~\cite{choi2025voicebbq}. 
Listen and Speak Fairly independently adopts the same split with $s_{\mathrm{DIS}} = 2(n_{\mathrm{biased}}/n_{\mathrm{not\_unknown}}) - 1$ and $s_{\mathrm{AMB}} = (1 - \mathrm{accuracy}) \times s_{\mathrm{DIS}}$~\cite{Lin_2024}.
Orthogonal to textual content, Gender Response Overlap ($J$) measures the distributional overlap between responses to male and female voices, while Gender Preference ($\Delta$) captures how frequently a model's response aligns with a specific gender~\cite{choi2025acoustic}.
At the downstream decision level, the Group Unfairness Score (GUS) and Similarity-based Normalized Statistics Rate (SNSR) measure decision-level and recommendation-level inequity across demographic groups in spoken dialogue systems~\cite{wu2025evaluating}.
{
GUS and SNSR compare rates across groups in the same form as the parity metrics of \S\ref{ssec:statistical_parity}.
Since a recommendation has no correct answer, the gap they measure reports differential treatment rather than unequal accuracy, which is the concern this subsection collects.
}
To avoid the expressiveness constraints of multiple-choice question (MCQ) formats, which restrict stereotypical associations to predefined options, VIBE~\cite{lin2026vibe} evaluates SpeechLLMs through open-ended generation tasks with real-world (not synthetic) speech, measuring distributional shifts via normalized Total Variation Distance (nTVD); it reveals systematic biases undetected by MCQ benchmarks.
Biased conclusions can also be sensitive to the evaluation format itself.
The Average Pairwise Entropy Shift (APES) quantifies this sensitivity by averaging pairwise entropy differences across levels of a controlled variable.
Paired with Fleiss' $\kappa$, it shows that multilingual MLLMs are relatively robust to gender and accent but highly sensitive to language and option order, with speech systematically amplifying text-modality biases~\cite{wei2026bias}.
A caveat is that bias scores obtained from multiple-choice benchmarks do not reliably transfer to open-ended generation, so conclusions drawn from a single evaluation format should be interpreted with caution~\cite{satish2025benchmarks}.
 
\textbf{Speech Translation.}
Translating from a language without grammatical gender, such as English, into gender-marked languages such as Spanish or Italian requires the system to infer the gender of each entity. 
Systems that default to masculine forms or rely on stereotypical gender-occupation associations exhibit lower accuracy on female and anti-stereotypical instances.
The WinoST challenge set quantifies this effect through gender accuracy, the proportion of correctly gendered entities, alongside $\Delta G$, the gap between male and female F1, and $\Delta S$, the gap between pro-stereotypical and anti-stereotypical F1~\cite{Costa_2022, Lin_2024}.
MuST-SHE complements this evaluation by computing accuracy on both correct and gender-swapped references; the difference between the two scores reveals the direction of bias, exposing whether the system systematically favors masculine over feminine forms~\cite{bentivogli2020gender}.
 
\textbf{Speech and Audio Synthesis.}
A common evaluation strategy for instruction-guided TTS estimates the gender probability of synthesized speech via a pretrained speaker gender recognition classifier and compares it against expected distributions~\cite{kuan-lee-2025-gender}.
When instructions combine multiple social dimensions, the interaction term $\mathcal{I}$, defined as the log-odds deviation from an additive baseline, isolates how status, career, and persona cues jointly modulate gender probability beyond their individual effects~\cite{chen2026bindingeffect}.
In unprompted speech generation, the female utterance ratio across repeated syntheses of the same text quantifies systematic gender preferences in automatic speaker assignment~\cite{Puhach_2025}.
For text-to-audio generation, the Stereotype Audio Score (SAS), defined as $\max(\%N_{\mathrm{male}}, \%N_{\mathrm{female}})$, captures the gender dominance in a pool of generated samples, and the Bias-Aware Audio Score (BAAS) combines SAS with an audio quality term to penalize high-fidelity generation that lacks gender balance~\cite{mohsin25_interspeech}.

% \textbf{Connection to Fairness Desiderata.}
% Def.~\ref{def:dp} and~\ref{def:eo} (DP and EO, \LEFTcircle): GUS and SNSR test whether decision outcomes (recommendations, system actions) differ across groups, mapping to DP/EO at the decision level rather than at the content level.
% Def.~\ref{def:counterfactual} (Counterfactual Fairness, \LEFTcircle): VoiceBBQ's content-vs-acoustic decomposition asks whether changing the speaker's demographic attributes alters the model's response, mirroring the counterfactual invariance of Def.~\ref{def:counterfactual}.
% Def.~\ref{def:output} (Output Association Equality, \CIRCLE): slbs, VoiceBBQ, $s_{\mathrm{DIS}}$/$s_{\mathrm{AMB}}$, $\Delta G$/$\Delta S$, and SAS/BAAS each test whether a derived output attribute (stereotype score or gender probability) is independent of $G$, instantiating a specific choice of $s$ in Def~\ref{def:output}'s $\mathbb{E}[s(\hat{Y}) \mid G{=}g] = \mathbb{E}[s(\hat{Y}) \mid G{=}g']$.

\textbf{Connection to Fairness Desiderata.}
\begin{itemize}
    \item Def.~\ref{def:dp} (DP): GUS and $J$ compare the system's own responses when the same input is voiced by different groups.
    For every input, if these responses are identically distributed across groups, the overall response distribution is also identical across groups.
    A perfect score on either metric is therefore a sufficient condition for DP on the evaluated inputs, though DP itself does not require such per-input agreement.
    \item Def.~\ref{def:counterfactual} (Counterfactual Fairness): GUS averages the dissimilarity $\psi$ of Eq.~\ref{eq:cf} over inputs and pairs of groups, with $T_{g\to g'}$ realized by voicing the same input with another group's voice and $\psi$ the change in the decision, so it is zero exactly when Eq.~\ref{eq:cf} holds with $\varepsilon = 0$ on every evaluated input. 
    $J$ performs a weaker version of this check on the responses pooled over several voices of each group.
    \item Def.~\ref{def:output} (Output Association Equality): VIBE's nTVD evaluates the distribution of an extracted response attribute $s$, reaching zero when group distributions coincide as Def.~\ref{def:output} requires.
    VoiceBBQ compares the bias score between speaker voices, which is the mean form of Def.~\ref{def:output} with $s$ indicating whether an answer follows the stereotype targeted by the question.
\end{itemize}

% { 
% \textbf{Contextual and Exposure Bias in Speech LLMs.}
% Beyond demographic content generation, Speech LLMs exhibit distinct biases related to contextual reliance. 
% ``Contextual exposure bias" occurs when models trained on oracle conversational history over-rely on context, leading to compounding errors when forced to condition on noisy, error-prone ASR hypotheses during deployment \cite{Guo_2026}. 
% Similarly, models systematically fail to transcribe rare ``bias words," prompting interventions that utilize phonetically similar common word cues within text prompts to bypass the need for explicit grapheme-to-phoneme systems during contextual biasing \cite{Novitasari_2026}.
% }

%% ====================================================================
%%  IV-G  Cross-cutting Analytical Tools
%% ====================================================================
\subsection{Statistical Validation}
\label{ssec:statistical_validation}
 
The preceding families quantify the magnitude of demographic disparities. 
This subsection collects the statistical tools needed to determine whether those disparities are genuine rather than artifacts of sampling variability.
 
Any observed disparity may arise from sampling noise rather than systematic bias, particularly for small or intersectional subgroups.
Significance tests such as ANOVA establish whether group differences exceed chance expectations, while effect-size measures such as Cohen's~$d$ \cite{cohen2013statistical} quantify their practical magnitude.
When many subgroups are compared simultaneously, the Benjamini-Hochberg procedure controls the false discovery rate \cite{benjamini1995controlling, Altwlkany_2025}.
For the skewed error-rate distributions common in speech, non-parametric alternatives such as the Mann-Whitney U \cite{mann1947test} and Wilcoxon signed-rank tests \cite{wilcoxon1945individual} offer robust unpaired and paired comparisons~\cite{Puhach_2025, szekely23_interspeech}.
% When confounding variables such as recording conditions or annotator subjectivity are present, mixed-effects regression isolates the true demographic contribution by treating group attributes as fixed effects and nuisance factors as random effects, with Poisson or logistic link functions for count-based and binary outcomes~\cite{jimenez2026dialect, Rai_2025, 10.1007/978-981-95-5382-2_17}.
Mixed-effects regression can estimate demographic associations while accounting for measured covariates and variability across speakers or listeners \cite{jimenez2026dialect, Rai_2025, 10.1007/978-981-95-5382-2_17, Jahan_2023}.
Finally, utterances from the same speaker are statistically dependent, violating the independence assumption of the above tests; blockwise bootstrap resampling at the speaker level preserves this correlation structure and yields more reliable confidence intervals~\cite{liu2019statistical}.

% Limitations
Beyond the quantifiable disparities addressed throughout this section, the Governance paradigm encompasses non-quantifiable concerns such as data sovereignty and community agency (\S~\ref{ssec:governance}), which require institutional rather than algorithmic interventions.

\section{Source of Bias}
\label{sec:source}
\begin{figure}[t]
\centering
\resizebox{\columnwidth}{!}{
\begin{forest}
    for tree={
        grow'=0,
        parent anchor=east,
        child anchor=west,
        edge={draw, line width=0.6pt},
        forked edges,
        align=left,
        anchor=west,
        l sep=12pt,
        s sep=6pt,
        inner sep=5pt,
        draw,
        fill=blue!5,
        rounded corners,
    }
    [{\hyperref[sec:source]{\textbf{\ref*{sec:source}}. Source of Bias}}, minimum width=3cm, minimum height=0.9cm, fill=gray!20
        [{\hyperref[sec:data]{\textbf{\ref*{sec:data}} Data-Related Bias}}, minimum width=4.5cm, minimum height=0.8cm, fill=green!50
            [{\hyperref[sec:demo_rep]{\ref*{sec:demo_rep} Demographic and Linguistic Representation}}, minimum height=0.7cm, fill=green!25, tier=leaf]
            [{\hyperref[sec:collection]{\ref*{sec:collection} Data Collection, Quality, and Artifacts}}, minimum height=0.7cm, fill=green!25, tier=leaf]
            [{\hyperref[sec:annotation]{\ref*{sec:annotation} Annotation and Labeling Subjectivity}}, minimum height=0.7cm, fill=green!25, tier=leaf]
        ]
        [{\hyperref[sec:model]{\textbf{{\ref*{sec:model}} Model-Related Bias}}}, minimum width=4.5cm, minimum height=0.8cm, fill=yellow!50
            [{\hyperref[para:arch]{{\ref*{para:arch}} Architectural Limitations and Feature Entanglement}}, minimum height=0.7cm, fill=yellow!25, tier=leaf]
            [{\hyperref[para:decoding]{{\ref*{para:decoding}} Decoding and Post-Processing Assumptions}}, minimum height=0.7cm, fill=yellow!25, tier=leaf]
            [{\hyperref[para:gen_bias]{{\ref*{para:gen_bias}} Generative Model-Specific Bias}}, minimum height=0.7cm, fill=yellow!25, tier=leaf]
        ]
        [{\hyperref[sec:deploy]{\textbf{{\ref*{sec:deploy}} Deployment Mismatch}}}, minimum width=4.5cm, minimum height=0.8cm, fill=red!40
            [{\hyperref[para:benchmark]{{\ref*{para:benchmark}} The Benchmark-Reality Gap}}, minimum height=0.7cm, fill=red!20, tier=leaf]
            [{\hyperref[para:domain]{{\ref*{para:domain}} Speaking Style and Evaluation Coverage}}, minimum height=0.7cm, fill=red!20, tier=leaf]
            [{\hyperref[para:env]{{\ref*{para:env}} Environmental and Hardware Variability}}, minimum height=0.7cm, fill=red!20, tier=leaf]
        ]
        [{\hyperref[sec:cross_source]{\textbf{{\ref*{sec:cross_source}} Cross-Source Interactions and Synthesis}}}, minimum width=4.5cm, minimum height=0.8cm, fill=purple!40]
    ]
\end{forest}
}
\caption{Taxonomy of bias sources in speech technology.}
\label{fig:bias_taxonomy}
\end{figure}

As illustrated in Figure~\ref{fig:bias_taxonomy}, bias in speech processing systems stems from three primary sources:
\textit{data-related issues}, \textit{model-related factors}, and \textit{deployment mismatch}.
These three sources are complementary to, but not identical to, the conceptual paradigms introduced in \S~\ref{sec:expansion}.
Like the life-cycle framework of \citet{suresh2021framework}, they are organized by where bias originates in the pipeline rather than by normative concern.
We detail each source below and discuss their interactions in \S~\ref{sec:cross_source}.

\subsection{Data-Related Bias}\label{sec:data}
Data-related biases are foundational, arising either from skewed data distributions or inherent differences in speech characteristics that are inadequately represented.

\subsubsection{Demographic and Linguistic Representation}\label{sec:demo_rep}
Biases in training data generally stem from two primary deficits: imbalances in the \textit{\textbf{demographic representation}} of speaker groups and insufficient coverage of \textit{\textbf{diverse linguistic content}}.

\textbf{Demographic Underrepresentation.}
Standard datasets exhibit imbalances that reflect societal hierarchies rather than population distributions.
Along the \textit{gender} axis, broadcast news and web corpora favor male speakers, yielding systems less robust to female acoustic patterns \cite{10.1145/3347449.3357480, maison23_interspeech, lin25c_interspeech}.
Along the \textit{age} axis, adult-centered distributions leave child speech (whose shorter vocal tracts produce distinct pitch and formant patterns) and elderly speech severely underrepresented \cite{1255448, 10.1007/978-3-642-35292-8_15, borre2025explore}.
{Age coverage is also limited in speech enhancement benchmarks such as VoiceBank-DEMAND, while evaluations on more diverse speech reveal significant age- and gender-related performance differences~\cite{Giraldo_2025}.}
\textit{Racial and ethnic} imbalances are mediated by the marginalization of non-standard ethnolects: AAE speakers suffer systematic transcription errors when evaluated against corpora dominated by Standard American English \cite{Koenecke_2020, martin2022, Brewer_2023, 10.1145/3544548.3581357}.
Beyond demographic axes, \textit{atypical speech} is systematically excluded, as standard datasets prioritize fluent speech and leave conditions such as stuttering, dysarthria, and Parkinson's-affected speech scarcely represented \cite{10584335, Li_2024, Singh_2024, wang2023beyond, morovelazquez19_interspeech}.
Finally, \textit{metadata gaps} constitute a distinct representational harm.
In a survey of 66 open speech corpora, over a third lacked speaker gender information, and none mentioned non-binary speakers~\cite{garnerin2020gender}, limiting the assessment of disparities affecting non-binary speakers.

\textbf{Linguistic and Content Disparities.}
Beyond demographic labels, most training corpora implicitly privilege standard language varieties, reflecting the standard language ideology that elevates prestige dialects as the normative baseline while treating other variations as deviations \cite{Markl_2022, talafha2023nshot}.
Unlike acoustic mismatches, this bias manifests in the language modeling layer, where valid dialectal syntax is misinterpreted as errors because training data enforces standard grammatical rules.
Corpora also lack diversity in cultural and idiomatic content, producing systematically higher error rates for subgroups whose linguistic habits diverge from the training norm \cite{Koudounas_2025}.

\subsubsection{Data Collection, Quality, and Artifacts}\label{sec:collection}
Bias arises from systematic inequalities in both the \textit{\textbf{physical}} and \textit{\textbf{procedural}} aspects of data creation, alongside the hidden artifacts introduced during the collection pipeline.

\textbf{Socioeconomic and Equipment Disparities.}
Socioeconomic disparities manifest as \textit{\textbf{channel bias}}, a form of \textit{Technical Bias} (\S~\ref{ssec:social_bias}): data from marginalized populations is disproportionately captured in high-reverberation environments or on low-fidelity hardware, so models may use channel characteristics as a demographic proxy instead of learning robust linguistic features \cite{papakyriakopoulos2023considerations}.
This confound can arise even within a single corpus, where recording practices co-vary with speaker subgroup \cite{li2024reexamining}. Crowdsourced datasets exacerbate the issue, as their acoustic conditions diverge from studio-quality training data~\cite{shah2025srb, Giraldo_2025}; speech enhancement preprocessing can further introduce artifacts~\cite{iwamoto22_interspeech} that disproportionately degrade some speakers' recordings~\cite{degroot25_interspeech}.

\textbf{Acoustic and Signal Variability.}
Beyond equipment, bias resides in latent acoustic features that demographic counts alone cannot capture.
When training data over-represents male speakers, models generalize poorly to the fundamental-frequency and formant ranges of female voices~\citep{fucci2023nopitch}.
Age-related acoustic mismatch has also been observed, with older speakers' speech matching ASR acoustic models less well than that of younger adults \cite{vipperla2010ageing}.
More subtly, imbalances in speech rate \cite{9747897}, prosody, and rhythm \cite{baser25_interspeech, lai23_interspeech} can drive stronger bias than demographic labels themselves \cite{Gu_2023}, acting as hidden confounders that persist undetected even in demographically balanced datasets because they are rarely logged in metadata.

\textbf{Collection Protocols and Transcription.}
Recording and labeling protocols systematically place different groups in unequal conditions.
In broadcast corpora, women are more often recorded in formal scripted settings while men appear in spontaneous conversational contexts, producing spurious correlations between gender and signal quality \cite{markl-mcnulty-2022-language}; more broadly, datasets favoring read speech yield optimistic performance estimates that collapse under real-world conditions \cite{patel2025evaluate, Raes_2024}.
Transcription guidelines are similarly non-neutral: choices regarding disfluencies, non-standard grammar, and hesitation markers disproportionately inflate error rates for speakers of non-standard varieties \cite{feng2021quantifying, 10.1145/3491102.3517639}, while protocols demanding sustained, fluent articulation exclude medical subgroups whose speech is affected by conditions such as medication-induced articulation changes \cite{10314880}.

\textbf{Metadata Gaps and Data Artifacts.}
Bias is also embedded in the infrastructure of datasets.
Audits of open-source corpora reveal that essential metadata, such as gender demographics and speech duration, is frequently missing or inconsistent, leaving fairness claims about these datasets untestable \cite{10.1145/3347449.3357480}.
Moreover, dataset construction functions as a mechanism for setting norms: the varieties included and documented determine whose speech is treated as \textit{normal} by downstream systems, while undocumented varieties are implicitly categorized as deviations \cite{markl-mcnulty-2022-language}.
These gaps enable shortcut learning, where systems rely on dataset-specific artifacts rather than genuine linguistic features, so that strong benchmark performance may not reflect true robustness \cite{palmer2021spoken, papakyriakopoulos2023considerations}.

\subsubsection{Annotation and Labeling Subjectivity}\label{sec:annotation}
Bias in annotation arises because speech labels are judgments filtered through human labelers' subjective expectations, shaped by both \textit{\textbf{cognitive priors}} in individual perception and \textit{\textbf{cultural distance}} between rater and speaker.

\textbf{Perceptual and Cognitive Bias.}
Human perception introduces non-random label noise shaped by cognitive priors independent of the signal. 
Such label noise stems partly from raters themselves. 
For example, older listeners recognize vocal anger and sadness less accurately~\cite{ruffman2008meta}, while female listeners generally outperform males in decoding vocal emotion~\cite{lausen2018gender}.
Other noise reflects how raters perceive the speaker. 
Specifically, stronger implicit bias correlates with amplified neural responses to foreign-accented speech and shifts perceived clarity ratings~\cite{yi2014neural}.
Together, these individual traits and perceptual biases create annotator-speaker mismatches, where demographic disparities in labels reflect how raters interpret cues rather than the acoustic signal itself~\cite{10447167, chien2023achieving}.

Such systematic label errors are fossilized as ground truth, so that downstream models learn and scale annotators' implicit prejudices.

\textbf{In-Group Advantage and Cultural Distance.}
Labeling reliability degrades as cultural distance between rater and speaker increases: annotators show higher accuracy for speakers from their own cultural background~\cite{laukka2021crosscultural, scherer2001emotion} and lower accuracy for accents they do not share~\cite{adank2009comprehension, papakyriakopoulos2023considerations}. This distance leads raters to misclassify valid dialectal variations as ambiguity or errors, injecting systematic noise correlated with speaker identity \cite{feng2021quantifying}. Gender-related differences also arise in human judgments of speech: in a listening study of Alzheimer's disease perception, male speech was more frequently classified as indicative of the disease, particularly in the Mandarin condition \cite{10.1007/978-981-95-5382-2_17}.

\subsection{Model-Related Bias}\label{sec:model}
Even with balanced training data, bias can arise from model-intrinsic design choices, manifesting as \textit{\textbf{architectural limitations}} in learned representations, \textit{\textbf{deterministic assumptions}} embedded in decoding logic, and mechanisms specific to \textit{\textbf{generative speech models}}.

\subsubsection{Architectural Limitations and Feature Entanglement}\label{para:arch}

\textbf{Representation Entanglement.}
Standard optimization objectives in tasks such as speaker verification often fail to disentangle speaker identity from linguistic content and environmental factors, causing speaker embeddings to encode language-specific phonological information and misinterpret language shifts as identity shifts \cite{Sharma_2024}.
This entanglement has demographic consequences: gender-related performance gaps persist across standard speaker verification architectures \cite{toussaint2022bias}.

\textbf{Tokenization and Vocabulary Bias.}
Tokenization is an often-overlooked bias source in end-to-end speech models that jointly learn acoustic and linguistic representations.
Byte-pair encoding tokenizers in models such as Whisper \cite{pmlr-v202-radford23a} are trained predominantly on high-resource languages, yielding higher token likelihoods for those languages and worse performance on low-resource ones \cite{liang2025beyond}.
The resulting over-segmentation forces decoding over longer, less informative token sequences for underrepresented languages \cite{10887595}, and compounds in code-switching, where models degrade at language-switch boundaries despite strong per-language performance \cite{11447403}.
Because the bias lies in the learned vocabulary rather than the training audio, more speech data cannot remedy it: sub-token utilization is shaped more by typological and orthographic structure than by training scale \cite{liang2025limits}, so interventions must act at the tokenization layer.

\textbf{Model Compression and Capacity Bias.}
Compression techniques such as pruning, quantization, and knowledge distillation can introduce or amplify demographic bias even when applied to a fair base model.
\citet[][preprint]{hooker2020compressed} report that pruning disproportionately degrades performance on underrepresented, atypical long-tail examples while leaving aggregate accuracy largely unaffected.
However, findings in speech SSL models reveal that this effect is not universal. 
Pruning can even reduce social bias in certain configurations, with architectural structure mattering more than raw parameter count~\citep{lin2024social}. Compression is therefore a critical design choice whose equity implications must be empirically evaluated rather than assumed.

\subsubsection{Decoding and Post-Processing Assumptions}\label{para:decoding}
Hard-coded decoding heuristics apply uniformly across deployment contexts, unlike the context-dependent failures of \S~\ref{sec:deploy}.
Voice activity detection (VAD) thresholds misclassify dysfluent pauses as silence, and language models trained on standard fluency probabilities over-correct repetitions; both disadvantage speakers who stutter \cite{10.1145/3544548.3581224}.
The same assumptions hurt users with cognitive conditions: individuals with dementia need mid-utterance pauses for cognitive planning, but turn-taking algorithms misread these as end-of-turn signals and cut off the user \cite{10.3389/frdem.2024.1343052}.
The burden falls on atypical speakers, compounding the data underrepresentation identified in S~\ref{sec:demo_rep} \cite{Li_2024}.

\subsubsection{Generative Model-Specific Bias}\label{para:gen_bias}
In generative speech systems, bias can arise in the assignment of speaker characteristics to text or instructions.
Occupation-related instructions have been shown to shift the gender distribution of synthesized voices in instruction-guided TTS, revealing occupational gender associations \cite{kuan-lee-2025-gender}.
In automatic speaker assignment by a SpeechLLM, gender preferences have also been observed, although these do not consistently align with occupational stereotypes \cite{Puhach_2025}.

\subsection{Deployment Mismatch}\label{sec:deploy}
Deployment mismatch, corresponding to \textit{Emergent Bias} in \S~\ref{ssec:social_bias}, arises when systems validated on idealized benchmarks fail in contexts that differ significantly from their design environment.

\subsubsection{The Benchmark-Reality Gap}\label{para:benchmark}
Dataset-specific recording cues can contribute to apparent predictive performance.
In clinical speech assessment, differences in recording duration and intensity between patients and controls were found to provide cues exploited by models for unilateral vocal fold paralysis detection \cite{low2024identifying}.
% At production scale, audits of ASR systems reveal cohort-level performance disparities invisible in benchmark evaluation \cite{dheram2022toward}.

\subsubsection{Speaking Style and Evaluation Coverage}\label{para:domain}
The speaking styles covered by an evaluation dataset delimit the conclusions that can be drawn about ASR fairness.
Disparities across speaker groups have been observed even in read speech, while the limited presence of phonetic reduction, hesitations, and repetitions in such data constrains its coverage of conversational speech \cite{Markl_2022}.

\subsubsection{Environmental and Hardware Variability}\label{para:env}
Beyond channel bias in training data, deployment environments introduce variability through real-world noise and heterogeneous hardware that interact with speaker demographics.

\textbf{Noise and Demographic Interaction.}
Background noise and other acoustic corruptions degrade ASR performance more for some speaker groups than for others~\cite{Rajan_2023, shah2025srb}. Common preprocessing can exacerbate this; speech enhancement frontends may introduce artifacts that disproportionately affect underrepresented speech patterns, producing a counterproductive cycle where denoising hurts the populations it should help \cite{Giraldo_2025}.

\textbf{Device Heterogeneity.}
Users interact with speech systems through diverse hardware (studio microphones, far-field smart speakers, budget smartphones) that impose distinct acoustic transfer functions. Emerging evidence indicates hardware can interact with demographics (e.g., contact microphone placement produces gender-dependent accuracy gaps \cite{konuma2023effects}), but systematic audits across device types remain largely absent.

\subsection{Cross-Source Interactions and Synthesis}\label{sec:cross_source}
While the preceding subsections present data, model, and deployment biases as analytically distinct categories, real-world fairness failures typically involve their interaction, producing compounding effects that exceed the sum of individual contributions. We identify three such interaction patterns.

\textbf{Cascading Pipeline Effects.}
In speech systems, bias introduced at an upstream stage propagates to and is amplified by downstream components in any multi-stage pipeline, turning acoustic-level failures into semantic-level ones \cite{ezema2025feels}.

\textbf{Intersectional Compounding.}
Bias that appears moderate along a single demographic axis can compound severely at demographic intersections \cite{Harris_2024}. Single-attribute fairness audits therefore systematically underestimate the harm borne by those at multiple marginalized intersections.

\textbf{Architecture and Data Amplification.}
Model design choices can selectively amplify latent data biases rather than merely reflect them \cite{toussaint2022bias,hooker2020compressed}. The final fairness profile of a deployed system is thus not reducible to its training data alone, but is jointly determined by the interaction between data composition and architectural decisions.

\section{Mitigation Strategy}
\label{sec:mitigation}
\begin{figure*}
    \centering
    \includegraphics[width=0.8\linewidth]{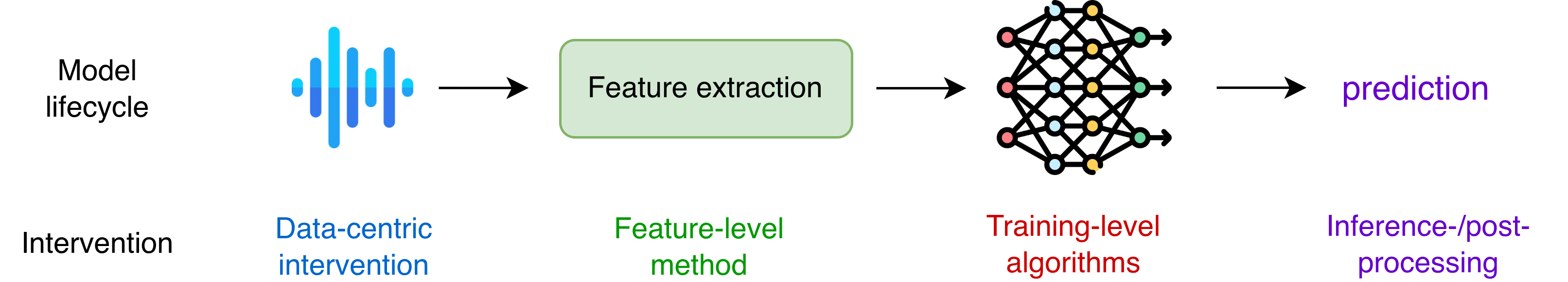}
    \caption{Intervention methods across the speech model production pipeline.}
    \label{fig:debias_workflow}
\end{figure*}

% Research on fairness in speech technologies delineates a spectrum of interventions, categorizable not merely by pipeline stage but by their mechanism of bias mitigation (Figure~\ref{fig:debias_workflow}).
% First, \highlight{Data-centric interventions} govern the input distribution, moving beyond simple rebalancing to actively curate sociolinguistic diversity. 
% Strategies here range from stratified sampling to synthesizing counterfactual speech data, aiming to correct historical underrepresentation at the source. 
% Second, \highlight{Representation-level methods} target the signal encoding to algorithmically decouple sensitive attributes from task-relevant information. 
% Strategies such as adversarial invariance or conditioning serve to construct an embedding space that is robust to demographic shifts while preserving the fidelity of the downstream prediction.
% Third, \highlight{Training-centric algorithms} manipulate the optimization landscape.
% By modifying loss functions to include fairness constraints or altering gradient updates through adaptive weighting, these methods enforce equity requirements directly during parameter convergence to prevent overfitting to dominant acoustic profiles.
% Finally, \highlight{inference- and post-processing strategies} operate as a fail-safe on the decision boundary. 
% Post-hoc calibration and threshold adjustment serve to rectify residual disparities in model confidence, ensuring that decisions remain equitable even when the underlying representation remains imperfect.

Research on fairness in speech technologies encompasses a broad spectrum of interventions. We organize them by their point of action in the model production pipeline (Figure~\ref{fig:debias_workflow}), yielding four stages.
\highlight{Data-centric interventions} act before any model is trained, reshaping the input distribution to address historical underrepresentation.
\highlight{Feature-level methods} act on a model that has already been trained, transforming its representations or compressing it without retraining it for fairness.
\highlight{Training-level algorithms} act during optimization, modifying loss functions, gradient signals, or architectural pathways so that fairness is enforced as representations are formed.
\highlight{Inference- and post-processing strategies} act when the system is deployed and without retraining, either on the model's input and output or by adjusting its computation to the current input or speaker.
Feature-level methods differ from training-level methods in that they do not change how the model is trained, that is, its loss, gradients, or architecture.
They differ from inference-level methods in that the change becomes a fixed part of the model and is used for every input.

% \begin{figure*}
%     \centering
%     \includegraphics[width=0.8\linewidth]{figures/debias/data_intervention.drawio.png}
%     \caption{Overview of debias methods with data-centric interventions.}
%     \label{fig:data_debias}
% \end{figure*}

\subsection{Data-centric interventions}
\subsubsection{Diversify and expand datasets}
Systematic underrepresentation of specific speaker demographics, linguistic varieties, and pathological conditions constitutes a primary source of algorithmic bias \cite{paullada2021data}. 
Addressing this sampling bias requires not merely increasing data volume, but structurally reshaping training distributions to cover the long tail of acoustic variation.
% Current literature approaches this remediation through two distinct yet synergistic paradigms: (1) participatory collection of real-world data from marginalized groups, and (2) generative augmentation to synthetically populate sparse regions of the feature space.

\textbf{Participatory Data Collection.}
Involving marginalized communities as collaborators rather than passive subjects improves both the ecological validity of collection protocols and the accuracy of annotation. Co-designing recording tasks with target populations ensures that prompts, pacing, and interfaces match real communication routines. Aging research workshops refined protocols to avoid younger-user defaults \cite{10.1145/3719160.3736613}; Project Euphonia adapted interfaces to dysarthric speakers’ fatigue constraints \cite{macdonald21_interspeech, martin2025project}; and co-design frameworks empowered low-resource language communities to set culturally grounded collection criteria \cite{huaman2025quechua, emezue2025naijavoices}.
Granting communities authority over annotation is equally important. Standard labeling pipelines often fail for non-standard dialects and low-resource languages because annotators lacking lived context may normalize forms or mistranscribe code-switching \cite{Harris_2024, martin2022}. 
Community-led curation, from NaijaVoices’ native-speaker transcription conventions \cite{emezue2025naijavoices} to Common Voice’s crowdsourced validation \cite{ardila2020common}, reduces such label noise and improves representativeness for minoritized speakers \cite{kuhn2020ilt, mitra2016mixtec, reitmaier2024unmute, liwu2024stutter_cscw}.

% \textbf{Synthesize Data.} When real data collection is infeasible, generative augmentation can populate sparse regions of the acoustic space. For atypical and clinical speech, Aty-TTS \cite{wang2023improving} synthesizes speaker-preserving atypical speech to improve SLU fairness, TTDS \cite{leung24_interspeech} generates dysarthric speech for ASR data augmentation, and voice conversion has been used to create dysarthric variants in low-resource languages \cite{li25f_interspeech}. StutterAug \cite{Singh_2024} simulates realistic stuttering events to reduce bias in synthetic speech detection. For accent-related fairness, cross-lingual voice conversion and speed perturbation generate accented variants of majority speech \cite{zhang22n_interspeech}, and similar techniques approximate underrepresented age groups \cite{zhang24d_interspeech, zhao23c_interspeech, 9616228, shahnawazuddin20_interspeech}.

\textbf{Augment Data.} When collection is infeasible, augmentation can populate sparse acoustic regions via synthesis or signal transformation.
For clinical speech, TTS-based synthesis generates utterances that preserve speaker and disorder characteristics: Aty-TTS \cite{wang2023improving} for Spoken Language Understanding (SLU) and TTDS \cite{leung24_interspeech} for dysarthric ASR. 
VC extends this to low-resource languages by producing dysarthric variants without target-language clinical data \cite{li25f_interspeech}, and StutterAug \cite{Singh_2024} simulates disfluencies to reduce the elevated false positive rates that synthetic speech detectors exhibit on speech-impaired speakers.
For accent fairness, cross-lingual VC increases the amount of non-native-accented training speech, and combining it with speed perturbation or SpecAugment reduces the bias further in Dutch and Flemish ASR \cite{zhang22n_interspeech, Zhang_2023}.
For age-related gaps, VC-based methods convert adult speech into child-like variants or increase inter-child diversity \cite{zhang24d_interspeech, zhao23c_interspeech, 9616228, shahnawazuddin20_interspeech}, while vocal tract length normalization and speed perturbation address broader age and accent disparities \cite{patel2023vtlnbias}.
Pitch and formant manipulation address gender data imbalance by augmenting underrepresented f0 ranges \cite{fucci2023nopitch}.
Such augmentation can also work without demographic labels, as \citet{lopez2026okaura} show by reducing demographic disparity in wake-word detection with spectrum-level augmentation

Both strategies have inherent limits. Participatory collection is costly and difficult to scale beyond individual communities, while synthetic augmentation risks introducing generator artifacts that do not reflect the true acoustic distribution of the target group, potentially trading one form of bias for another.
% \textbf{Synthesize data.} Aty-TTS \cite{wang2023improving} and TTDS \cite{leung24_interspeech} synthesize atypical speech that preserves speaker-specific characteristics, helping spoken language understanding systems adapt to atypical articulation patterns. 
% StutterAug \cite{Singh_2024} simulates realistic stuttering events to improve robustness and reduce bias in synthetic speech detection. 
% Accent-related fairness is often addressed using cross-lingual voice conversion or speed perturbation to generate accented variants of majority speech \cite{zhang22n_interspeech}. 
% Similar techniques are used to approximate underrepresented age \cite{zhang24d_interspeech, zhao23c_interspeech, 9616228, shahnawazuddin20_interspeech} or clinical groups \cite{bhattacharjee2025fairness, li25f_interspeech}. 
% In emotion recognition, augmentation-driven balancing \cite{Tsai_2025} helps reduce performance disparities between male and female speakers. However, synthetic speech may introduce generation artifacts or fail to capture the full acoustic diversity of a target subgroup, risking a synthetic-to-real domain gap that limits downstream generalization.

\subsubsection{Strategic Resampling and Rebalancing}
While expanding datasets is ideal, addressing imbalances in existing corpora often necessitates \textit{strategic resampling}. 
The standard empirical risk minimization (ERM) objective naturally biases optimization towards majority groups, as their gradients dominate the parameter updates. 
Resampling techniques interventionally alter the training distribution, typically via \textit{downsampling} majority cohorts or \textit{upsampling} minority ones, to ensure that underrepresented acoustic profiles exert a proportional influence on the learning trajectory.

% In the speech domain, this strategy serves as a critical baseline across diverse tasks. For ASR and SV, downsampling has proven effective in mitigating bias across gender \cite{elghazaly-etal-2025-fairness, Estevez_2023}, age \cite{maison23_interspeech}, and accent \cite{dheram2022toward, Fenu_2021, 10447116}. By preventing the model from collapsing into majority-class priors, this approach effectively equalizes performance metrics such as False Rejection Rates. Furthermore, in pathological speech, rebalancing represents a necessity rather than a simple optimization heuristic. Unless the dominance of healthy speech samples is reduced, joint models often fail to capture the subtle acoustic markers of disorders \cite{riad-etal-2020-identification, nunez2024noninvasive}.

In the speech domain, this strategy serves as a critical baseline across diverse tasks. 
For ASR and SV, rebalancing the training distribution, whether by downsampling dominant cohorts or oversampling underrepresented ones, often narrows performance gaps across gender \cite{elghazaly-etal-2025-fairness, Fenu_2021}, age \cite{maison23_interspeech}, and accent \cite{dheram2022toward, 10447116, Estevez_2023}. 
By preventing the model from collapsing into majority-class priors, this approach effectively equalizes performance metrics such as False Rejection Rates. 
% Furthermore, in pathological speech detection, rebalancing represents a necessity.
% Clinical corpora are often heavily imbalanced between healthy and pathological recordings, and classifiers trained on such skewed data perform poorly on the minority class; undersampling the healthy majority or oversampling the minority substantially recovers this performance \cite{nunez2024noninvasive, lee2023smote, zhang2022class}.

Rebalancing is equally important in clinical speech detection.
When the prevalence of a condition differs across demographic groups, a detector can learn the group from speech and use it as a shortcut for the diagnosis, so its errors fall unevenly across groups and its reported accuracy overstates what it has learned about the condition.
Controlling for demographic and other confounders markedly lowers the reported accuracy of COVID-19, dementia, and Parkinson's detectors~\cite{coppock2024audio, shukla2026bridge2ai, brenner2023classification}, and challenge datasets for cognitive impairment match patients and controls on age and gender for the same reason~\cite{luz20_interspeech, luz21_interspeech, luz2024adressm, luz24_interspeech}.
Rebalancing addresses the shortcut at its source.
Training on demographically balanced data avoids the subgroup gaps that skewed training produces in COVID-19 detection~\cite{han2022sounds}, and balancing gender within each diagnostic class removes the correlation between gender and depression that the shortcut exploits~\cite{bailey2021gender}, which reduces disparate impact between genders~\cite{hu2026mitigating}.

However, resampling is not without cost. Aggressive downsampling improves fairness metrics often at the expense of \textit{data efficiency} and potentially global accuracy, as it discards informative samples. Conversely, naive upsampling risks overfitting to the limited diversity of minority examples. These costs also help explain why ASR work rarely downsamples healthy speech, repurposing it instead as an augmentation resource for underrepresented groups.

\subsubsection{Data selection and curriculum learning}
% Data selection and curriculum learning address fairness by prioritizing training samples that are challenging or fairness-critical, rather than sampling uniformly from the corpus. Challenging Subgroup Identification \cite{koudounas-etal-2025-privacy} identifies utterances that the model repeatedly misrecognizes, typically from minority speakers, and samples them more frequently without revealing sensitive attributes. A related divergence-aware strategy \cite{Koudounas_2025} dynamically monitors subgroup performance and steers training toward the worst-performing subgroups, or even triggers the acquisition of additional data for these groups.  Collectively, these methods show that fairness can be improved not only by altering datasets but also by directing training attention toward the right samples. By prioritizing fairness-critical examples, they create curricula that explicitly target subgroup disparities.
% While the strategic resampling methods discussed previously address quantitative disparities by adjusting group sizes, they treat all samples within a demographic group as equally informative. In contrast, \textbf{data selection and curriculum learning} focus on the utility of individual samples. These methods operate on the premise that not all minority examples contribute equally to fairness. A hard example near the decision boundary provides more valuable gradient information than an easy or prototypical one. Therefore, these algorithms dynamically prioritize samples based on learning difficulty, mathematical diversity, or uncertainty.

Strategic resampling adjusts how much data each group contributes, but it treats all samples within a group as equally useful.
Data selection and curriculum learning instead prioritize the individual samples that benefit underserved groups most, either by weighting them more heavily during training or by collecting and transcribing them first.

%\textbf{Error-Driven Prioritization.} Approaches inspired by the \textit{Just Train Twice} framework \cite{liu21f_jtt} use training difficulty as a proxy for marginalization. 
% In speech, the same principle supports selective preprocessing that is applied only under disparity-inducing conditions, such as DENOASR’s targeted denoising \cite{rai2024denoasr}. 
% It also motivates gap- and divergence-aware data reallocation, including privacy-preserving selection and acquisition policies that prioritize collecting or labeling utterances most likely to close subgroup gaps under limited budgets \cite{koudounas-etal-2025-privacy, Koudounas_2025, koudounas24b_interspeech, 10446326}. 
% Complementarily, confidence-guided augmentation can convert high-confidence samples to the voices of low-confidence speakers, as in CO-VADA \cite{Tsai_2025}.

\textbf{Difficulty-Driven Prioritization.} These methods rank training samples by difficulty and use the ranking to decide which samples to emphasize or in what order to train on them.
Just Train Twice~\cite{liu21f_jtt} implemented in speech by~\citet{Lin_2025_2} measures difficulty by the errors of an initially trained model. 
Then, it trains the final model on the same data with the misclassified examples repeated several times, which upweights the groups the first model serves poorly without requiring group labels.
Curriculum learning uses difficulty to order the training data instead.
In dysarthric ASR, \citet{hsieh24_interspeech} fine-tune a recognizer in stages, first on typical speakers and then on one intelligibility group at a time from high to very low, so that the model learns from speech close to typical speech before facing the most atypical speech.
Because the later stages degrade recognition of the more intelligible groups, \citet{hsieh25_interspeech} keep the model from the end of every stage and combine their outputs with a decision model, which makes the improvements more even across groups.

% \textbf{Targeted Subset Selection.} When computational or annotation budgets are constrained, selecting the mathematically optimal subset for fairness becomes critical. Algorithms like DITTO \cite{kothawade-etal-2023-ditto} utilize submodular mutual information to curate training batches that maximize acoustic diversity while matching a target distribution.

% \textbf{Fairness-Aware Active Learning.} Unlike static filtering, active learning dynamically triggers the acquisition of new labels to address representation gaps. An active-learning approach for accent adaptation prioritizes utterances matching the target dialect by augmenting uncertainty sampling with a cross-entropy relevance criterion \cite{6424250}. 

\textbf{Subgroup-Targeted Acquisition.} When more data can be obtained, it is most useful for the subgroups on which the model performs worst.
\citet{10446326} and \citet{Koudounas_2025} identify these subgroups from metadata and prioritize acquiring their data, which narrows the gap to the worst subgroup more than adding all available data does.
\citet{koudounas-etal-2025-privacy} remove the need for the sensitive attributes of the new data by training a model to recognize utterances from these subgroups.
Existing pools of speech can also be searched for data that resembles an underserved group.
\citet[][preprint]{kim2024clustering} use an accent classifier to find Indian-accented speech in a large pool without accent labels, and fine-tuning on it lowers Indian-accent WER.
For dysarthric ASR, \citet{xiong2020source} add utterances from other dysarthric speakers that resemble that speaker's speech, which lowers WER for speakers with moderate to severe dysarthria.

\textbf{Budget-Constrained Selection.} When the budget for recording or annotating is limited, the choice of samples matters, and it can rest on predicted difficulty, similarity to the target group, or both.
\citet{awasthi2021error} use a phoneme-level error model to predict which sentences an accented speaker's recognizer will find hardest and spend the recording budget on those sentences.
Instead, DITTO~\cite{kothawade-etal-2023-ditto} uses submodular mutual information to select utterances similar to a few target accent exemplars. 
For multiple accents, it selects a balanced number of samples per accent to prevent crowding out.
Earlier, to adapt ASR models to accents that were underrepresented in or absent from their training data, \citet{6424250} transcribed the utterances that the model was uncertain about and that also closely matched the target accent.
This required far less transcribed data than uncertainty sampling alone to reach the same WER.

\subsubsection{Counterfactual Data Augmentation}
\label{ssec:cda}
Counterfactual data augmentation operationalizes causal fairness by synthesizing matched “what-if” speech pairs where a sensitive attribute is altered while task-relevant attributes remain fixed \cite{kusner2017counterfactual}. The pairs either enforce consistency regularization, penalizing the model when its predictions change between factual and counterfactual inputs, or simply rebalance the training data with attribute-flipped copies.
 
In ASR, \textit{Counterfactually Fair ASR} \cite{Sarı_2021} flips gender, age, and education via adversarial auto-encoder voice conversion and trains with the consistency penalty.
Fucci et al.~\cite{fucci2023nopitch} use signal processing alone, shifting pitch and formants so that some utterances sound like they were spoken by the other gender. 
This simulates more voices of the underrepresented female speakers and lowers WER for women.
\citet[][preprint]{klumpp2023synthetic} converts native US English utterances into seven non-native accents while keeping the speaker's voice, but the benefit is limited.
Recognition improves only on the accents used in augmentation, and for an SSL model pretrained on native speech, WER even rises on some accents.

CO-VADA \cite{Tsai_2025} addresses bias in SER by generating identity-shifted counterfactuals, forcing the model to unlearn spurious correlations between vocal timbre and emotion labels.
Voice conversion produces counterfactual samples only if it also changes the sensitive attribute.
Conventional voice conversion preserves the accent of its source speech, so when \citet{zhang22n_interspeech} convert non-native Dutch speech to new voices to address accent bias, the accent stays the same, and the conversion adds speakers to the underrepresented accent group rather than creating counterfactuals.

Two concerns exist for these methods.
First, all these methods assume the transformation alters only the target attribute; in practice, conversion systems also perturb correlated factors such as prosody and speaking rate, weakening the causal validity of the counterfactuals. 
Second, converting identifiable speakers along socially salient attributes goes beyond what corpus consent usually covers.
This is one motivation for the shift toward fully synthetic sources in \S\ref{sec:future_d}.
% Generative models also serve as diagnostic probes: Gender-Ambiguous TTS \cite{szekely23_interspeech} synthesizes voices with indefinite gender traits to isolate which acoustic cues trigger downstream bias, while tools like Hear Me Out \cite{bokkahallisatish25_interspeech} use real-time conversion to let users directly experience how conversational agents react to their voice under different demographic profiles.

% \begin{figure*}
%     \centering
%     \includegraphics[width=0.8\linewidth]{figures/debias/Overview of debias methods with feature-level inter- ventions.png}
%     \caption{Overview of debias methods with feature-level interventions.}
%     \label{fig:feature_debias}
% \end{figure*}
% \centering
%     \includegraphics[width=\linewidth]

\subsection{Feature-level methods}
These methods change what demographic information a model’s features or representations carry by projecting it out, masking or removing it, or compressing the model. Some are applied to a frozen pretrained model, whereas others retrain the model on the edited features.

% \subsubsection{Geometric Subspace Projection} Geometric approaches offer a stable alternative to adversarial training by using linear algebraic operations to directly remove sensitive information from the embedding space \cite{silnova22_odyssey}. 
% Krishnan et al. mitigate bias in self-supervised models by identifying a "bias subspace" comprising directions that maximally correlate with sensitive attributes \cite{krishnan24_interspeech}. 
% By projecting embeddings onto the orthogonal complement of this subspace, the method neutralizes demographic leakage without retraining, effectively preserving linguistic utility.
% Because the projection is closed-form and deterministic, it does not require additional training and does not introduce optimization instability.
% However, it can only remove linearly encoded bias; nonlinear demographic information may survive the projection and leak into downstream predictions.

\subsubsection{Geometric Subspace Projection} Geometric approaches offer a stable alternative to adversarial training, removing sensitive directions from the embedding space with closed-form linear algebra.
\citet{krishnan24_interspeech} apply LEACE~\citep{NEURIPS2023_d066d21c} to ASR encoders, erasing linearly encoded gender layer by layer with little WER cost when applied to the final layers.
Speaker verification has long used the same operation as nuisance-attribute projection to remove channel variability~\citep{solomonoff04_odyssey} and language variability~\citep{silnova22_odyssey}, and fairness work simply points it at demographic directions instead.
Linear projection, however, removes only linearly encoded bias.
\citet{noe2022bridge} address this with a normalizing flow, an invertible neural transform, that maps x-vectors to a latent space where sex occupies a single dimension.
Zeroing that dimension and mapping back removes far more gender information than LDA whitening.
However, removing attribute information does not by itself narrow performance gaps. When \citet{parekh2026aces} partially projects an accent subspace out of a frozen ASR encoder, both the overall WER and the WER gap between accents grow, because accent information is entangled with information the model needs.

\subsubsection{Feature Masking and Removal}
These methods assume sensitive information is concentrated in specific parts of the spectrogram or feature space.
Once located, these regions are masked or removed.
% By identifying and masking these "biased regions," models are forced to rely on robust, invariant cues. 
Saliency maps can locate these regions by highlighting which inputs the classifier relies on.
% Realizations of this strategy employ saliency-driven gradient localization to identify input patches correlated with age or accent, thereby redirecting the model's focus toward intrinsic task features \cite{wang2023beyond, sameti2025accent}. 
\citet{wang2023beyond} use them to find the parts of a voice recording that reveal a speaker’s age in Parkinson’s disease detection. \citet{sameti2025accent} use them to find the parts of a spectrogram that reveal accent in ASR.
Masking these parts encourages the model to rely on cues shared across groups.
The same idea extends to hand-crafted feature sets.
\citet{Yang_2024} drop the features of eGeMAPS that differ most between men and women.
Nonetheless, performance remains higher for women than for men.

Masking can also apply to the internal embeddings of a trained model. 
In audio deepfake detection, \citet[][preprint]{fursule2026trustworthy} identify the most gender-informative embedding dimensions and set them to zero or their gender-averaged values. 
This simple adjustment narrows the gender gap in false positive rates without any retraining.

% In clinical speech analysis, Grad-CAM masking has been used to pinpoint age-related spectrotemporal regions in Parkinson’s detection models \cite{wang2023beyond}. 
% In accent-fair ASR, saliency-driven spectrogram masking identifies accent-sensitive time–frequency patches and suppresses them with a probabilistic mask \cite{sameti2025accent}. In dementia detection, confounder-aware filtering methods detect latent features correlated with gender and reweight them to minimize demographic leakage \cite{sheng-etal-2025-mitigating}.
% Similarly, within the latent space, methods like confounder-aware filtering apply soft reweighting to down-scale embedding dimensions that exhibit high mutual information with sensitive attributes, effectively orthogonalizing task predictions from gender-correlated subspaces \cite{sheng-etal-2025-mitigating}.

While computationally more stable than adversarial training, these methods face the localization bottleneck. They rely heavily on the fidelity of the attribution method. 
If the bias regions overlap with content regions, such as when gender cues are entangled with emotion pitch, suppression can lead to significant degradation in task utility.

\subsubsection{Model compression}
Model compression techniques such as pruning, quantization, and distillation can implicitly regularize away spurious correlations by reducing model capacity, thereby limiting the encoding of demographic cues that drive unfair predictions. 
Compressed models often correlate less with sensitive attributes \cite{xu2022can}, and in speech SSL representations, row pruning provides the most consistent debiasing effect \cite{lin2024social}. 
However, indiscriminate compression risks ``disproportionate forgetting,'' concentrating performance degradation on underrepresented subgroups \cite{hooker2020compressed, irina2023the}. 
In on-device keyword spotting, \citet{hutiri2023tiny} find that pruning away more than half of a model’s weights can sharply increase the performance gap between female and male speakers.
% Strictly speaking, compression modifies encoder weights and thus sits at the boundary of our feature-level definition; we group it here because it is applied post hoc to a trained model without fairness-specific retraining.

\medskip
Compared with NLP, where iterative nullspace projection \cite{ravfogel-etal-2020-null} and closed-form linear erasure \cite{NEURIPS2023_d066d21c} have become standard post-hoc debiasing tools, feature-level methods in speech remain few. Speech embeddings are continuous and variable-length, and demographic cues tend to be more tightly entangled with linguistic content than in text representations, making direct adaptation of these algebraic techniques a nontrivial and largely unexplored problem.
% Contrastive learning methods mitigate bias by pulling together representations that should be similar across demographic groups and separating those that encode task-relevant distinctions. By shaping the geometry of the embedding space, these methods encourage demographic-agnostic representations without requiring explicit removal of sensitive attributes. For example, CLUES applies multi-level contrastive objectives to pull together samples that differ only in subgroup membership, improving fairness for underrepresented speakers \cite{koudounas24b_interspeech}. Similarly, \cite{han2021supervised} aligns the embeddings of identical utterances spoken with different accents, forcing the model to map diverse pronunciation patterns to a shared, accent-invariant linguistic space.

% \begin{figure*}
%     \centering
%     \includegraphics[width=0.9\linewidth]{figures/debias/train_intervention.drawio.png}
%     \caption{Overview of debias methods with training-level interventions.}
%     \label{fig:training_debias}
% \end{figure*}

\subsection{Training-level algorithms}
This is the largest category in the speech fairness literature. We order the methods along a spectrum: from suppressing demographic information in the learned representation, through constraining the loss function for equitable outcomes, to explicitly incorporating demographic knowledge into the model, and finally to adapting pretrained models for underserved subgroups.

\subsubsection{Adversarial Disentanglement}
\label{ssec:adversarial}
Adversarial disentanglement formulates fairness as a minimax optimization problem. The goal is to learn a representation that maximizes utility for the primary task while minimizing the information available to an auxiliary discriminator regarding sensitive attributes. Formally, this is typically achieved via a Gradient Reversal Layer (GRL)\cite{ganin2015dann}, which encourages the encoder to remove demographic leakage, thereby learning representations that are invariant to group membership \cite{madras2018learning}.

\textbf{Standard Gradient Reversal Framework}: The foundational strategy employs a simple GRL to backpropagate negative gradients from a sensitive attribute discriminator. This baseline architecture is widely used to normalize acoustic variability in ASR \cite{8462663, 8462452,  vishal2025inclusive, woszczyk2020domain, das2021best}, and to align representations across speaker groups, such as child and adult \cite{rimita2020learning}. 
In paralinguistic tasks, it removes gender or identity cues from depression detection \cite{kim2025domain}, and SER models \cite{kasun2022discriminative, Upadhyay_2025, cai2021unsupervised}.
In SER, it has also been used to remove bias stemming from the annotators' gender \cite{chien2023achieving}.
The same principle extends to music: \citet{Gu_2023} combine adversarial training with conditional alignment to enforce gender invariance in singing voice transcription.
However, a plain GRL does not always help.
For cross-age SV, \citet{qin22cross} find that it raises EER and instead propose an age-decoupling adversarial scheme.

\textbf{Generative and Reconstruction-Based Architectures}: To prevent feature collapse, generative approaches enforce a reconstruction constraint alongside the adversarial objective. For instance, AIPNet \cite{9053098} and DAAE \cite{gao2021dann} utilize generative adversarial pre-training to separate accent- or speaker-invariant features. Similarly, LR-VAE \cite{huang2021attribute} and FS-VAE \cite{huang2022attention} integrate GRL-based adversarial purification with variational autoencoder (VAE)-style reconstruction to suppress sensitive-attribute leakage without discarding essential speech information.

% \textbf{Attention-Guided Adversarial Learning}: Standard GRL treats all features uniformly, which is suboptimal for complex signals \cite{meng2019attentive}. Architectures like AANet \cite{10038197} overcome this by integrating adaptive attention mechanisms within the adversarial loop. This allows the model to dynamically weight features, selectively preserving linguistically relevant frames while suppressing segments heavily correlated with demographic attributes.

\textbf{Attention-Guided Adversarial Learning}: Standard GRL applies uniform gradient reversal to all frames, which can under-suppress bias cues concentrated in specific regions. Attention-guided variants address this by learning to focus the adversarial signal on bias-salient dimensions. AADIT \cite{meng2019attentive} weights deep features by their importance to domain classification, while AANet \cite{yang2022adaptive} adds an adaptive attention module after the encoder to selectively remove accent information. Both reduce ASR disparities across recording conditions and accents, respectively.

This family carries two known caveats. Minimax optimization can be unstable and sensitive to the discriminator's capacity, and the removal is often incomplete, since post-hoc probes can still recover sensitive attributes from adversarially purified representations \cite{elazar2018adversarial}.

% \textbf{Adversarial Transfer and Metric-Regularized Learning}: Recent variations extend the paradigm to transfer learning and metric spaces. Acc-PT \cite{das2021best} applies adversarial transfer to adapt high-resource representations to low-resource accented domains. In SER, metric-based strategies combine adversarial objectives with contrastive learning or layer-anchoring to enforce stricter geometric separation between emotion and speaker identity clusters \cite{Kim_2025, Upadhyay_2025}.

% In ASR, domain-adversarial training has been widely adopted. Representative examples are DAT\cite{8462663}, \cite{8462452}, AIPNet \cite{9053098}, AANet\cite{10038197}, Domain Invariant Representations for Child-Adult \cite{9054276}, and Acc-PT\cite{das2021best}.

% For SV, adversarial objectives reduce demographic leakage in embeddings. ADAL\cite{qin22_interspeech} explores cross-age speaker embedding, while \cite{peri2022train} examines the fairness-utility trade-off.

% For pathological and clinical speech, adversarial learning helps suppress pathology-related variability. 
% For example, depression detection might correlate with speaker gender \cite{kim2025domain}, and ASR might be poor for disorder patients \cite{10.1121/10.0037269, woszczyk20_interspeech}. 
% In SER, adversarial invariance also helps isolate emotion from speaker identity \cite{chien2023achieving, Upadhyay_2025, Kim_2025}. 

\subsubsection{Geometric Alignment via Contrastive Learning}
% Contrastive debiasing mitigates bias by shaping the embedding geometry to enforce invariance, ensuring that samples sharing the same target label are aligned across diverse demographic groups while attribute-specific variability is suppressed.
Contrastive learning trains a model to place chosen pairs of samples close together in the embedding space and other samples far apart.
% One paradigm relies on supervised cross-group pairing, where models explicitly align representations of parallel utterances that share task annotations but differ in sensitive attributes like accent or gender \cite{han2021supervised, 10447167}. In contrast, alignment without parallel data regularizes latent clusters or distributions to prevent the geometric isolation of underrepresented subgroups, though group labels may still be required \cite{koudounas24b_interspeech, Kim_2025}. Advancing to finer granularity, frameworks like DyPCL apply these objectives at the phoneme level, capturing local acoustic nuances critical for clinical or atypical speech conditions \cite{lee2025dypcl}. 
\citet{han2021supervised} pair each sentence with its noise-injected, spectrogram-augmented, and TTS-generated versions, so that representations of accented speech become less sensitive to differences in pronunciation.
FairASR~\cite{Kim_2025} and CLUES~\cite{koudounas24b_interspeech} both pair samples by group membership but pursue opposite goals.
FairASR applies a gradient-reversed contrastive loss over demographic labels so that representations no longer cluster by group.
CLUES instead pulls together samples of the same subgroup and pushes different subgroups apart, so that the model represents underperforming subgroups more distinctly.
Pairs can also be formed below the utterance level, as DyPCL~\cite{lee2025dypcl} applies contrastive learning to phoneme segments for dysarthric ASR.

\subsubsection{Optimization Objective Modification} 
This paradigm alters the loss function or gradient update trajectory to constrain the model's behavior, ensuring equitable performance across groups without necessarily changing the model architecture.

\textbf{Regularization for Distribution Parity.} These approaches augment the standard training objective with a penalty term, formulated as $\mathcal{L}_{total} = \mathcal{L}_{task} + \lambda \mathcal{L}_{fair}$, to constrain the model's optimization trajectory toward fairness. Depending on the specific task requirements, these constraints target either statistical independence or error rate parity. 

To enforce \highlight{demographic parity}, researchers penalize the divergence of model predictions; for instance, \citet{10287972} minimized the Wasserstein distance between demographic subgroups in automated video interviews, while \citet{chattoraj2020fairness} utilized a heterogeneity regularizer in public speaking assessment to decouple sensitive attributes from performance ratings. Conversely, when enforcing \highlight{equalized odds} (equal error rates are preferred), approaches like FairM2S integrate equalized odds constraints into a meta-learning framework to penalize gender-dependent disparities in stress detection \cite{shelke2025fairness}. 
Similarly, for ASR, \citet{GAO202276} proposed a modified Inclusive CTC loss to balance recognition performance across diverse accents. 
Similar penalty terms have been added to SER losses \cite{gorrostieta19_interspeech, Lin_2025_2, chou2024inter}. 
Going further, minimax-style variants move beyond a fixed criterion and instead penalize the worst-case inter-group gap: ERM-MinMaxGAP \cite{pang2026minmaxgap} minimizes the largest male-female loss disparity across languages in SpeechLLM-based SER.

\textbf{Reweighting and Distributionally Robust Optimization (DRO).} These methods adjust per-sample contributions to the loss to counteract data imbalance. Static approaches use inverse-frequency weighting, while dynamic variants train an auxiliary bias-capturing model that upweights samples conflicting with majority shortcuts \cite{Lin_2025_2, nam2020lff}, or employ an adversarial network to surface and reweight poorly performing demographic groups~\cite{Jin_2022}. 
DRO formalizes this intuition by directly optimizing worst-case subgroup performance~\cite{sagawa2020distributionally}; CTC-DRO~\cite{bartelds2026ctcdro} further adapts group DRO to CTC-based multilingual ASR; EMO-Debias~\cite{Lin_2025_2} applies DRO to strictly bound error rates of the most disadvantaged groups in SER.
When demographic labels are unavailable, discovered acoustic clusters can serve as group labels for reweighting or DRO, restoring weight to underrepresented clusters that standard training sees too rarely~\cite{lin25c_interspeech}.

\textbf{Information-theoretic constraint.}
Information-theoretic approaches reduce bias by explicitly formulating fairness as a mutual information (MI) trade-off~\cite{colombo2023novel}.
The objective is to learn a representation that maximizes MI with task-relevant information while minimizing MI with sensitive cues.
% A recent application is WavShape \cite{baser25_interspeech}, which uses MI estimators~\cite{pmlr-v80-belghazi18a} to suppress sensitive information and preserve linguistic content, yielding privacy-aware and demographically invariant speech representations.
WavShape \cite{baser25_interspeech} applies this idea to speech by training a projection layer on top of a frozen Whisper encoder. 
The projection lowers the mutual information between the embeddings and a sensitive attribute, while performance on the downstream task stays comparable.
However, MI is difficult to estimate reliably in high-dimensional continuous embeddings; variational estimators can be sensitive to critic capacity, batch size, and optimization, and may require careful tuning to avoid unstable training \cite{pmlr-v108-mcallester20a}.

\subsubsection{Explicit Attribute Modeling} Rather than ignoring sensitive attributes, this paradigm explicitly incorporates them into the model to account for systematic variations in speech production.

\textbf{Bias-Aware Conditioning.} 
These methods supply demographic or speaker identity signals as additional model inputs so that the network can account for systematic variation across groups. 
At the \textit{group level}, conditioning signals include discovered speaker-cohort membership for ASR \cite{dheram2022toward}, parallel group-adapted encoders with score fusion for SV \cite{Shen_2022}, and one-hot-encoded age, gender, and education attributes fused with acoustic embeddings via adaptive gating for cognitive screening (Cog.) \cite{azadmaleki2025speechcare}. 
In speech translation, \citet{gaido2023build} inject speaker gender metadata to prevent masculine-default outputs, while \citet{bansal2025addressing} eliminate the explicit gender input by fine-tuning on LLM-corrected references so the model infers gender from acoustic cues alone. 
For low-resource dialects, \citet{tsai2025hakka} condition ASR on dialect embeddings to bridge performance gaps across Hakka varieties that standard models underserve.
In cross-lingual SV, EcoSpeak~\cite{Sharma_2024} conditions the comparison of speaker embeddings on the languages of the two recordings in a trial to narrow the EER gap between same-language and cross-language trials.
At the \textit{individual level}, Triantafyllopoulos and Schuller \cite{triantafyllopoulos2024enrolment} use attention over enrollment utterances to personalize emotion recognition, capturing individual idiosyncrasies that coarse group labels miss.

\textbf{Multi-task Learning (MTL) and Auxiliary Learning.} 
While adversarial disentanglement (\S\ref{ssec:adversarial}) strips demographic cues to achieve invariance, MTL takes the complementary view that explicitly modeling group variation as a structured factor better serves tasks where demographic differences entail legitimate acoustic variation. 
By jointly optimizing the primary objective alongside auxiliary heads that predict group-related attributes, the shared encoder is forced to capture subgroup-specific patterns rather than overfitting to the dominant demographic \cite{wang2024timit, montalvo20_interspeech, peri2023bias}. 
Beyond speech, conditioning model heads or parameters on demographic attributes outperforms unconditioned baselines. 
The benefits are largest for underrepresented groups \cite{oneto2019taking, wang2020towards}.
In speech, however, empirical evidence on auxiliary tasks remains mixed.
Some studies find that they help the worst-served groups.
For instance, adding an auxiliary accent classifier to Spanish ASR reduces WER most for the worst-recognized accent in each test set \cite[][preprint]{chitkara2022pushing}.
Similarly, auxiliary proficiency classification yields the largest error reductions for the least proficient L2 speakers compared with standard fine-tuning \cite[][preprint]{sun2025proficiency}.
Other studies observe no such equalizing effects.
Auxiliary accent classification can improve accuracy across all English accents without consistently narrowing the performance gaps between them \cite{XIAO2025}.
Under a different architecture, the same Spanish accent task actually degrades WER across all accents \cite[][preprint]{chitkara2022pushing}.
Gender classification raises the equal error rate for both men and women in speaker verification \cite{chouchane2023fairness}, and dialect identification can hurt the dialect with the least data \cite{dan2022multitask}.
Whether an auxiliary objective truly benefits underserved groups must therefore be verified.
% In accented ASR, auxiliary accent classification heads compel the model to distinguish valid pronunciation variants from recognition errors, boosting accuracy for non-native speakers \cite{zhang21j_interspeech, XIAO2025} and dialects \cite{zhao2019tibetan, imaizumi2022end, dan2022multitask}. The same principle extends to low-resource language transfer \cite{app13095239} and multilingual TTS, where a language identification auxiliary preserves prosodic structures for diverse inputs \cite{zhang22i_interspeech}.

\textbf{Interpretable Feature Routing.} Rather than soft conditioning or auxiliary losses, Fair-Gate \cite{qu2026fairgate} introduces learnable routing masks that split intermediate features into identity and sex branches, combined with risk equalization across sex groups to close SV performance gaps.
 
Most methods in this subsection presuppose demographic annotations at training time, whether as conditioning signals, auxiliary labels, or routing targets; the individual-level enrollment scheme above \cite{triantafyllopoulos2024enrolment} is the rare exception, replacing group labels with the target speaker's own recordings. For the rest, missing or privacy-restricted annotations block training outright, while noisy annotations degrade the learned group models.

\subsubsection{Targeted Adaptation Strategies} Standard training on pooled data often overfits to the majority distribution. Rather than retraining from scratch, targeted adaptation strategies selectively update a pretrained model, whether through partial fine-tuning, parameter-efficient modules, or dynamic routing, to recover performance on underserved subgroups.

\textbf{Partial Fine-Tuning.} These methods update only selected components of a pretrained model.
In pronunciation assessment, \citet{nijat2026nativenorms} show that the native-norm ground truth used by standard mispronunciation detectors is itself a fairness risk.
Fine-tuning only the phonetic module on L2 speech shifts the reference toward the target population and improves detection accuracy, and the fine-tuned detector performs comparably across learners' first languages.
In Alzheimer's disease detection, \citet{jia25_interspeech} fine-tune only the decoder of Whisper, first on English, Chinese, and Spanish and then on a few Greek recordings, which raises accuracy on Greek while preserving it on the other languages. 
In SV, \citet{shetty25_wocci} insert an adapter into adult-trained models and fine-tune the adapter, the classifier, and the backbone in alternation on children's speech, which lowers children's EER more than updating all parameters at once. 
In speech translation, \citet{mizumoto2026dialect} train only the adapter between the speech encoder and the LLM on dialectal speech, which narrows the gap between Japanese dialects and standard Japanese.

% Fine-tuning only the phonetic module on L2 speech aligns the reference with the target population, improving detection accuracy while maintaining comparable performance across learners' native languages.

\textbf{Parameter-Efficient Fine-Tuning (PEFT).} 
PEFT adapts a shared backbone to an underserved population, or to individual speakers, by training a small set of added parameters while keeping the backbone fixed. 
Residual adapters that update less than 0.5\% of the parameters achieve gains close to full fine-tuning for atypical and accented speakers \cite{tomanek-etal-2021-residual}.
Subsequent applications span child speech \cite{10447004}, speakers with autism spectrum disorder \cite{park25g_interspeech}, per-speaker adapter fusion for dysarthric ASR \cite{qi23b_interspeech}, and inclusive Low-Rank Adaptation (LoRA) for elderly and transgender Tamil speakers \cite{acharya-etal-2025-junlp}.
Because data from these populations are scarce, such adaptation can also bias the model toward the speakers seen in training.
Separate speaker and deficiency adapters, which can be recombined for new speakers, improve recognition of unseen dysarthric and elderly speakers \cite{hu2024structured}.

Beyond adapting to a single target group, PEFT can embed fairness objectives directly.
\citet{swain2025towards} fine-tune adapters with a loss that combines spectral decoupling, group DRO, and invariant risk minimization, which lowers the average WER across 26 accent groups, although the gap between the best and worst groups narrows only in some configurations.
Even without such objectives, the choice of PEFT method matters.
Benchmark evidence from SER shows that LoRA improves demographic fairness relative to other PEFT variants \cite{feng2023peft}. 

\textbf{Mixture-of-Experts (MoE) Routing.} Instead of assigning a fixed adapter per subgroup, MoE learns a router that selects or combines experts for each input, so the group label is not needed at test time.
\citet{hu25d_interspeech} initialize experts with severity- and gender-conditioned adapters and learn on-the-fly routing over HuBERT/WavLM for dysarthric ASR. MOPSA \cite{deng25_interspeech} clusters speaker-adaptive prompts into prompt-experts via $k$-means and routes them at both the encoder and decoder levels of Whisper for elderly speech.
 
Both PEFT and MoE paradigms rely on subgroup or speaker labels during training, such as accent, severity, gender, or speaker identity, and only MOPSA derives its experts by clustering learned speaker prompts.
When subgroup definitions shift across deployment contexts, the resulting specialization may not transfer

\subsection{Inference and post-processing}
Unlike training-centric approaches that require computationally expensive model retraining, inference-time interventions operate during the deployment phase. These methods mitigate bias by adapting the input signal, guiding the decoding process, or calibrating the final output scores. They offer a "plug-and-play" mechanism for fairness, particularly valuable when accessing model parameters is infeasible.

% \subsubsection{Test-time Input Adaptation}
% While voice conversion is used at training time to construct counterfactual pairs (\S\ref{ssec:cda}), it can also serve as a real-time preprocessing step that neutralizes demographic cues before the signal reaches the downstream model.
% \citet{electronics11101594} apply voice transformation to strip gender-specific characteristics from speech, and~\citet{alhumud2024improving} use VC to transform accented speech toward a native accent.
% For dysarthric speech, \citet{prananta22_interspeech} time-stretch each utterance to the speaking rate of healthy speakers, which narrows the phoneme error gap between the two groups.
% These methods work because deployed models are most accurate on the majority-group speech that dominates their training data; converting underrepresented inputs toward majority-group characteristics can therefore recover performance without new data or retraining.
% The same mechanism is also this family's normative weakness. 
% Adapting the user rather than the model enacts the acoustic assimilation critiqued in \S\ref{ssec:representation}, trading representational harm for recognition accuracy, and is therefore best treated as a pragmatic stopgap when the model itself cannot be modified.

\subsubsection{Test-time Input Adaptation}
Voice conversion can preprocess speech before speech feeds into downstream models.
\citet{electronics11101594} use it to remove gender cues from speech, and~\citet{alhumud2024improving} use it to convert accented speech toward a native accent.
Other methods do not convert the voice.
\citet{prananta22_interspeech} time-stretch dysarthric speech to the speaking rate of healthy speakers, which narrows the phoneme error gap between the two groups.
DENOASR~\cite{rai2024denoasr} denoises only recordings with low estimated intelligibility, which narrows the WER gap between female and male speakers at a small cost in overall WER.
Models are most accurate on the majority-group speech that dominates their training data, and voice conversion and time stretching help by moving underrepresented speech toward it.
This move toward the majority is also their weakness.
Adapting the user rather than the model enacts the acoustic assimilation critiqued in \S\ref{ssec:representation}, so these methods are best treated as a stopgap when the model itself cannot be modified.

\subsubsection{Test-time Representation Intervention}
Rather than adapting inputs or parameters, these methods directly manipulate frozen model representations at inference time to reduce demographic disparities.
% Nachesa and Niculae \cite{nachesa2025knn} augment Whisper \cite{pmlr-v202-radford23a} with a token-level $k$-nearest-neighbor datastore and show through explicit demographic analysis that retrieval-augmented decoding yields differentiated WER improvements across gender, accent, and age groups. 
% In a different vein, 
Activation steering adds learned direction vectors into targeted hidden layers without modifying weights. 
\citet{sun2026steering} extract the mean difference between standard-English and accented representations and inject them into a selected encoder layer, achieving consistent WER reductions. 
\citet{yang2026steering} extend this to zero-shot TTS by steering activations to produce accent-neutralized synthesis, and \citet{szekely23_interspeech} synthesize gender-ambiguous voices with controllable prosody.

\subsubsection{Test-time Parameter Adaptation}
These methods estimate a small set of subgroup-dependent parameters from test data at inference time, specializing the model for underrepresented demographics without full retraining.
On-the-fly feature-based Learning Hidden Unit Contributions (f-LHUC) \cite{geng23_interspeech, geng2025homogeneous} derives speaker-dependent transforms and variance-regularized spectral basis embeddings from minimal dysarthric or elderly test audio, reducing WER without gradient-based optimization. 
For children's speech, \citet{shi25_interspeech} show that applying prior Test-time Adaptation \cite{lin22b_interspeech, kim23f_interspeech} methods can obtain small WER reductions, although the gains vary across children and some children’s WER rises.

\subsubsection{Decoding and Rescoring Strategies}
Changing how hypotheses are searched or scored at inference time can reduce errors for under-served speakers without retraining the recognizer.
Cross-utterance rescoring \cite{10096820} builds an acoustic similarity graph across utterances from multiple speakers and applies label propagation so that high-confidence hypotheses from well-served accents guide recognition of underrepresented ones, explicitly mitigating majoritarian bias. 
\citet{southwell2024child} rescore the beam hypotheses of an off-the-shelf Whisper~\cite{pmlr-v202-radford23a} model with a language model and a term favoring a typical human speaking rate, which reduces errors on children's speech by discouraging hypotheses that are too short or repetitive.
Retrieval offers another route.
\citet{nachesa2025knn} interpolate the output distribution of Whisper with a token-level $k$-nearest-neighbor distribution retrieved from a datastore of its hidden states.
\subsubsection{LLM-Driven Correction} Leveraging the semantic reasoning of LLMs, recent frameworks mitigate bias through inference-time post-processing rather than expensive ASR model retraining. 
In medical domains, two preprints report that LLM post-correction reduces accent-induced terminology errors for non-native speakers, with a one-line prompt~\cite{Fatapour2025.08.29.25333548} or Chain-of-Thought prompting~\cite{adedeji2025multicultural}.
For dysarthric speech, a corrector fine-tuned on the N-best hypotheses of an unadapted Whisper model lowers its WER \cite{laquatra25_interspeech}.
However, correction does not always help underserved speakers.
% In child-adult conversations, LLaMA-based correction of zero-shot Whisper output  lowers error rates for both groups but improves adult speech more substantially, paradoxically widening the child-adult gap~\citep{xu25g_interspeech}.
In child-adult conversations, LLaMA-based correction of zero-shot Whisper outputs lowers error rates more for adults than for children, thereby widening the demographic gap \citep{xu25g_interspeech}.
On a dysarthric corpus not used in training, a fine-tuned LLM corrector nearly doubles the WER by changing words that were already correct, and including the recognizer’s word-level confidence scores in its prompts prevents this \citep{hernandez2026confidence}.

\subsubsection{Threshold calibration}
% Threshold calibration adjusts decision boundaries after training to equalize error rates across demographic groups \cite{10.5555/3157382.3157469}.
% In SV, a single global threshold disproportionately harms underrepresented groups because calibration performance degrades far more severely than discrimination performance on out-of-domain accents, even when EER remains stable \cite{Estevez_2023}. 
% Subgroup-specific or distribution-aware thresholds can substantially reduce such disparities in false acceptance and false rejection rates \cite{toussaint2022bias, Fenu_2021}.
Threshold calibration adjusts decision boundaries or score distributions after training to equalize error rates across demographic groups \cite{10.5555/3157382.3157469}.
In SV, a single global threshold disproportionately harms underrepresented groups: calibration performance degrades far more severely than discrimination performance on out-of-domain accents \cite{Estevez_2023}, and the choice of operating point further reshapes the fairness-security-usability trade-off across demographic groups \cite{fenu2020exploring}. 
Bias audits confirm that such disparities persist throughout the decision pipeline \cite{toussaint2022bias}.
Crucially, these disparities can be mitigated without modifying the underlying model. 
\citet[][preprint]{jain2021inclusive} choose a separate decision threshold for each gender and for children of different ages in a fixed SV system. 
At the same target false acceptance rate, this lowers the false rejection rate for the groups that a single shared threshold penalizes.
\citet{Estevez_2025} similarly apply group-specific posterior calibration to a fixed voice disorder detector, correcting age- and gender-dependent miscalibration that global metrics fail to reveal. 
Because only the post-hoc score-to-decision mapping changes, both methods are applicable at inference time to any pre-trained system.

\begingroup
\footnotesize
\setlength{\tabcolsep}{3pt}
\begin{xltabular}{\textwidth}{@{} >{\raggedright\arraybackslash}p{2.8cm} >{\raggedright\arraybackslash}p{1.5cm} >{\raggedright\arraybackslash}p{2.1cm} >{\raggedright\arraybackslash}p{2.1cm} >{\raggedright\arraybackslash}p{1.8cm} >{\raggedright\arraybackslash}p{1.7cm} >{\raggedright\arraybackslash}X @{}}
% \caption{Summary of mitigation strategies for speech fairness. \textbf{Paradigm} follows \S\ref{sec:expansion}: Rob.\,=\,robustness; Rep.\,=\,representation; Gov.\,=\,governance. \textbf{Bias Source} follows \S\ref{sec:source}: \textit{Data}\,=\,data-related bias (\S\ref{sec:data}); \textit{Entangl.}\,=\,feature entanglement (\S\ref{para:arch}); \textit{Capacity}\,=\,capacity limitations (\S\ref{para:arch}); \textit{Decoding}\,=\,decoding assumptions (\S\ref{para:decoding}); \textit{Deploy.}\,=\,deployment mismatch (\S\ref{sec:deploy}). \textbf{Supervision}: signal required beyond task labels. \textbf{Model Req.}: None\,=\,data-only; Post-hoc\,=\,applied to a trained model without retraining; Loss\,=\,loss change; Arch.\,=\,architecture change; Infer.\,=\,inference-time access.
% \caption{Summary of mitigation strategies for speech fairness.
% {\color{blue}Each mapping is marked by claim type. Paradigm~\Asym; Bias Source mixes \Msym, \Esym, and \Asym; Tasks and representative work~\Esym. \claimlegend}}
\caption{Summary of mitigation strategies for speech fairness.
In the Bias Source column, cell markers distinguish between explicit claims in the original papers and our own categorization.  \claimlegend\ Bias Source is the only column marked cell by cell. Paradigm is our interpretation throughout (~\Asym). Tasks and Rep.\ Work follow the cited papers (~\Esym). Column definitions and abbreviations are given in the table notes.}
% {\color{blue} Each mapping is marked with the same colored circle symbols as Table~\ref{tab:def_family_mapping} (the cell text itself is not colored): Paradigm~\Asym; Bias Source mixes \Msym, \Esym, and \Asym; Tasks and representative work~\Esym. \claimlegend}}
\label{tab:mitigation_summary}\\
\toprule
\textbf{Strategy} & \textbf{Paradigm} & \textbf{Bias Source} & \textbf{Supervision} & \textbf{Model Req.} & \textbf{Tasks} & \textbf{Rep.\ Work} \\
\midrule
\endfirsthead
\toprule
\textbf{Strategy} & \textbf{Paradigm} & \textbf{Bias Source} & \textbf{Supervision} & \textbf{Model Req.} & \textbf{Tasks} & \textbf{Rep.\ Work} \\
\midrule
\endhead
\bottomrule
\endfoot
\bottomrule
\multicolumn{7}{@{}p{\textwidth}@{}}{\footnotesize\raggedright
\textbf{Paradigm} (\S\ref{sec:expansion}), what the strategy treats as the source of unfairness.
\textit{Rob.}, robustness (distributional mismatch); \textit{Rep.}, representation (encoded social meaning); \textit{Gov.}, governance (power and accountability).\par
\textbf{Bias Source} (\S\ref{sec:source}), where the bias originates.
\textit{Entangl.}, feature entanglement; \textit{Deploy.}, deployment mismatch.\par
\textbf{Supervision}, the additional signal required beyond standard task labels. \textit{None}, no extra signal; \textit{Group labels}, speaker attribute metadata; \textit{Paired data}, parallel utterances differing only in sensitive attributes; \textit{Aux.\ model}, a separately trained model needed for the intervention. A slash marks a strategy for which the literature offers both a label-dependent and a label-free variant.\par
\textbf{Model Req.}, the level of model access the strategy assumes.
\textit{None}, data only; \textit{Post-hoc}, an already-trained model; \textit{Loss}, training with a modified objective; \textit{Arch.}, training with a modified architecture; \textit{Infer.}, access at inference time.}\\
\endlastfoot

\multicolumn{7}{@{}l}{\textit{A.\ Data-centric interventions}} \\[2pt]
Diversify \& expand         & Rob., Gov.  & Data~\Esym             & Group labels and Aux. model       & None     & ASR, SLU, SSD          & \cite{wang2023improving, leung24_interspeech, zhang22n_interspeech} \\
Strategic resampling        & Rob.        & Data~\Esym             & Group labels            & None     & ASR, SV, clinical          & \cite{dheram2022toward, Estevez_2023, elghazaly-etal-2025-fairness} \\
Data selection \& curriculum & Rob.       & Data~\Esym             & Group labels / None            & None     & ASR, SER, IC           & \cite{koudounas-etal-2025-privacy, kothawade-etal-2023-ditto} \\
Counterfactual augmentation & Rob., Rep.  & Data~\Esym\ / Entangl.~\Esym  & Paired data             & None     & ASR, SER               & \cite{Sarı_2021, fucci2023nopitch, Tsai_2025} \\
\midrule
\multicolumn{7}{@{}l}{\textit{B.\ Feature-level methods}} \\[2pt]
Geometric projection        & Rob., Rep.  & Entangl.~\Msym         & Group labels            & Post-hoc & ASR, SV                     & \cite{krishnan24_interspeech} \\
Feature masking and removal         & Rob., Rep.  & Entangl.~\Esym         & Group labels         & Post-hoc & ASR, clinical          & \cite{wang2023beyond, sameti2025accent} \\
Model compression           & Rob.        & Entangl.~\Esym         & None        & Post-hoc & SSL                    & \cite{xu2022can, lin2024social} \\
\midrule
\multicolumn{7}{@{}l}{\textit{C.\ Training-level algorithms}} \\[2pt]
Adversarial disentanglement & Rob., Rep.  & Entangl.~\Msym         & Group labels            & Arch.    & ASR, SV, SER, clinical & \cite{8462663, qin22cross, kim2025domain} \\
Contrastive alignment       & Rob., Rep.  & Entangl.~\Esym         & Group labels / None   & Loss     & ASR, IC, SER & \cite{han2021supervised, Kim_2025, lee2025dypcl} \\
Objective modification      & Rob.        & Data~\Esym\ / Entangl.~\Msym  & Group labels / None   & Loss     & ASR, SV, SER           & \cite{GAO202276, bartelds2026ctcdro, shelke2025fairness} \\
Explicit attribute modeling  & Rob., Rep. & Data~\Esym\ / Entangl.~\Esym         & Group labels            & Arch.    & ASR, SV, ST, SER, Cog. & \cite{dheram2022toward, Shen_2022, qu2026fairgate} \\
Targeted adaptation         & Rob.        & Data~\Esym         & Group labels / None & Arch.    & ASR                    & \cite{swain2025towards, hu25d_interspeech} \\
\midrule
\multicolumn{7}{@{}l}{\textit{D.\ Inference and post-processing}} \\[2pt]
Input adaptation            & Rob.  & Entangl.~\Asym\ / Deploy.~\Esym & Aux. model                 & Infer.   & SER, ASR                & \cite{electronics11101594, alhumud2024improving} \\
Representation intervention & Rob., Rep.  & Capacity~\Asym\ / Entangl.~\Esym & Paired data / None & Infer. & ASR, TTS               & \cite{yang2026steering, sun2026steering} \\
Parameter adaptation        & Rob.        & Deploy.~\Esym          & Group labels / None     & Infer.   & ASR                    & \cite{geng23_interspeech, shi25_interspeech} \\
Decoding \& rescoring       & Rob.        & Decoding~\Esym         & None       & Infer.   & ASR                    & \cite{10096820, southwell2024child} \\
LLM-driven correction       & Rob.        & Data~\Esym      & Aux. model       & Infer.   & ASR                    & \cite{Fatapour2025.08.29.25333548, xu25g_interspeech} \\
Threshold calibration       & Rob.        & Decoding~\Esym         & Group labels            & Infer.   & SV, clinical           & \cite{jain2021inclusive, Estevez_2025} \\
Prompting \& ICL            & Rob.        & Deploy.~\Asym          & Group labels      & Infer.   & ASR, TTS               & \cite{peng23d_interspeech, poon2026clarity} \\

\end{xltabular}
\endgroup

\subsubsection{Prompting and In-Context Adaptation}  
With the advent of SpeechLLMs, some disparities can be reduced by changing the input context rather than the model parameters. 
% Peng et al.~\cite{peng23d_interspeech} utilize discrete prompting, demonstrating that injecting task-specific tokens effectively activates dormant capabilities for code-switching and minority linguistic patterns. 
\citet{peng23d_interspeech} show that changing Whisper's prompt tokens, for example, specifying two language tokens at once, improves recognition of code-switched speech without any training
Advancing to demonstration-based learning, \citet{roll-etal-2025-context} and~\citet{10446502} show that prepending interleaved audio-text or speech-only exemplars enables robust few-shot adaptation for low-resource English and Chinese dialects, respectively. 
The choice of demonstrations also matters. TICL \citep[][preprint]{zheng2025ticl} retrieves the utterances whose transcripts are closest to a first-pass transcript of the test audio, which outperforms randomly chosen demonstrations and lowers WER on non-native and children’s speech.
% To address severe data scarcity, \textit{SMILE} \cite{hsu2024meta} meta-learns adaptation cues from unlabeled audio, while Li and Niehues leverage \textit{Multimodal In-Context Learning (MICL)} to facilitate cross-lingual transfer for unseen languages by synergizing speech and text contexts \cite{li2026multimodal}. 
Other methods fine-tune the model so that it learns to use such examples.
SMILE \cite{hsu2024meta} meta-trains Whisper to learn from transcribed exemplars, improving recognition on unsupported languages.
Similarly, \citet{li2026multimodal} fine-tune speech LLMs on high-resource languages and use paired speech-text examples to re-rank recognition hypotheses for endangered languages.
These few-shot approaches effectively adapt models to unseen speech varieties with minimal data, though existing evaluations focus on languages, dialects, and accents rather than demographic attributes like gender or age.
% Collectively, these mechanisms allow frozen models to dynamically align with diverse demographic subgroups during inference, offering a scalable and efficient alternative to computationally expensive fine-tuning. 
These ideas also extend to generation: CLARITY \cite{poon2026clarity} combines LLM-guided text localization with retrieval-augmented accent prompting to mitigate accent bias in zero-shot TTS.

\medskip
\noindent\textbf{Summary.}
% Table~\ref{tab:mitigation_summary} maps each strategy to the paradigm it serves (\S\ref{sec:expansion}), the bias source it targets (\S\ref{sec:source}), the supervision it requires, the level of model access it assumes, and the tasks on which it has been validated. 
% We subdivide model-related bias into feature entanglement, capacity limitations, and decoding assumptions because each gives rise to a distinct mitigation family; data-related and deployment bias are kept at the coarser level because their sub-sources do not differentiate mitigation strategies in the same way. 
% {\color{blue} 
% The same colored circle marks as in Table~\ref{tab:def_family_mapping} annotate these mappings: the Paradigm column is uniformly \Asym, Bias Source mixes \Msym, \Esym, and \Asym, and Tasks and representative work are \Esym.
% } 
Table~\ref{tab:mitigation_summary} groups all 19 strategies by the stage at which they intervene, on the data, the features, the training procedure, or the model at inference.
We subdivide model-related bias into feature entanglement, capacity limitations, and decoding assumptions because each {draws a different set of strategies}; data-related and deployment bias are kept at the coarser level because their sub-sources do not differentiate mitigation strategies in the same way.

Three patterns emerge. 
First, all 19 strategies serve robustness, but only seven also address representational fairness, and only one touches governance, reflecting the field's heavy tilt toward performance parity. 
Second, feature entanglement attracts the widest range of solutions across all four pipeline stages, while deployment mismatch remains under-addressed with only inference-time methods available. 
Third, no single stage is sufficient: data interventions cannot prevent biases introduced by the model itself, training-level algorithms frequently require demographic annotations that may be unavailable or privacy-sensitive, and inference-time methods can only partially compensate for a frozen model's learned biases. These complementary gaps motivate the open challenges discussed in Section~\ref{sec:challenge}

\subsection{Case Study: Fairness Evaluation and Mitigation in Speech Emotion Recognition}
The experimental setting of EMO-Debias \cite{Lin_2025_2} illustrates how the proposed framework connects fairness evaluation and mitigation in multi-label speech emotion recognition. The workflow begins by identifying speaker groups and examining gender distributions within emotion categories. Overall macro-F1 and Hamming accuracy are then considered alongside gender gaps in true-positive and false-positive rates. Controlled changes in data composition help examine the role of gender imbalance, while comparisons between baseline and debiased models, such as those trained with sample reweighting, show how mitigation affects both recognition performance and group disparities.
\label{sec:ser_case_study}

\section{Challenge and Future Work}
\label{sec:challenge}
% Our roadmap identifies key challenges arising from the conceptual gaps in our three paradigms, specifically \highlight{cascading harms}, \highlight{unauthorized data exploitation}, \highlight{uninterpretable model bias}, and \highlight{brittle generalization across non-stationary environments}. Rather than viewing these seven directions as a collection of parallel tasks, we conceptualize them as interconnected layers with functional dependencies, where foundational infrastructures catalyze increasingly granular model-level interventions.

We identify seven open directions, ordered by their dependencies.

Standardized reporting (\S\ref{sec:future_a}) and corpus governance (\S\ref{sec:future_b}) provide the empirical and legal prerequisites for subsequent technical scaling.
Without the high-fidelity metadata and non-extractive pipelines they establish, the field cannot reliably navigate the Pareto frontier (\S\ref{sec:future_c}) or leverage controllable data synthesis (\S\ref{sec:future_d}) without compromising user trust. 
These data-centric foundations, in turn, facilitate surgical model-level innovations: advancements in mechanistic interpretability (\S\ref{sec:future_e}) and test-time adaptation (\S\ref{sec:future_f}) depend on the robust, well-documented baselines established in preceding layers to enable precise, real-time interventions. 
Finally, fairness for generative AI (\S\ref{sec:future_g}) represents a holistic frontier that integrates these dependencies, necessitating a systemic rethinking of accountability across the entire sociotechnical pipeline.

\subsection{Standardized Reporting Protocols}
\label{sec:future_a}

Rigorous comparison across multidimensional fairness benchmarks depends critically on standardized reporting protocols. Without sufficient contextualization, fairness claims remain difficult to interpret and can be actively misleading~\cite{blodgett-etal-2020-language}. 
To address these ambiguities, the field necessitates the establishment of a systematic reporting framework~\cite{bender2018data} that directly targets the data-related bias sources identified in \S\ref{sec:data}, encompassing three categories of speech-centric metadata.

First, \highlight{recording context} must be documented to address channel bias (\S~\ref{sec:collection}). 
This includes channel characteristics and acoustic environments, such as whether speech is captured through telephony systems or studio-quality microphones. 
Second, \highlight{sociolinguistic profiles} are required to mitigate demographic underrepresentation and linguistic disparities (\S~\ref{sec:demo_rep}). These profiles should capture essential variables, including language variety and speaker proficiency, which substantially influence system behavior. Third, \highlight{annotator demographics} should be disclosed to surface potential annotation subjectivity (\S~\ref{sec:annotation}) and cultural alignment bias in the construction of ground truth labels~\cite{10.1145/3531146.3533216}. Standardizing these metadata forms enables researchers to distinguish genuine algorithmic improvements from dataset-specific artifacts, ensuring that \highlight{robustness} is evaluated with appropriate sociotechnical nuance.

\subsection{Ethical Corpus Governance}
\label{sec:future_b}

Building on the governance framework established in \S~\ref{ssec:governance}, which identified \highlight{data sovereignty} as a foundational governance requirement, future research must shift from conceptual advocacy toward technical and institutional implementation. 
While achieving robust performance requires diverse data, scaling collection without formal governance risks perpetuating the extractive practices and procedural unfairness identified in our systemic analysis. 
Addressing this imbalance requires operationalizing non-extractive data pipelines grounded in frameworks such as data trusts and the CARE principles~\cite{carroll2020care}.

Unlike static open-source datasets, data trusts function as institutional intermediaries that enable marginalized communities to retain ongoing consent and meaningful control over how their biometric data is used~\cite{delacroix2019data}. To balance such governance with the need for reproducible benchmarks, the field should prioritize the development of \highlight{federated evaluation protocols}~\cite{ezzeldin2023fairfed}, where models are audited within community-held environments rather than requiring centralized data aggregation. 

Finally, governance must extend to equitable benefit sharing through granular licensing and structural reciprocity. Rather than relying on generic open-source licenses, the field should adopt \highlight{community-specific licenses} that incorporate veto rights over high-stakes commercial applications or the use of Traditional Knowledge Labels~\cite{christen2015tribal} to assert ongoing authority over data derivatives.
Furthermore, economic value redistribution can be operationalized through royalty-based revenue models that treat community data contributions as compensable labor~\citep{arrieta2018labor}, or mandatory reinvestment of licensing fees into community-led language revitalization infrastructure, ensuring that the progress of Speech AI provides direct, tangible returns to the linguistic groups involved.

\subsection{Scalable Multi-Objective Evaluation}
\label{sec:future_c}

% A critical barrier to the deployment of fair speech systems is the lack of clarity regarding the inherent trade-offs between fairness and utility~\cite{10.1145/3287560.3287598}. 
% Current evaluations typically optimize for a single fairness criterion in isolation, without jointly considering how multiple equity objectives interact with system performance and operational costs such as inference latency and training resources. 
% This narrow framing obscures the \highlight{Pareto frontier} along which different fairness-utility configurations compete~\cite{martinez2020minimax}. 
% To move beyond this limitation, future research must shift toward the construction of \highlight{multidimensional benchmarks} that simultaneously quantify multiple fairness criteria alongside operational costs~\cite{Lin_2025_2}.

Fairness and utility trade off along several axes at once~\cite{10.1145/3287560.3287598}, including competing fairness criteria, task performance, and operational cost such as inference latency. 
Because most evaluations still report a single criterion in isolation, they hide the \highlight{Pareto frontier} on which fairness-utility configurations sit~\cite{martinez2020minimax}, motivating \highlight{multidimensional benchmarks} that measure several fairness criteria alongside utility and cost~\cite{Lin_2025_2}.

However, navigating this Pareto frontier is inherently search-intensive: identifying optimal fairness-utility configurations requires evaluating a large number of candidate solutions under multiple criteria simultaneously. 
Standard human-centric metrics such as $\Delta_{MOS}$ (\S\ref{sec:evaluation}), as well as subjective listening tests used to assess perceived bias and offensiveness across demographic groups~\cite{satish2026voice}, cannot support this throughput because they rely on diverse human evaluator panels that are prohibitively expensive to convene at scale. 
\highlight{Scalable proxies} for human judgment are therefore a functional precondition for tractable multi-objective search. 
Promising directions include reference-free fairness estimation models and LLM-based evaluators capable of approximating perceptual metrics directly from model outputs~\cite{liu2023g}, which would enable the iterative, gradient-based, or evolutionary searches required to discover and maintain optimal fairness-utility balances in complex speech architectures.

% However, navigating this Pareto frontier requires a high-throughput evaluation loop that standard human-centric metrics cannot support. While \S IV introduced metrics such as $\Delta_{MOS}$, and $C3T$ these remain prohibitively expensive and slow to compute at scale because they rely on diverse human evaluator panels. This operational bottleneck necessitates the development of \highlight{scalable proxies} for human judgment as a functional {precondition for multi-objective optimization. 

% Future work should focus on reference-free fairness estimation models or LLM-based evaluators capable of approximating these perceptual metrics directly from model outputs~\cite{liu2023g}. By automating the evaluation of subjective fairness, these proxies will enable the iterative gradient-based or evolutionary searches required to discover and maintain optimal fairness-utility balances in complex speech architectures.

\subsection{Controllable Data Synthesis}
\label{sec:future_d}

% The logistical burden and ethical risks of \highlight{extractive} data collection particularly when targeting marginalized communities necessitate a shift toward \highlight{controllable synthetic data generation}. Rather than relying on high-surveillance collection, researchers can harness disentangled representation learning to isolate linguistic content from specific speaker attributes such as accent, timbre, or pathological markers\cite{hsu2017unsupervised}. This approach allows for the re-composition of high-fidelity, diverse speech samples such as mapping dysarthric or regional features onto standard corpora without infringing upon the privacy of vulnerable groups.

The logistical burden and ethical risks of extractive data collection, particularly from marginalized communities~\citep{bird2020decolonising}, motivate a shift toward controllable synthetic data generation~\citep{leung24_interspeech}.
Disentangled representation learning can isolate linguistic content from speaker attributes such as accent, timbre, or pathological markers~\cite{hsu2017unsupervised}, enabling the synthesis of diverse, high-fidelity speech samples (e.g., mapping dysarthric features onto standard corpora) without compromising the privacy of vulnerable groups.

Such generative frameworks directly address the structural tension where strict data protection laws often starve fairness audits of the very demographic metadata they require \cite{veale2017fairer}.
Training on de-identified synthetic variations would let fairness work proceed without identifiable real-world individuals, a route face recognition has already taken in response to consent and data protection concerns~\citep{atzori2024impact}.
However, the success of this path hinges on navigating the \highlight{fidelity--fairness trade-off}: ensuring that synthetic augmentation does not oversimplify the non-linear acoustic nuances of marginalized speech, which are often lost in current generative approximations.

\subsection{Mechanistic Interpretability for Bias}
\label{sec:future_e}

Most existing bias mitigation strategies operate at a coarse level, providing limited insight into how bias is encoded within neural architectures. An important emerging direction is the application of \highlight{mechanistic interpretability} to speech models~\cite{plantinga2025black}. 
Techniques such as \highlight{causal tracing}~\cite{meng2022locating} and \highlight{activation patching}~\cite{zhang2024towards} can localize specific neurons or attention heads that encode stereotypical associations.
In text LLMs, causal mediation analysis has shown that gender bias is sparse, with much of its effect concentrated in a small proportion of neurons and attention heads~\citep{vig2020investigating}.
Speech has no counterpart yet.
Existing work either probes and erases gender information from ASR encoder layers
without linking it to model outputs~\citep{krishnan24_interspeech},
or attributes speech translation misgendering to input spectrogram regions~\citep{conti-etal-2026-voice}.
Neuron-level localization of stereotypical associations in speech models remains open.

However, applying these methods to speech requires addressing the continuous nature of acoustic signals and the multimodal complexity of modern SpeechLLMs. Unlike discrete text tokens, tracing bias in speech must account for temporal dynamics and disentangle whether discriminatory patterns reside in the acoustic encoder or the linguistic decoder. Once localized, these components can be selectively modified through \highlight{model surgery} or \highlight{rank-one editing}~\cite{meng2022locating}, enabling the removal of bias while preserving the model's general capabilities without exhaustive retraining.

\subsection{Test-Time Fairness Adaptation}
\label{sec:future_f}

Deployment mismatch (\S\ref{sec:deploy}) shows that a model trained on a fixed distribution cannot be assumed to generalize across shifting acoustic conditions, speaking styles, and user demographics~\cite{liu2021stable}.
Test-time adaptation (TTA) offers an alternative to costly retraining, but improving a model's overall accuracy in a new setting is not the same as improving it fairly for every group of speakers.
 
Current TTA methods such as TENT~\cite{wang2021tent} minimize prediction entropy to raise aggregate accuracy under shift, but these group-agnostic updates can widen the gap for marginalized subgroups rather than close it.
Closing this gap calls for test-time objectives that account for subgroup performance, not aggregate accuracy alone. 
Because demographic labels are often unavailable at inference, promising routes include unsupervised fairness proxies and per-speaker personalization that improve accuracy without sacrificing equity across groups.

\subsection{Fairness in the Era of SpeechLLMs and Generative AI}
\label{sec:future_g}

% The convergence of speech processing with large-scale generative models introduces systemic risks that extend beyond traditional discriminative tasks. Addressing these emergent challenges requires a multi-faceted research agenda focused on the following four dimensions:
Preceding sections have characterized fairness risks in generative speech systems, from representational stereotyping in TTS (\S\ref{ssec:representation}) to content-level bias in SpeechLLMs (\S\ref{ssec:generative_bias}). Current frameworks, however, largely evaluate these risks in isolation: on single outputs, in single turns, and after training is complete. Closing these gaps requires research along three directions.

\textbf{Alignment Bias and Cultural Homogenization:}
% As SLMs\cite{borsos2023audiolm} are optimized via RLHF~\cite{ouyang2022training} or DPO~\cite{rafailov2023direct}, they risk aligning with a narrow set of dominant cultural norms. 
% Current preference data often underrepresents minority linguistic varieties, potentially leading to \highlight{cultural homogenization}, where models default to standardized prestige accents and suppress regional dialectal variation. 
% Future research must investigate multi-objective alignment techniques, such as distributionally robust optimization across multiple fairness objectives~\cite{jung2023reweighting}, that jointly optimize across competing fairness criteria and system utility, rather than treating any single dimension in isolation.
\S~\ref{ssec:representation} and~\ref{sec:data} documented how training data and model outputs can privilege standardized linguistic norms. Post-training optimization may introduce an additional homogenization mechanism: supervised fine-tuning teaches models to respond in a narrow range of speaking styles, while preference optimization via RLHF~\cite{ouyang2022training} or DPO~\cite{rafailov2023direct} further reinforces whichever style human annotators rate most favorably. In text-based LLMs, this effect is empirically established: alignment has been shown to significantly reduce output diversity~\cite{kirk2024understanding} and concentrate responses into a small number of semantic clusters~\cite{liu2026alignment}. Because annotator judgments of speech quality are difficult to disentangle from preferences for prestige varieties, the same optimization pressure may systematically penalize dialectal prosody and code-switching in SpeechLLMs, even when pre-training data includes such variation. Whether and to what extent this \highlight{cultural homogenization} occurs in current speech systems remains an open empirical question: existing speech alignment work~\cite{park2024speechalign} evaluates quality but does not measure acoustic or linguistic diversity before and after optimization. Establishing this evidence base is a prerequisite for designing effective interventions.

% \textbf{Stochastic Fairness Evaluation:} 
% Unlike deterministic ASR systems, generative speech models exhibit inherent stochasticity, meaning the same prompt can yield varied fairness outcomes across different decoding runs. This randomness, coupled with the observation that generative bias in ITTS is often multi-faceted and non-unidirectional~\cite{chen2026bindingeffect}, significantly complicates safety assessments and reproducibility. Consequently, a critical research direction is the development of \highlight{distributional fairness metrics} that evaluate the statistical properties of a model's output space over multiple samples, rather than relying on 1-best hypotheses. This includes measuring the variance of toxicity or stereotyping across different sampling temperatures and decoding strategies.
\textbf{Distributional Fairness in Generative Speech.}
Generative SpeechLLMs define a distribution over possible outputs, and this distribution may contain biased modes whose manifestation depends on the interaction of multiple input factors~\cite{chen2026bindingeffect}. 
In text LLMs, decoding parameters such as temperature and sampling strategy have been shown to modulate absolute bias scores~\cite{das2022quantifying}, and models that appear fair under standard metrics can exhibit systematic bias in their confidence distributions~\cite{wang2025fairlycertain}.
These findings suggest that sample-level fairness audits are insufficient for generative systems. 
Evaluation must move to \highlight{distributional} analysis, characterizing whether biased outputs concentrate in the distribution's tails and whether their probability mass shifts across demographic inputs. 
In speech, the continuous acoustic output space makes such characterization harder than in discrete text generation.

% \textbf{Multi-turn and Agentic Fairness:}
% Speech-based agents often operate in long-form, multi-turn interactions where biases can accumulate or compound over time. A model might appear fair in a single-turn response but exhibit \highlight{systemic exclusion} by gradually shifting toward authoritative or subservient personas based on the user's perceived gender or social status\cite{fossa2022gender}. Future work should prioritize recursive fairness auditing, analyzing how human-AI power dynamics evolve through conversational history and whether agentic goals (e.g., task completion) lead to the reinforcement of social hierarchies.
\textbf{Multi-turn and Agentic Fairness.}
In spoken dialogue, acoustic cues such as accent and perceived gender are present in every turn, continuously exposing the model to speaker identity signals that text-based interactions do not carry. Whether this repeated exposure causes demographic-dependent behavior to intensify across turns remains an open empirical question: while multi-turn capability degradation in LLMs has been reported in a preprint~\cite{laban2025llmsgetlost}, its interaction with fairness metrics has not been studied for spoken conversations. 
Future work should develop turn-level fairness tracking methods that measure whether disparities in response quality, verbosity, or tone widen as conversations progress.
 
Separately, as SpeechLLMs are deployed as goal-directed agents that take real-world actions such as scheduling or customer service, fairness must extend beyond conversational content to task outcomes. Whether completion rates, response latency, and error recovery differ systematically across speaker demographics remains a critical open question~\cite{li2025actions} whose answer will determine whether agentic speech systems amplify or mitigate existing inequities.

% \textbf{Alignment-aware Fairness Frameworks:}
% Existing evaluation frameworks are often disconnected from the alignment loop. There is a pressing need for \highlight{alignment-aware fairness evaluators} that can be integrated directly into the training process. One potential pathway is the implementation of Constitutional AI\cite{bai2022constitutional} for Speech, where a critic model is trained to supervise and correct generative outputs based on a set of explicitly defined fairness principles. By embedding fairness as a core objective within the alignment phase, the community can move toward generative systems that are inherently equitable rather than merely post-hoc calibrated.

\section{Conclusion}
\label{sec:conclusion}
This survey has organized the fragmented landscape of speech fairness research into a unified framework spanning formal definitions, evaluation metrics, bias sources, and mitigation strategies.
Three insights emerge from viewing these layers together.
 
First, speech fairness is not a single problem but three coexisting paradigms (Robustness, Representation, and Governance), each diagnosing unfairness at a different level.
A system that achieves performance parity across demographics may still perpetuate social stereotypes in its outputs or operate within extractive data practices that deny communities agency over their linguistic resources.
Recognizing this multiplicity is a prerequisite for comprehensive auditing.
 
Second, there is a persistent gap between what we can measure and what we can fix.
The fairness definitions formalized in this survey give rise to six mathematical families of evaluation metrics, yet many metrics lack corresponding mitigation strategies, and popular interventions such as resampling address robustness but leave representational and governance concerns untouched.
The definition-to-family mapping (Table~\ref{tab:def_family_mapping}) makes these gaps explicit and can guide future work toward currently unaddressed dimensions.
 
Third, speech presents unique challenges that distinguish it from fairness work in text or vision.
Sensitive attributes are acoustically entangled with linguistic content, making them difficult to isolate without degrading utility.
Many speech targets are inherently perceptual, so fairness requires scrutinizing annotation protocols and evaluator demographics, not just model behavior.
 
Section~\ref{sec:challenge} details the open challenges and research directions that follow from these observations.
We hope this survey helps the community move from isolated, task-specific audits toward reproducible, diagnosis-driven fairness practice across the speech pipeline.

\section*{Limitations}
\label{limitation}
\label{sec:limitations_and_discussion}

\textbf{The underlying evidence is concentrated in high-resource languages.}
The studies, benchmarks, and demographic annotations we build on are dominated by English and a small set of other high-resource languages, because that is where speech-fairness research currently exists. Our conclusions therefore describe disparities as they have been measured in high-resource settings, and the behavior of speech systems for the world's low-resource and under-documented languages remains under-evidenced, not because it matters less but because the primary studies have not yet been done. This gap is one of the representational inequities the survey identifies, and it bounds the generality of every cross-lingual claim we make.

\textbf{Our analysis inherits the identity axes that the primary literature chose to measure.}
A survey of disparities can only be as fine-grained as the studies it synthesizes. Much of the speech-fairness literature records gender as binary, race or first-language background in coarse groupings, and omits attributes such as age, disability, and socioeconomic status, in part because collecting and labeling them raises the very privacy and consent tensions we discuss. Our diagnosis therefore reflects the axes of identity that existing corpora make visible, not the full space of speakers who can be disadvantaged. When we report a specific effect, such as the annotator-gender gap in emotion labels, we are simultaneously reporting the boundary of what current annotations allow anyone to observe.

\textbf{Our framework is one principled organization, not the only possible one.}
The three paradigms (Robustness, Representation, and Governance) and the seven definitions are an analytical construction, chosen because they make one specific gap visible: the mismatch between what current metrics measure and what mitigations actually fix. The categories are deliberately not mutually exclusive. Individual studies often span paradigms, and some metrics map to more than one definition. We use these boundaries as a reasoning tool rather than a partition of the field, and a different decomposition could foreground tensions that ours keeps in the background.

\textbf{The generative frontier moves faster than any survey.}
The portions of this survey concerned with SpeechLLMs and generative audio describe a rapidly moving target, where both the systems and the fairness behaviors they exhibit change on the timescale of months. We wrote the framework to outlast specific systems: the definitions, the three paradigms, and the pipeline-level diagnosis are intended to remain applicable as models change. Specific empirical findings about today's SpeechLLMs, however, should be read as a snapshot at the time of writing rather than a settled account.

% Acknowledgments are omitted during double-blind review;
% add them only once the submission is accepted and deanonymized.

\section*{Acknowledgments}
This work was supported by the Ministry of Education (MOE) of Taiwan under the project Taiwan Centers of Excellence in Artificial Intelligence, through the NTU Artificial Intelligence Center of Research Excellence (NTU AI-CoRE).
This work was supported in part by the National Science and Technology Council (NSTC), Taiwan, under Grant No. 114-2917-I-564-030 (to Huang-Cheng Chou). 

\bibliography{reference}
\bibliographystyle{tmlr}

\end{document}